\documentclass[acmsmall,nonacm]{acmart}
\usepackage[utf8]{inputenc}
\usepackage{enumitem}
\usepackage{colortbl}
\usepackage{soul,xcolor}
\usepackage{amsmath,bm}
\usepackage{lipsum}
\usepackage[htt]{hyphenat}
\usepackage{cuted,tcolorbox,lipsum}
\definecolor{gray}{rgb}{0.83, 0.83, 0.83}
\definecolor{LightCyan}{rgb}{0.88,1,1}
\usepackage{array}
\newcommand{\PreserveBackslash}[1]{\let\temp=\\#1\let\\=\temp}

\newcolumntype{C}[1]{>{\PreserveBackslash\centering}p{#1}}
\newcolumntype{R}[1]{>{\PreserveBackslash\raggedleft}p{#1}}
\newcolumntype{L}[1]{>{\PreserveBackslash\raggedright}p{#1}}

\definecolor{gray}{rgb}{0.83, 0.83, 0.83}
\definecolor{LightCyan}{rgb}{0.88,1,1}

\newcommand\PaperTotal{98 }

\newcommand\PapaerSourceCodeDL{94 }

\usepackage{longtable}
\usepackage{booktabs}
\usepackage{multirow}
\newcommand{\dgray}{\cellcolor{gray!50}}

\usepackage{array}

\title{AI-Based Software Vulnerability Detection: \\A Systematic Literature Review}

\acmJournal{CSUR}

\author{Samiha Shimmi}
\affiliation{%
  \institution{Northern Illinois University}
  \streetaddress{100 Normal Rd}
  \city{DeKalb, IL}
  \country{USA}}
\email{sshimmi@niu.edu}

\author{Hamed Okhravi}
\affiliation{%
  \institution{MIT Lincoln Laboratory}
  \streetaddress{244 Wood Street}
  \city{Lexington, MA}
  \country{USA}}
\email{hamed.okhravi@ll.mit.edu}
\author{Mona Rahimi}
\affiliation{%
  \institution{Northern Illinois University}
  \streetaddress{100 Normal Rd}
  \city{DeKalb, IL}
  \country{USA}}
\email{rahimi@cs.niu.edu}

\date{January 2022}

\begin{document}

\begin{abstract}

Software vulnerabilities in source code pose serious cybersecurity risks, prompting a shift from traditional detection methods (e.g., static analysis, rule-based matching) to AI-driven approaches. This study presents a systematic review of software vulnerability detection (SVD) research from 2018 to 2023, offering a comprehensive taxonomy of techniques, feature representations, and embedding methods. Our analysis reveals that 91\% of studies use AI-based methods, with graph-based models being the most prevalent. We identify key limitations, including dataset quality, reproducibility, and interpretability, and highlight emerging opportunities in underexplored techniques such as federated learning and quantum neural networks, providing a roadmap for future research. 

\end{abstract}




\maketitle

\section{Introduction}
\label{sec:introduction}
Software vulnerabilities have been an ever-growing problem over the past decades~\cite{cvedetails}, and despite the vast body of work on software vulnerability detection (SVD) techniques (a.k.a. sanitization techniques)~\cite{song2019sok}, new and more effective techniques are still needed in this area. Software vulnerabilities can have large-scale, damaging impacts as demonstrated, among others, by the infamous vulnerabilities in Apache Log4j2 \cite{CVElog1, CVElog2} that affected more than 35K Java packages \cite{zhang2023comparing}. Such vulnerabilities are often identified and enumerated in databases such as the Common Vulnerabilities and Exposures (CVEs) \footnote{\url{https://cve.mitre.org/}} for individual vulnerabilities, and Common Weakness Enumeration (CWE) \footnote{\url{https://cwe.mitre.org/}} for vulnerability categories. According to the CVE database, the number of reported CVEs has grown substantially over time. In 1999, only 321 CVE records were reported, whereas in 2023, the total reported record has increased to 29k \cite{cve}. This substantial increase demonstrates, despite decades of research in this area, effective and scalable SVD techniques are still a major need for cybersecurity.

Perhaps unsurprisingly, with the recent significant growth in artificial intelligence (AI)-based techniques and their application to security, AI has been applied to the SVD as well. Transformers, first introduced in 2017, revolutionized the field by enabling AI models to learn patterns in sequential data, such as text or code, and handle tasks like translation or summarization \cite{vaswani2017attention}. BERT, introduced in 2018, builds on transformers by understanding the meaning of words within their surrounding context, making it particularly powerful for natural language processing tasks \cite{devlin2018bert}. Large language models (LLMs) have pushed this capability even further, generating human-like text and answering complex questions \cite{brown2020gpt3}. Graph Neural Networks (GNNs) \cite{scarselli2008graph} provide another dimension by learning from data structured as graphs, such as code dependency networks or social connections, enabling the detection of intricate patterns and relationships . Self-supervised learning allows AI models to identify useful structures in data without relying on manually labeled examples, making it possible to leverage vast, unlabeled datasets \cite{ibm_self_supervised}. Automated Machine Learning (AutoML) \cite{feurer2015efficient} simplifies the process of building and tuning models, enabling non-experts to harness machine learning’s power. Explainable AI (XAI) \cite{ribeiro2016should} ensures that the decisions made by AI systems are transparent and understandable, fostering trust and enabling humans to interpret results. Federated Learning (FL) \cite{mcmahan2017communication}, introduced in 2017, offers a privacy-preserving way to train AI models directly on users’ devices rather than on centralized data servers, reducing the need to share sensitive information \cite{mcmahan2017communication}. 

Although several conventional and systematic literature reviews exist in the SVD domain \cite{eberendu2022systematic,nazim2022systematic,6405650,MALHOTRA2015504,10.1145/2187671.2187673,7866201}, and were insightful in their time, they do not fully capture this significant, recent body of literature on AI-based SVD. According to Hanif et al. \cite{p126}, traditional approaches such as static analysis, dynamic analysis, pattern matching, taint analysis, and statistical methods represented roughly 25.6\% of the reviewed studies. This share is relatively modest, and since earlier surveys have already addressed these conventional methods \cite{pistoia2007survey,liang2018fuzzing,zaazaa2020dynamic}, our focus is on more recent advancements, particularly those leveraging cutting-edge ML and DL-based SVD techniques.

Recent surveys like \cite{p126,liu2019survey,zeng2020software,nazim2022systematic,eberendu2022systematic} exist but not without limitations. Some existing surveys \cite{zeng2020software,liu2019survey,nazim2022systematic} are conventional and do not follow systematic literature review principles. They often lack specific inclusion and exclusion protocols, resulting in missing critical information such as the number of papers reviewed, the year range considered, and potentially overlooking relevant studies. Consequently, these surveys may not provide comprehensive data necessary for drawing robust conclusions. Even for systematic reviews, the number of papers analyzed is small in some cases. For instance,  Nazim et al.~\cite{nazim2022systematic} only examined 10 papers, and thus the data provided, although helpful, might not depict the whole picture.

Another gap in the existing surveys is a systematic taxonomy. Although several existing surveys \cite{eberendu2022systematic,p127} provided different types of basic categorization, in order to know the current practices and find the gap in the existing research, it is essential to have a comprehensive categorization based on several dimensions such as techniques used, feature representation method used, embedding methods used, etc. A comprehensive taxonomy would allow researchers to grasp the landscape of AI techniques and methodologies, facilitating the identification of trends, innovations, and under-explored areas. Additionally, it provides a structured framework for comparing different approaches, enhancing the clarity and depth of future research endeavors. One existing work worth mentioning is Hanif et al. \cite{p126} where the authors provided a detailed taxonomy. However, the perspectives they selected were based on research interest and also the categorization provided for ML-based approaches only cover four basic categories which are insufficient for more comprehensive analysis.  Similarly, Harzevili et al. \cite{10.1145/3699711} provide a broad taxonomy covering various data types and model categories. However, a more detailed and fine-grained classification is still needed, particularly for deep learning–based approaches. Our three-dimensional analysis—spanning DL techniques, feature representation methods, and embedding strategies—extends prior work by introducing a more structured and comprehensive taxonomy tailored to DL-based vulnerability detection in source code. This framework also enables us to highlight insights and challenges for future research in the field.

Considering these gaps, we conducted a systematic literature review with specific inclusion and exclusion criteria, analyzing \textbf{\PaperTotal} recent papers published between 2018 and 2023 focused on \textbf{AI-based source code vulnerability detection}. Our initial analysis involved characterizing the approaches along several dimensions: the availability of their data and models, their primary detection objective, whether they aimed solely to classify code as vulnerable or non-vulnerable or pursued fine-grained analysis, the programming languages targeted, and the level of granularity employed in the detection process.

We then created a detailed taxonomy that categorizes the reviewed approaches based on several key factors, including the specific techniques used, such as neural networks or graph-based methods, the ways the key features of code are identified and represented (e.g., as sequences or graphs), and the embedding methods employed to translate the code into numerical formats that can be processed by machine learning models. The chosen time range in this paper, targets the most recent papers in the domain, addressing the rapid advancements in AI-based techniques. Our taxonomy aids in understanding the current approaches, identifying their limitations, and providing future guidelines. Additionally, we systematically characterize dataset information, documenting all datasets used in the collected papers, along with recently released data within this time frame, and highlight dataset issues for future researchers to address. Unlike existing surveys, our approach is more systematic, covering a broad range of recent papers and offering a well-defined taxonomy that aligns with the latest advancements in AI-based techniques for vulnerability detection in source code. Providing this structured overview aims to facilitate future research directions in this area, particularly for emerging researchers in the field. A brief overview of our key contributions include:

\begin{itemize}
\item \textbf{Comprehensive Taxonomy:} We have developed a comprehensive taxonomy that considers various aspects of applying AI models for the purpose of vulnerability detection, including the techniques used, feature representation methods, and embedding strategies. This taxonomy aids in understanding the application of AI-based approaches highlighting their limitations and guiding future research directions.

\item \textbf{Systematic Dataset Characterization:} We have investigated and analyzed the datasets used in each of the collected papers. We documented the characteristics of each dataset, including the most recently released ones, and highlighted their potential challenges, such as their limitations and interpretability of the models developed based on these datasets, with the purpose of guiding future research in this domain.

\item \textbf{Focus on Recent Advancements:} By focusing on recent years (2018–2023), our review captures the surge in AI-based vulnerability detection techniques and offers a more up-to-date perspective on the field, addressing a gap that was missing in previous literature reviews.

\item \textbf{Identification of Basic Characteristics:} We have collected and analyzed key characteristics of existing approaches, including data and model availability, detection outcome, whether the models are designed to classify vulnerable and non-vulnerable code or aim for more fine-grained analysis, the programming languages targeted, granularity levels, and other relevant properties. This detailed examination highlights gaps and limitations in current methods, paving the way for more focused and effective future research.

\end{itemize}

The research questions explored in this survey paper are:

\begin{itemize}[leftmargin=*]
\item \bm{$RQ_{1}$} \textbf{(Datasets): 
What datasets are prevalent in current vulnerability detection research and what are their primary characteristics?}

Identifying prevalent datasets is crucial as it highlights the most commonly used benchmarks in the field. Exploring the essential characteristics of datasets, such as their size, diversity, origin, and format, provides insights into their strengths and limitations. Understanding these attributes can guide researchers in selecting appropriate datasets, ensuring their studies are robust and generalizable.

  \vspace{3pt}

\item \bm{$RQ_{2}$} \textbf{(Basic Characteristics): 
What are the main characteristics of source code-based vulnerability detection approaches?}
    
     Understanding the principle characteristics of code-based vulnerability detection approaches, such as their granularity level, programming languages used, type of classifiers (binary or multi-label), and other relevant attributes, provides a foundation for categorizing and comparing multiple solution in this problem domain. This RQ can be further broken down into the following RQs:
     \begin{itemize}[leftmargin=*]
         \item \bm{$RQ_{2a}$} \textbf{(Data/Model Availability):} What is the availability of the datasets and models used in source code-based vulnerability detection approaches?
\item \bm{$RQ_{2b}$} \textbf{(Detection Objective):} Do these approaches focus solely on identifying whether code is vulnerable or do they aim for a more detailed, fine-grained analysis?

\item \bm{$RQ_{2c}$} \textbf{(Programming Languages Used):} Which programming languages are targeted by these vulnerability detection approaches?

\item \bm{$RQ_{2d}$} \textbf{(Granularity Level):} What levels of granularity (e.g., statement-level, function-level, file-level, etc.) are considered in these approaches for the purpose of identifying the vulnerabilities?

     \end{itemize}

  \vspace{3pt}
\item \bm{$RQ_{3}$} \textbf{(Taxonomy): How can we construct a comprehensive taxonomy of existing source code-based vulnerability detection approaches?}
       
Constructing a comprehensive taxonomy of vulnerability detection approaches helps to systematically categorize and compare different methods. This is important for understanding the landscape of recent techniques, identifying the impact of recent advancements in ML, DL, and AI, and uncovering trends and gaps in the literature.
  \vspace{3pt}
 \item \bm{$RQ_{4}$} \textbf{(Limitations and Future Directions): 
    What are the limitations of current vulnerability detection methodologies, and what recommendations can be proposed to address these shortcomings and guide future research efforts?}
        
          Identifying the limitations of current methodologies is essential for recognizing areas needing improvement. By understanding these shortcomings, such as issues with datasets, models, researchers can focus on addressing these challenges. Proposing recommendations and future research directions fosters innovation and guides the development of more effective and efficient vulnerability detection techniques. 
\end{itemize}

  \vspace{3pt}

The rest of the paper is organized as follows. In Section \ref{sec:related}, we discuss the current research work in the SVD domain and differences between these approaches and ours. Section \ref{sec:ReviewMethod} describes our methodology, where we explain how we conducted our survey, including the inclusion and exclusion criteria for selecting papers. Section \ref{sec:dataset} explains all the detailed characteristics of the existing datasets and answers RQ1. In Section \ref{sec:detection}, we describe the basic characteristics of the primary studies and answer RQ2. In Section \ref{sec:taxonomy2}, we present a taxonomy of the existing research areas related to code-based detection and and thoroughly examine the research contributions in this category to address RQ3. Section \ref{sec:Future} highlights the limitations and challenges of current approaches and outlines future research directions to address RQ4. Finally, we summarize the key findings of our survey in Section \ref{sec:conclusion}.
\section{Related Work}
\label{sec:related}

This section provides an overview of related literature reviews and other related studies in this area.

A brief overview is provided in Table \ref{tab:Related1} where each paper contains the corresponding category, the period for which the study was conducted (if provided by the authors), the total number of papers surveyed (if provided by the authors), and the key concept of the paper. We  categorize the existing work into two categories, namely literature reviews and other Analytical and Comparative Approaches, based on their focus and methodology. Each category is described in the following subsections.

\subsection{Literature Reviews}
This subsection includes papers that either conducted systematic literature reviews with defined inclusion and exclusion criteria, paper counts, year ranges, and other specific details, or conventional literature reviews where such statistics are not explicitly provided.

Hanif et al. \cite{p126} conducted their literature review based on 90 papers ranging from 2011-2020, where they developed two independent taxonomies. The first taxonomy categorized existing works based on different types of research interests, grouping them into categories with respect to their methods, detection, features, code, and datasets. In essence, this taxonomy reflects how the reviewed studies were structured around these high-level categories of research focus. In the second taxonomy, they categorized ML-based approaches into four main categories namely supervised, unsupervised, ensemble, and deep learning. Additionally, Ghaffarian and  Shahriari \cite{p127} reviewed ML and data mining techniques in SVD domain and provided a high-level categorization. In their conventional survey, they provided a summary of several papers across various categories, including machine learning and data mining techniques, software metric-based approaches, anomaly detection methods, and others.

Harzevili et al. (2024) \cite{10.1145/3699711} recently conducted a comprehensive systematic literature review, offering broad insights into the use of machine learning for software vulnerability detection across diverse data types—including source code, binaries, and commit metadata—over an extended period (2011–2024). In contrast, our work differs in both scope and depth. We focus exclusively on AI-driven techniques for source code vulnerability detection, specifically within the dynamic and rapidly advancing period of DL-based methods from 2018 to 2023. This narrower focus enables us to construct a fine-grained taxonomy across three key dimensions: detection methods, feature representation techniques, and embedding strategies. Additionally, we provide a detailed analysis of dataset characteristics such as programming language, granularity, metadata availability, and detection objectives. As such, our study complements and extends the prior review by delivering a more focused and in-depth classification, along with actionable insights to guide future research in AI-based software vulnerability detection.

\begin{table}[]
\renewcommand{\arraystretch}{1.05}
 \small
  \caption{A Brief Overview of Related Literature Reviews}
 \vspace{-10pt}
  
  \centering

 \begin{tabular}{p{1.9cm} p{2cm} p{1.35cm}p{0.9cm}p{1.15cm}p{4.6cm}}

 \hline

\textbf{Paper}& 
\textbf{Category} &
\textbf{Period} &
\textbf{\#Papers} &
\textbf{Published} &
\textbf{Key Concept}\\\hline

 Hanif et al. \cite{p126} & Vulnerability Detection    & 2011-2020 &90&2021& Presented two separate taxonomies in SVD based on research interest and approaches
\\

Ghaffarian and  Shahriari \cite{p127}  & Vulnerability Detection   &  -&-&2017& Provided bassic categorization of efforts in SVD domain \\

Zeng et al. \cite{zeng2020software} & Vulnerability Detection & -&-&2020& Identified and discussed four game-changing papers and discussed there impact  \\

Liu et al.~\cite{liu2019survey} & XSS Vulnerability Detection  &-&-&2019&   Discussed classification of XSS attack \\

Lomio et al.~\cite{lomio2022just}  & Vulnerability Detection &- &-&2022&    Investigated how Machine Learning is assisting developers to detect vulnerabilities  \\

 Zhu et al. \cite{zhu2022application}& Vulnerability Detection   & 2016-2017&48&2022&Studied how perception gap can be reduced    \\

   Chakrobrty et al. \cite{9448435} & Vulnerability Detection  &-&-& 2022&  Conducted a survey to see how existing DL-based vulnerability detection methods work in a real word dataset \\

   Zheng et al. \cite{zheng2021representation} &- &-&-&2021& The authors studied how ML strategies influence vulnerability detection in source code   \\

   Harzevili et al.\cite{10.1145/3699711} &Vulnerability Detection  & 2011-Jun'24 &138   &2024  & Studied ML-based approaches to uncover publication trend, understand the dataset, along with representation, provide architectural classification of models, discover popular vulnerability type explored etc.   \\
\hline
\dgray{This work}  & \dgray{Vulnerability Detection}  & \dgray{2018- 2023}   & \dgray{\PaperTotal} &  \dgray{TBD}&\dgray{Using a systematic approach, documents basic characteristics of existing datasets, provides a comprehensive taxonomy of the recent source code based vulnerability approaches}\\
   \hline

\end{tabular}
\label{tab:Related1}
 %\vspace{-25pt}
\end{table}

Zeng et al. \cite{zeng2020software} conducted their survey based on four recent efforts which they referred to as “game changer”.
While their approach identifies critical research directions emerging from these "game changers," their survey is distinct from ours in terms of scope and methodology. They primarily focused on highlighting key innovations from these four papers, offering insights into potential future directions. In contrast, our study aims to create a more systematic literature review, encompassing a broader range of research within the SVD domain. 

Additionally, some surveys focus on specific types of attacks, such as the work by Liu et al.~\cite{liu2019survey}, which compares and analyzes techniques solely aimed at detecting XSS vulnerabilities.

Compared to the above-mentioned surveys, we provide a more comprehensive analysis. We deliberately target recent advancements in AI within the timeframe of 2018 to 2023 (inclusive), capturing the surge in AI-powered methods that are reshaping the SVD landscape. Furthermore, we contribute a novel and detailed taxonomy that goes beyond the existing classifications. This taxonomy analyzes AI-based approaches from multiple perspectives, considering the techniques used, feature representation methods for vulnerability characterization, and the role of embedding techniques in transforming code for DL models.

\subsection{Analytical and Comparative Approaches}
\label{sec:relatedEmp}

This section includes papers that are not strictly systematic literature reviews but compare various approaches or conduct studies to gain insights into current methods. While these papers contribute to a deeper understanding of existing approaches, they differ from our work in that their primary goal was not to conduct a more systematic literature review with specific inclusion and exclusion criteria for a defined range.
 
For example, Lomio et al.~\cite{lomio2022just}  investigated how the existing ML-based SVD mechanism supports the developers in commit-level detection by considering only 9 projects, a selection whose justification is not well explained. \cite{zhu2022application} conducted a study to re-execute a set of DL-based methods and reported a significant drop in their performance compared to their original experimental results. They referred to this as a perception gap and explored ways to reduce it. Similarly, Chakraborty et al.~\cite{9448435} conducted a study to evaluate the effectiveness of existing DL-based vulnerability detection methods on real-world datasets. Their major finding is that the existing DL-based vulnerability detection methods, while promising in controlled settings, often struggle to maintain their effectiveness when applied to real-world datasets.

Zheng et al.~\cite{zheng2021representation} conducted an experiment to evaluate the impact of various ML strategies on vulnerability detection outcomes. The results indicated that the attention mechanism played a crucial role in detecting vulnerabilities, while transfer learning did not improve model performance. Their analysis revealed that composite code representations achieved the best results. In contrast to our work, their study had a more specific focus, analyzing the effects of different ML strategies within the SVD domain.

\section{Methodology}
\label{sec:ReviewMethod}

We followed the widely used PRISMA (Preferred Reporting Items for Systematic Reviews and Meta-Analyses) \cite{primsma} principle to conduct the systematic literature review. 
PRISMA contains a set of guidelines designed to help researchers report systematic reviews thoroughly, ensuring reproducible reporting of the review process. It has emerged as a widely adopted methodology for conducting surveys, with the paper being cited in over 12,000 articles at the time of writing this paper. 

According to the original PRISMA statement, there are four primary phases in a systematic review process, namely \textit{identification}, \textit{screening}, \textit{eligibility}, and \textit{included} phase. The identification phase involves searching for potential studies from various sources (e.g., databases, reference lists). During screening duplicate records are removed, and the remaining studies are assessed based on predefined inclusion and exclusion criteria, typically focusing on titles and abstracts. Full-text articles of studies that passed the screening phase are reviewed in detail, during the eligibility phase,  to confirm their relevance and compliance with the study criteria. The final phase lists the studies that meet all criteria and are included in the systematic review for further detailed analysis and synthesis.

\begin{figure*}[t!]
\centering
    \includegraphics[scale=0.45]{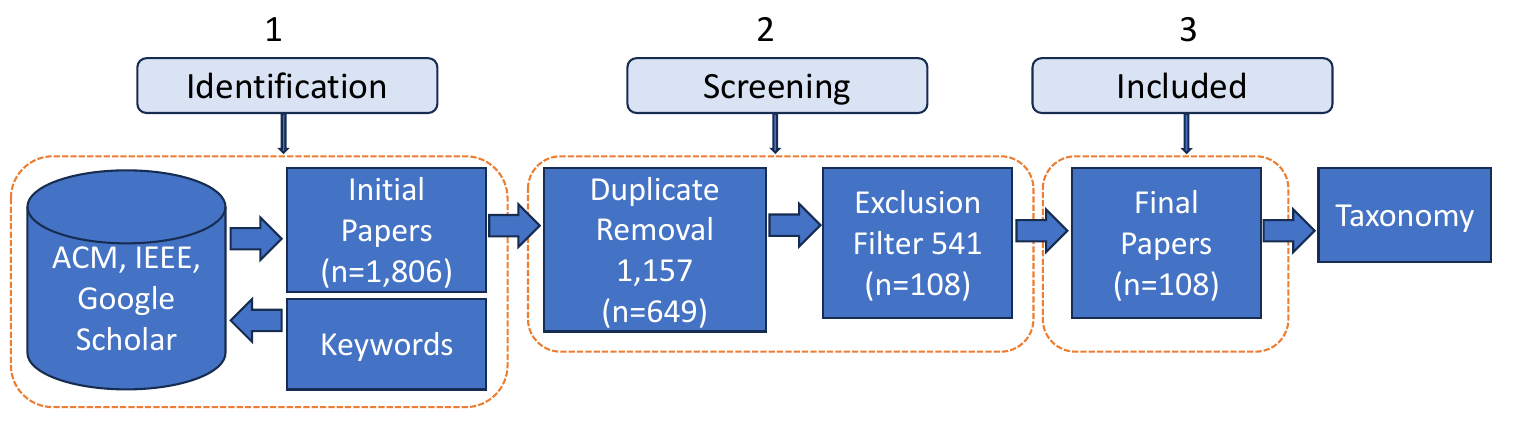}
\vspace{-10pt}
    \caption{Basic workflow of the proposed approach based on PRISMA method}
\vspace{-5pt}
    \label{fig:overview}
\end{figure*}

In the revised PRISMA methodology\footnote{\url{https://www.eshackathon.org/software/PRISMA2020.html}}, the ``eligibility'' phase is integrated into the screening process. We have adopted this updated version for our study for simplicity and clarity. The detailed steps of our methodology are illustrated in Figure~\ref{fig:overview}. In the Figure, Step 1 illustrates the PRISMA \textbf{identification} phase, where we collected a total of 1,806 papers.

We began this phase by systematically documenting a set of \textbf{inclusion criteria} for our search below:  

\begin{itemize}

    \item \textbf{Search Keywords:} Since we are only interested in papers on source-code-based software vulnerability detection,
    we formulated the keywords: \textit{software vulnerability detection} and \textit{source code vulnerability detection} based on our experience. 
    We additionally experimented with additional keywords, including \textit{software vulnerability identification} and \textit{software vulnerability discovery}; however, they did not identify any new relevant papers that were not already included in our existing pool. 

     \item \textbf{Timeframe:} We focused on papers published in the past six years (2018–2023) to capture recent efforts, given the rapid advancements in AI-based approaches.
    
    \item \textbf{Databases:} For each keyword, we searched three databases ACM, IEEE, and Google Scholar that are also used in other surveys \cite{semasaba2020literature, eberendu2022systematic, p126,senanayake2023android,ashawa2019analysis}. These databases are well-regarded in the research community for their comprehensive coverage of high-quality, peer-reviewed literature.  
\end{itemize}

 As shown in Figure \ref{fig:overview}, we initiated the process by downloading papers from the three specified databases for each keyword. For each database, the search concluded when no further relevant papers could be identified based on their titles. The list of papers selected during this search is presented in Table \ref{tab:Stat}. The initial search yielded a total of 1,806 papers across all three databases, with a detailed breakdown by keyword and database also provided in the table. 

Once collected the papers from all three resources, we removed the duplicates using Mendeley \cite{Mendeley}, an open-source reference management tool. This removed 1,157 papers, resulted in 649 remaining papers. We then read the abstract, and introduction.

\begin{table}[]
 \caption{Statistics of the total number of papers}
\begin{tabular}{llll|cc}
\hline
Keyword & \multicolumn{1}{l}{IEEE} & \multicolumn{1}{l}{GScholar} & ACM & \multicolumn{1}{l}{Removing } & \multicolumn{1}{l}{Removing} \\ \cline{1-4}
SVD     & \multicolumn{1}{l}{624}  & \multicolumn{1}{l}{515}        & 372 & Duplicate              &        Exclusion    \\ \hline
SCVD    & \multicolumn{1}{l}{102}  & \multicolumn{1}{l}{147}        & 46  &       \multirow{2}{*}{649}                            &        \multirow{2}{*}{\PaperTotal}                                \\ \cline{1-4}
Total   & \multicolumn{3}{c|}{1,806}                                         &                                   &                                    \\ \hline
\end{tabular}
\label{tab:Stat}
\end{table}

 We then reviewed the abstract and introduction of each remaining paper to assess whether it should be filtered based on our exclusion criteria. In  

 cases, when we were uncertain about the paper's relevance from the abstract alone, we read the entire paper for further assessment. The \textbf{exclusion criteria} for our work are outlined below:

\begin{itemize} 
\item Papers outside the scope of our study, primarily were related to network vulnerabilities, embedded and/or IoT vulnerabilities, cloud computing vulnerabilities, Android vulnerabilities, web vulnerabilities, smart contract vulnerabilities, binary-based vulnerabilities, and fuzzing-based approaches, and the one those addressed SVD using conventional approaches rather than DL or ML. 
\item Papers that were not peer-reviewed were excluded because their validity could not be verified, with the exception of VulDeePecker paper~\cite{li2018vuldeepecker}. Although this paper was not peer-reviewed, it has garnered 977 citations at the time of writing, indicating its significant impact on the community.
\item Work-in-progress papers, vision papers, posters, and case study papers were excluded due to incomplete results or insufficient data for further analysis. 
\item Papers lacking clarity or containing ambiguous explanations. Some papers could not be fully evaluated due to missing basic information (e.g., dataset, programming language, granularity level, methodology) resulting from unclear or vague descriptions. 
\item We identified 66 papers that were literature reviews, study papers, or papers that compared multiple approaches/techniques in the SVD domain. Although relevant, these papers were excluded from our primary list because they did not introduce new techniques. However, we examined these papers to extract valuable insights, concerns, and useful statistics, which were incorporated into our analysis and were previously shared in Section \ref{sec:relatedEmp}.
\item Papers not written in English. 
\end{itemize}

\begin{figure}[tbh!]
\centering
\vspace{-5pt}
    \includegraphics[scale=0.3]{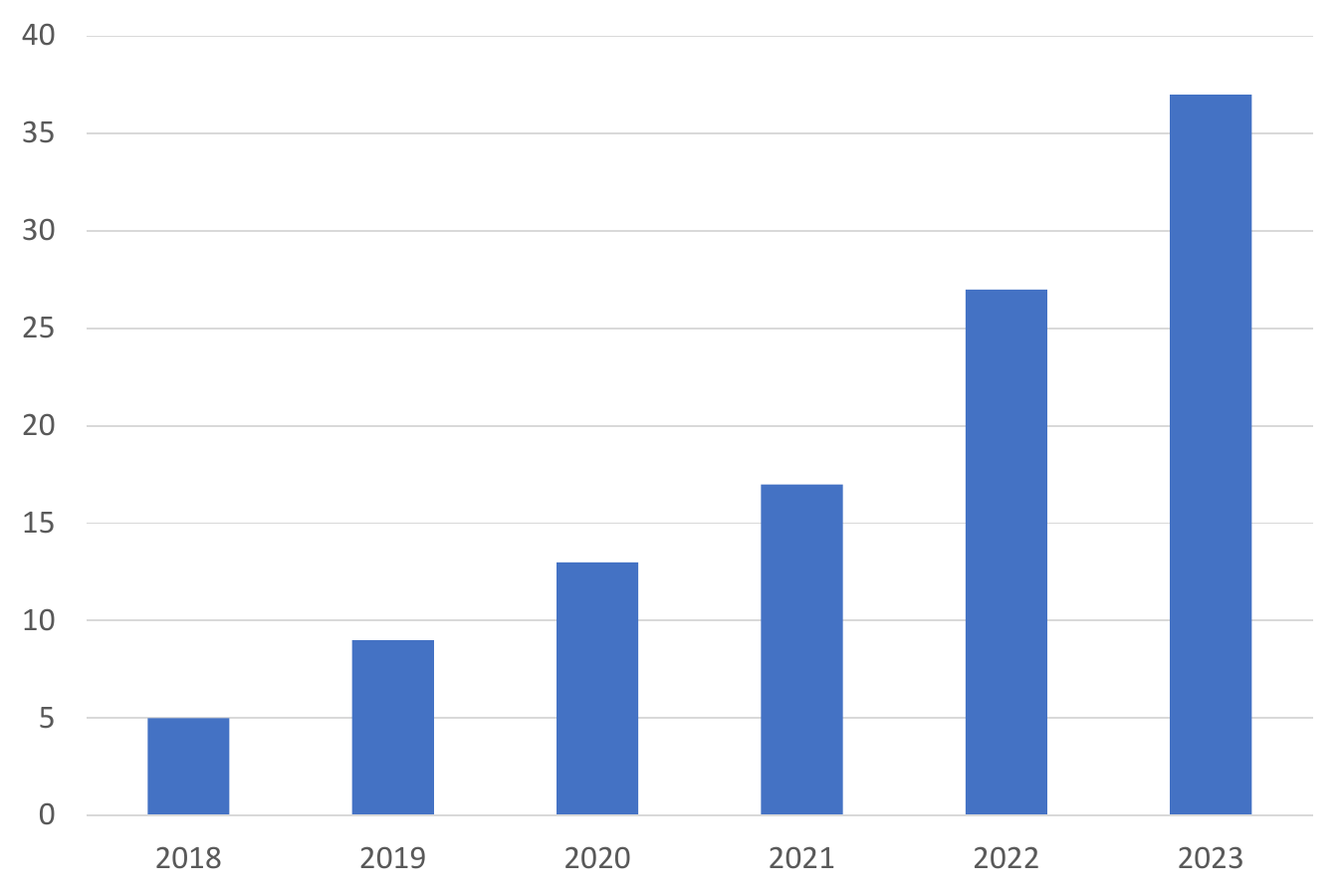}
\vspace{-5pt}
\vspace{-5pt}
 \caption{Year-wise distribution of all Papers}
    \label{fig:DistOverall}
\end{figure}

Following the exclusion criteria, 541 papers were removed, leaving a total of \textbf{\PaperTotal} papers for further review. A summary of these statistics is presented in Table \ref{tab:Stat}. We captured the year-wise distribution of the remaining papers, as illustrated in Figure \ref{fig:DistOverall}, revealing an upward trend. Notably, in 2018, only five papers met our selection criteria, while this number has increased to 37 in 2023.

\section{Datasets}
\label{sec:dataset}

In this section, we address \textbf{RQ1} by documenting \textit{reused datasets} identified in the papers we studied. By reused datasets, we refer to those that have been employed in more than one paper. Datasets used only once by their original authors and those not made publicly available were excluded from our analysis. While we have not listed the excluded datasets here due to space constraints, a comprehensive list of all datasets is provided in our supplementary material \footnote{\url{https://drive.google.com/drive/folders/1QY1iLadRe9YYwW7bAQMIH4y_I8aO6QPf?usp=sharing}}.

Our analysis resulted in the identification of eight reused datasets introduced within the timeframe of our study. Additionally, we included datasets utilized by these reused datasets, arriving at a total of 27 datasets for inclusion.

Table~\ref{tab:datasetInPapers} lists these datasets, providing a basic description, granularity level, language support, and download link for each dataset. This table enables an analysis of key dataset characteristics. For instance, it conveys that 16 out of 27 datasets are specifically designed for C/C++ code, 7 support multiple languages, 3 of them contain Java code and the remaining 1 focuses on JavaScript code. Regarding granularity, 15 datasets offer function-level analysis, 5 utilize code gadget/code slice granularity, referring to code segments, which directly influence (or are influenced by) a specific computation or variable of interest or perform a specific task, 1 employs statement-level granularity, and 2 provide multi-level analysis, 3 focus on commit-level code. 

Datasets marked with an asterisk (*) contain not only the standard labels that indicate whether a given sample is vulnerable or safe, but also they provide information about the specific type of vulnerability associated with each sample, such as CVE  or CWE identifiers—that categorizes the nature of the vulnerability. Having this additional information allows researchers and practitioners to better understand the exact security issues present in the data and tailor their detection and remediation strategies accordingly.

\small

\begin{longtable}{ p{2cm}p{3.5cm} p{1.6cm}  p{1.6cm} p{3.7cm} }

\caption{Characteristics of datasets used in papers (source code-based detection)} \label{tab:datasetInPapers} \\
\hline
\textbf{Name} & \textbf{Description} & \textbf{Granularity} & \textbf{Language} & \textbf{Download Link} \\
\hline
\endfirsthead
\caption{Characteristics of datasets used in papers (source code-based detection) (continued)} \\
\hline
\textbf{Name} & \textbf{Description} & \textbf{Granularity} & \textbf{Language} & \textbf{Download Link} \\
\hline
\endhead
\hline
\multicolumn{5}{r}{\textit{Continued on next page}} \\
\endfoot
\hline
\endlastfoot

 Devign \cite{devign}  &  Manually labeled dataset from FFmpeg and QEMU used in CodexGlue \cite{codexGluePaper} leaderboard containing 12,460 vulnerable and 14,858 safe functions &Function &  C/C++     &   \url{https://sites.google.com/view/devign}                                     \\

 VulDeePecker$^{*}$  \cite{li2018vuldeepecker} &  Based on NVD, contains 17,725  vulnerable and 43,913 not vulnerable code gadgets  & Code Gadget& C/C++   &     \url{https://github.com/CGCL-codes/VulDeePecker}                                       \\
    NVD$^{*}$  &  Vulnerability management database maintained by US government    &Slice& Multiple   &  \url{https://nvd.nist.gov/}                                         \\
        CVE$^{*}$  &   Publicly disclosed cybersecurity vulnerability list that contains relevant information about vulnerabilities.    &Slice& Multiple   &  \url{https://cve.mitre.org/}                                         \\
        Reveal \cite{chakraborty2021deep} &   Labeled dataset from Chromium and
Debian projects. Labeling was done with the help of issue tracking system  &Function   &    C/C++& \url{https://github.com/VulDetProject/ReVeal}                                       \\
         Russel et al.$^{*}$ \cite{russell2018automated} &  Contains examples from Debian Linux distribution, public Git
repositories on GitHub (labeling was done by static analyzers), also labeled synthetic data from Juliet test suite   &Function& C/C++   &                  \url{https://osf.io/d45bw/}                                            \\

Juliet/ SARD${^*}$    & Contains production software applications with known vulnerabilities. Artifacts contain designs, source code, and binaries     &Function& Multiple   &  \url{https://samate.nist.gov/SARD/test-suites}                                        \\
         VulDeeLocator \cite{li2021vuldeelocator}&  Samples collected from NVD and SARD. Labeling of NVD records were done by their \textit{"diff"} files before and after patches. The dataset contains 14,511 programs, including 2,182 real-world
programs and 12,329 synthetic and academic programs   &Function& C/C++   & \url{https://github.com/VulDeeLocator/VulDeeLocator}                                          \\
              
                  Wang et al.$^*$ \cite{wang2020combining} &  Vulnerable sample collected based on CVE and SARD. Labeling is done using a set of predictive
models or experts.   &Function& C/C++   & \url{https://github.com/HuantWang/FUNDED\_NISL}                                          \\
                  Project KB$^*$ \cite{ponta2019manually}&   Collected both from NVD and from project-specific Web resources
that the authors monitored on a continuous basis  & Statement&   Java&          \url{https://github.com/SAP/project-kb/tree/main/MSR2019}                              \\
                SySeVR$^*$  \cite{li2021sysevr}   & Contains data from NVD and SARD     & Slice&  C/C++  &   \url{https://github.com/SySeVR/SySeVR}                                     \\
             
                  Lin et al.$^*$  \cite{lin2018cross} &  contained manually labeled
457 vulnerable functions and collected 32,531 non-vulnerable
functions from 6 open-source projects based on CVE and NVD    &Function & C/C++& \url{https://github.com/DanielLin1986/TransferRepresentationLearning}                                       \\
                 Cao et al.$^*$ \cite{cao2022mvd}  & Covers 13 common memory-related vulnerabilities based on SARD and CVE. Labeling was done automatically using \textit{"diff"} files     & Statement&C/C++ &     \url{https://github.com/MVDetection/MVD}                                   \\
                  ApacheCrypto\-API-Bench \cite{afrose2022evaluation}& Consists of 86 real vulnerabilities
from 10 Apache open-source projects      & Snippet &Java & \url{https://github.com/CryptoAPI-Bench/}                                       \\
                Kluban et al.$^*$ \cite{kluban2022measuring}&     Dataset is curated based on Snyk vulnerability database [59] and Google
VulnCode-DB project \cite{vulcode-db} & Function & Javascript &  \url{https://github.com/Marynk/JavaScript-vulnerability-detection}                                      \\
      Alves et al.$^*$  \cite{alves2016software}     & Data is collected based on 2875 security patches of Linux Kernel, Mozilla, Xen Hypervisor, httpd
and glibc      & Function, File, Class& C/C++&    \url{https://eden.dei.uc.pt/\%E2\%88\%BCnmsa/metrics-dataset}                                    \\  Cao et al.$^*$  \cite{cao2021bgnn4vd}& Based on NVD and Github, they labeled function based on \textit{"diff"} file. Contains 3867 vulnerable and 92,058 vulnerable functions      &  Function&C/C++ &               \url{https://github.com/SicongCao/BGNN4VD}      \\ 
        Ponta et$^*$ al.\cite{ponta2019manually}  & Maps 624 publicly disclosed vulnerabilities   affecting 205 distinct open-source Java projects& Commit level  & Java                                                  & \url{https://github.com/eclipse/steady}              \\ 
 
                 Reis and Abreu$^*$ \cite{reis2021ground}  &   The authors scrapped the CVE details database for GitHub references and
augmented the data with 3 other security-related datasets (Bigvul, Secbench \cite{rei2017database,reis2017secbench}, Pontas et al. \cite{ponta2019manually}). The dataset contains natural language artifacts (commit messages,
commits comments, and summaries), meta-data, and code changes   & Commit level&20 languages  & \url{https://github.com/TQRG/security-patches-dataset}                                       \\

 Tian et al.$^*$ \cite{tian2024enhancing} &  Based on SARD. Contains 264,822 labeled 
 synthetic functions, where 166,641 are non-vulnerable and
98,181 are vulnerable across 118 CWE-IDs.    &Function  & C/C++&        \url{https://github.com/XUPT-SSS/TrVD}          

\\
 
 Bhandari et al.$^*$ \cite{bhandari2021cvefixes}& Contains vulnerable and corresponding patched code, Programming language used, five levels of abstraction-commit-, file-, and function levels, repository- and CVE levels. In total, this dataset contains 18,249 files and 50,322 methods      & Commit, file, function&Independent       & \url{https://github.com/secureIT-project/CVEfixes}                                       \\ 
Big-Vul$^*$ \cite{fan2020ac}   &  Extracted from 348 Github projects, BigVul data contains functions before and after fixing the vulnerability. In total, there are 11,823 vulnerable and 253,096 non-vulnerable functions in this dataset    & Function & C/C++                                                                                  & \url{https://github.com/ZeoVan/MSR_20_Code_Vulnerability_CSV_Dataset} \\ 
xVDB$^*$ \cite{hong2022xvdb}     & Contains 12,432   CVE      patches from repositories and issue trackers, and 12,458 insecure posts   from Q\&A sites    & Commit& C, C++, Java, JavaScript, Python and Go & \url{https://iotcube.net/}                                                   \\ 
Cross-vul$^*$ \cite{nikitopoulos2021crossvul}  & File-level granularity dataset that contains a) The directory with all source code files, b) A JSON file containing commit and file information associated with CVE IDs and CWE IDs, c) the commit message dataset that consists of three separate CSV files. Contains 13,738 vulnerable and 13,738 non-vulnerable files                    & Function                             & 40   programming languages                                               & \url{https://doi.org/10.5281/zenodo.4734050},   \url{https://doi.org/10.5281/zenodo.4741963}                                     \\ 

D2A \cite{zheng2021d2a}        & Contains a) Bug Reports (Trace), b) Bug function source code (Function),
c) Bug function source code, trace functions source code, and bug function file URL (Code).  Labeled by static analysis and differential analysis                                                   &    Function   & C/C++                                                                            & \url{https://github.com/IBM/D2A}                                                    \\ 

 NVD-new$^*$ \cite{zhang2023comparing}&  Obtained patch files from the reference links to the NVD, and
commits on the Github repository. Functions before and after the fixing patches are generated by VulnDBGen. Contains 9947 and 9947 non vulnerable functions       &Function& C/C++   & \url{https://github.com/CGCL-codes/DissectVulDetection/tree/main/Dataset} \\ 
DiverseVul$^*$ \cite{chen2023diversevul} & Contains 18,945   vulnerable functions spanning 150 CWEs and 330,492 non-vulnerable functions   extracted from 7,514 commit records        &Function& C/C++   & \url{https://github.com/wagner-group/diversevul}  
               \\ \cline{1-5}
\end{longtable}

\section{Primary characteristics of Code-based Detection  }
\label{sec:detection}

In this section, we analyze and document the key characteristics of the 
reviewed papers to address \textbf{RQ2}. Our focus includes, but is not limited to, identifying the detection objectives, exploring the granularities of vulnerability detection, and examining the targeted programming languages. Understanding these characteristics is essential for identifying trends and gaps in current research, providing insights to guide future studies in enhancing model design and implementation.

\subsection{Data Availability}
\begin{figure*}[hbt!]
\centering
\vspace{-5pt}
    \includegraphics[scale=0.55]{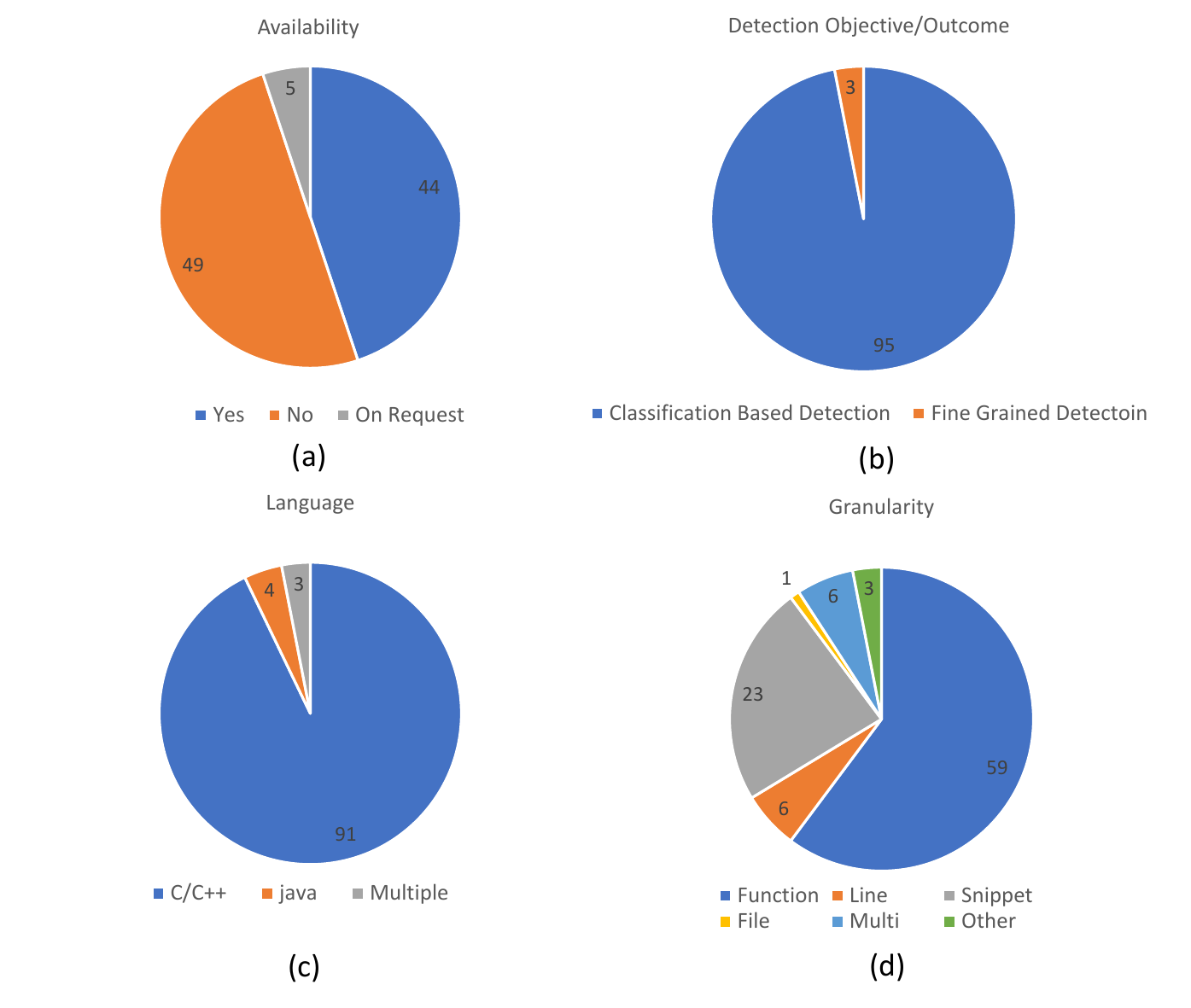}
\vspace{-5pt}
\vspace{-5pt}
 \caption{Characteristics of source code-based classification approaches}
    \label{fig:pieChartSourceCode}
\end{figure*}

To address \textbf{RQ2a}, we assessed the availability of code and data in the collected papers. As shown in the pie chart in Figure \ref{fig:pieChartSourceCode}(a), artifacts were available for around 50\% of the papers. The lack of available artifacts in the remaining papers can hinder future model improvements and dataset utilization. Please note that availability does not guarantee reproducibility, raising further questions about replicating the results. 

\subsection{Detection Objective/Outcome}
Accurately identifying software vulnerabilities is a important step in improving code security. To address RQ2b, depending on the chosen approach, detection efforts may focus on broad classifications of code as vulnerable or non-vulnerable, or they may focus on pinpointing specific lines of code that contribute to security issues. Each of them is described in the following subsections.

\subsubsection{Classification-Based Detection}
The primary goal of the efforts in this category is to determine whether a code segment, such as a code snippet, function, or file is vulnerable. Models in this category perform a binary (vulnerable/non-vulnerable) or multi-label classification at various levels of granularity, such as file-level, function-level, or slice-level. Classification-based detection is well-suited for scanning large codebases, quickly identifying areas of potential concern, and helping security teams prioritize further investigation. Although efficient, this method often provides limited insight into the exact location or cause of vulnerabilities, as it only labels code segments as a whole. As demonstrated in the pie chart in Figure \ref{fig:pieChartSourceCode}(b), in our analysis, the vast majority of papers (95 out of 98) employed this approach. Among those papers, 92 developed a binary classifier where the main objective was to detect security flaws irrespective of vulnerability type. We found 3 papers that developed multi-class classifiers for specific types of vulnerability. These particular efforts, while capable of identifying specific types of vulnerabilities, were limited in scope, addressing only a small number of known vulnerability categories.

\subsubsection{Fine-Grained Detection}
Compared to the previous set of approaches, fine-grained detection seeks to pinpoint the exact statements or lines within a code snippet that lead to vulnerabilities. By highlighting these specific elements, fine-grained approaches allow developers to quickly focus their attention on the root causes of security bugs. Only 3 papers followed this path in our analysis, which indicates that pinpointing vulnerability in a specific code segment is yet to be explored more thoroughly.

\subsection{Programming Languages}
To address \textbf{RQ2c}, we analyzed the programming languages targeted for vulnerability detection in the reviewed studies. In most cases, the models were programming language specific. The pie chart 
in Figure \ref{fig:pieChartSourceCode}(c) illustrates that C/C++ dominated, being the focus of 91 papers. Java was the target in 4 papers while only 3 studies addressed multiple languages (e.g. C/C++, Java, Swift, PHP at the same time).

\subsection{Code Granularity }
To address \textbf{RQ2d}, we identified, granularity levels of the solution proposed in each paper, such as file, function, line, and code gadget-based levels.

As demonstrated in the pie chart in Figure \ref{fig:pieChartSourceCode}(d), 59 papers focused on function-level granularity, making it the most common granularity level. The second most common level of granularity being used was code snippet/code gadget/program slice (23 papers), collectively referred to as "snippet." At this granularity level, the released dataset did not include a specific function but rather several lines of code, which could be either consecutive or non-consecutive, with the authors detecting whether each particular code snippet is vulnerable or not. 
Six papers examined code vulnerabilities at the granularity level of line or statement 
, one at file level, six papers considered multi-level granularity such as both function and statement levels, and finally three papers considered other levels of granularity, such as commit-level, component-level, multiple functions and minimum intermediate representation (simplified version of the code).

\section{Taxonomy: Code-Based Detection}
\label{sec:taxonomy2}

\begin{figure}[]
\centering
\vspace{-5pt}
    \includegraphics[scale=0.13]{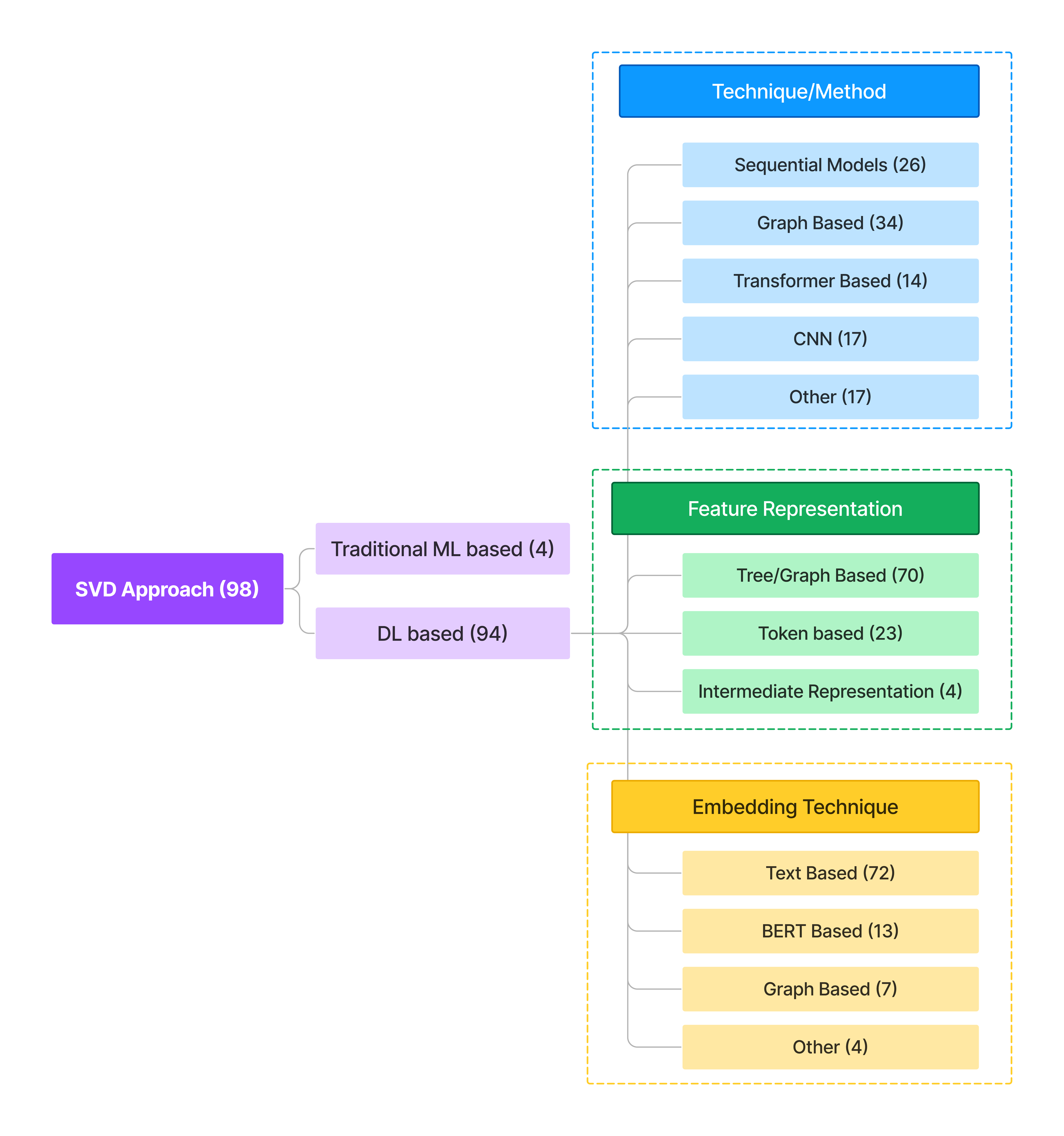}
\vspace{-5pt}
    \caption{Taxonomy of source code-based vulnerability detection methods. Each colored rectangle contains classification based on different aspects and thus classifies the same papers based on different characteristics}
\vspace{-5pt}
    \label{fig:taxonomy2}
\end{figure}

In this section, we propose a taxonomy developed based on the papers reviewed for code-based software vulnerability detection, addressing \textbf{RQ3}. The taxonomy is depicted in Figure \ref{fig:taxonomy2}. At the first level of the hierarchy, the existing approaches are categorized into two main groups: Machine Learning (ML)-based approaches, and Deep Learning (DL)-based approaches.

Among the papers analyzed, a substantial majority (96\%, or \PapaerSourceCodeDL papers) employed DL techniques, such as graph neural network and transformer-based models. In contrast, only 4\% (4 papers) utilized solely ML-based approaches, such as those using code metrics as features for classification purposes. 
We begin by presenting several examples of existing alternative approaches to ML- and DL-based methods, followed by a discussion of ML- and DL-based approaches. Finally, we provide a detailed taxonomy of DL-based approaches in Section \ref{sec:taxanomy}.

\subsection{Other Approaches}

While ML- and DL-based models dominate the field of vulnerability detection, several alternative methods are also employed, encompassing both static and dynamic analysis techniques. A few examples of the static approaches include \textbf{Code Similarity-Based Methods}, such as clone matching \cite{akram2021sqvdt, bowman2020vgraph}, which identifies vulnerable code clones by tracing patch files and utilizing graph-based components; pattern-based similarity matching \cite{kluban2022measuring, mosolygo2022line, zhang2022example}, which leverages pattern recognition and textual similarity to detect vulnerabilities in real-world applications; and graph-based similarity matching \cite{cui2020vuldetector, wu2020graph}, which compares function-level graphs and code property graphs to assess vulnerabilities. Another example is \textbf{Rule and Logic-Based Methods}, such as rule-based techniques \cite{rahaman2019cryptoguard}, which rely on predefined detection rules to identify cryptographic and SSL/TLS API misuses, and propositional functions \cite{han2019optimized}, which categorize software vulnerabilities using logical models based on CWE classification. In contrast, dynamic approaches, such as \textbf{Data Flow and Program Analysis-Based Methods}, analyze runtime behaviors, such as taint analysis \cite{kang2022tracer} which tracks how tainted data propagates through programs and representing vulnerability signatures as traces of inter-procedural data dependencies.

\begin{figure}[tbh!]
\centering
\vspace{-5pt}
    \includegraphics[scale=0.3]{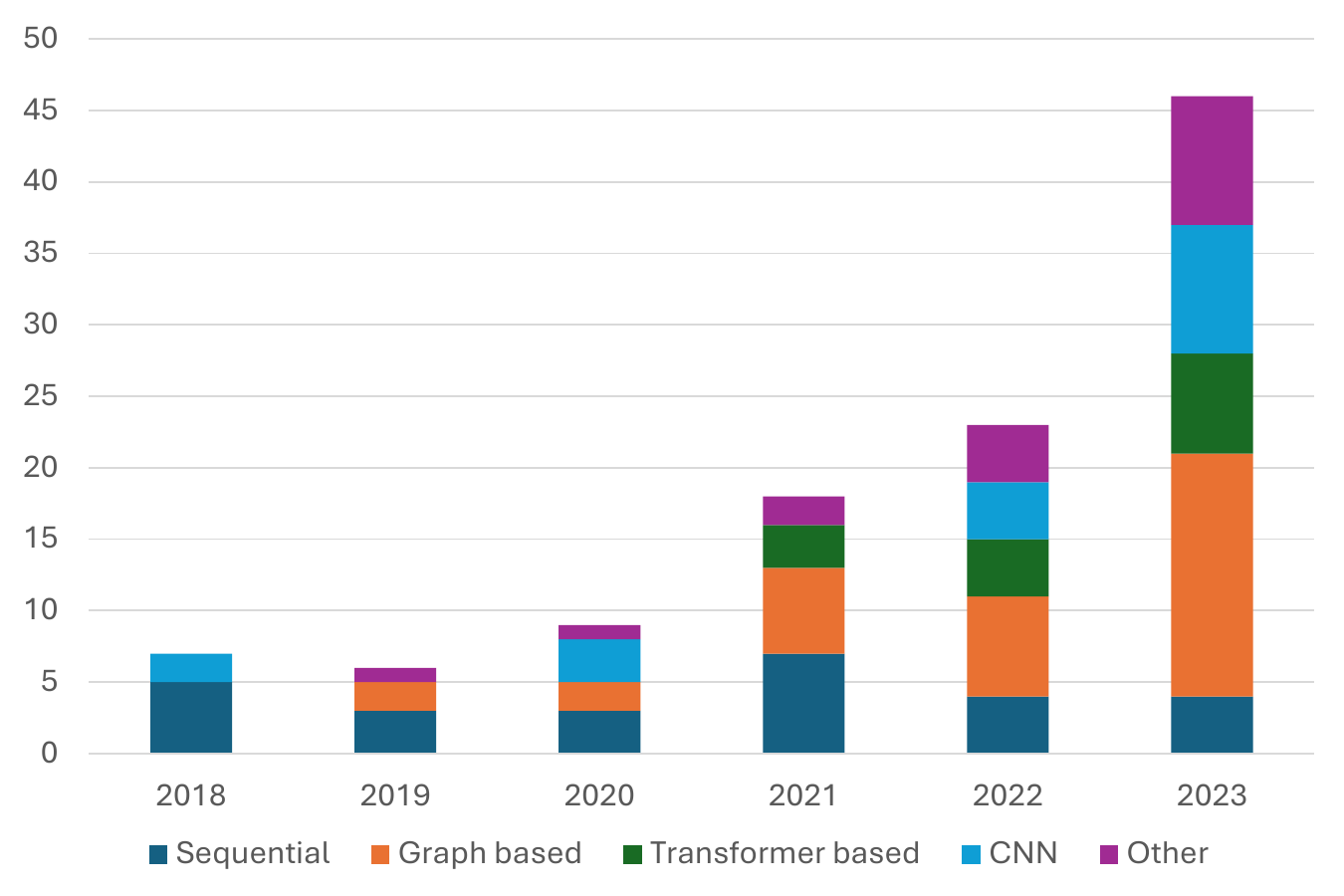}
\vspace{-5pt}
\vspace{-5pt}
 \caption{Year-wise distribution of the techniques of papers for DL-based detection}
    \label{fig:yearWiseTechniqueWise}
\end{figure}

\subsection{DL-based Approaches}
\label{sec:taxanomy}

Although DL-based methods were used in most of the efforts, basic ML-based approaches were also used in 4 papers gathered in our study. These efforts used different software metrics to detect vulnerabilities in source code. For instance, Medeiros et al. \cite{medeiros2020vulnerable} aimed to understand how the information provided by software metrics, such as Cyclomatic Complexity, Lines of Code, and Coupling Between Objects, can be utilized by ML-based approaches, including decision tree (DT), random forest (RF), extreme gradient boost (EGB), and linear support vector machine (SVM), to differentiate between vulnerable and non-vulnerable code. Similarly, Zagane et al. \cite{zagane2020new} employed code metrics, such as the number of total lines, cyclomatic complexity, and the number of distinct operators, to quantify extracted pieces of code. This quantification provided insights into the presence of vulnerabilities at a fine granularity level, using random forest (RF), decision trees (DT), and K-nearest neighbor (KNN) approaches. Salimi et al. \cite{salimi2020improving} introduced the concept of vulnerable slices, referring to vulnerable code units, to measure software using both structure-based and statement-based metrics. These newly measured metrics were subsequently used to characterize vulnerable codes, with SVM-based classification applied for vulnerability detection.

Some other approaches used AST-Based clustering methods for the identification of vulnerabilities in code. For instance, Debeyan et al. \cite{al2022improving} compared AST-based approaches with existing metric-based approaches using SVM, RF, and naive Bayes (NB) classifiers. They found that AST-based approaches achieved higher predictive performance, particularly in multiclass classification tasks, by leveraging AST n-grams as features.

Out of a total of \PaperTotal papers, \PapaerSourceCodeDL 
papers utilized DL-based techniques for detecting software vulnerabilities. As illustrated in Figure \ref{fig:taxonomy2}, we further categorized these papers according to three key aspects. First, the \textbf{DL technique employed} (represented by the blue rectangle), second, the \textbf{feature representation technique used} (indicated by the green rectangle), and third, the \textbf{embedding methods} utilized to feed data into the neural network (shown in the yellow rectangle). Each rectangle corresponds to a distinct classification criterion, with papers being grouped according to their approach for each aspect. 

It is important to note that if a paper employs more than one approach, it is placed into all relevant sub-categories. As a result, the total number of papers in the main category may be smaller than the sum of papers across all sub-categories within each rectangle. This classification structure provides a more comprehensive understanding of the diverse ways in which DL techniques, feature representations, and embedding methods are applied in the context of software vulnerability detection.

\begin{table}[tbh!]
\renewcommand{\arraystretch}{1.10}
\footnotesize

\caption{Models used in DL-based techniques}
%\vspace{-15pt}

\centering

\begin{tabular}{ p{2.3cm} p{1.3cm} p{5.4cm} p{2cm} }
\hline
\textbf{Category} & \textbf{\# Unique Papers} & \textbf{Models (Total Papers)} & \textbf{Most Popular Model} \\
\hline
Sequential models & 26 & LSTM (6), BiLSTM (19), tree-LSTM (1), RNN (1), BiRNN (1), GRU (1), BiGRU (2), SeqGAN (1) & BiLSTM (19) \\
%\hline
Graph-based models & 34 & GNN (13), GGNN (5), RGCN (2), GCN (3), SAR-GIN (1), behavior graph (1), GAT (1), BGNN (1), DGCNN (1), HetGNN (1), context-aware graph (1), hierarchical embedded (1), jump GAT (1), UCPG (1) & GNN (13) \\

Transformer-based models & 14 & BERT (4), RoBERTa (1), transformer-Based LM (1), HGT (1), transformer (2), dubbed VulD-transformer (1), SAT (1), hierarchical compression EM (1), CodeBERT (1), LLM (1) & BERT (4)
\\

Convolutional neural networks (CNN) & 17 & basic CNN (13), DGCNN (1), TextCNN (2), convolutional pooling layers (1), AlexNet (1), Lenet (1), Tcnn (1) & basic CNN (13) 
\\

Other & 17 & Quantum neural network (1), MTLF (1), Hierarchical attention neural network (1), serialization-based NN (1), MLP (3), Serialization-based (1), Sequence \& structure Fusion (1), curriculum learning (1), horizontal federated learning (1), meta-learning (1), PU learning (1), IGS (1), pathfinding \& pruning (1), path-flow based (1), attention neural network (1) & MLP (3) \\
\hline
\end{tabular}
\label{tab:summaryTechniqueWise}
\end{table}

\subsubsection{\textbf{Method/Technique}}

This section classifies the papers based on the deep learning (DL) model employed in their work. We categorize them into five major groups: sequential models, graph-based models, transformer-based models, convolutional neural networks (CNN), and others. For each major category, we further refine the classification into more specific subcategories based on the specific architecture in the neural model. Table \ref{tab:summaryTechniqueWise} provides an overview of each major category along with their subcategories and the number of related papers in each subcategory. The following sections describe each category in detail:

    \textbf{(1) Sequential Models}: In this category, models are specifically designed to process sequential data, typically one element at a time, which makes them particularly effective for tasks involving natural language or time series data. Out of the \PapaerSourceCodeDL papers utilizing DL-based approaches, 25 studies employed sequential models as the foundation for vulnerability detection. Several distinct models with a wide variety of neural network architectures were identified. For instance, \textbf{Long short-term memory (LSTM)} is a type of recurrent neural network (RNN) capable of capturing long-term dependencies in sequential data, and it was used in several efforts~\cite{dam2018automatic, cao2020ftclnet, ziems2021security, jeon2021autovas, saccente2019project, xiaomeng2018cpgva}. Similarly, \textbf{Bidirectional long short-term memory (BiLSTM)} is an extension of the LSTM model that processes input in both forward and backward directions, and it was utilized in many technique~\cite{li2018vuldeepecker, liu2019deepbalance, li2021sysevr, lin2019software, li2021vulnerability, zou2019mu, chen2022hlt, guo2022hyvuldect, mao2020explainable, tian2021bbreglocator, lin2018cross, zhang2021isvsf, liu2020cd, ziems2021security, nguyen2021information, chang2023vdda, du2023cross, jeon2021autovas, xiaomeng2018cpgva}. In addition, \textbf{Tree-structured long short-term memory (Tree-LSTM)} is a variant of LSTM designed to process hierarchical or syntactic structures, such as trees in abstract syntax trees (ASTs), making it particularly suitable for tasks involving structural data~\cite{wang2023deepvd}. On the other hand, \textbf{Basic RNN} focuses on simple sequential data processing but is limited by vanishing gradients in long sequences, and it was utilized in only one work in our survey \cite{russell2018automated}. Furthermore, \textbf{Bidirectional RNN (BiRNN)} captures contextual information by processing data in both directions and was used in one effort~\cite{li2021vuldeelocator}. Moreover, \textbf{Gated recurrent unit (GRU)} is a simplified version of LSTM that reduces computational complexity while retaining performance, and it was employed in two studies \cite{jeon2021autovas, xiaomeng2018cpgva}. Likewise, \textbf{Bidirectional gated recurrent unit (BiGRU)} extends GRU to process input in both forward and backward directions and was utilized in two additional studies \cite{feng2020efficient, jeon2021autovas}. Finally, \textbf{SeqGAN} is a generative adversarial network tailored for sequential data generation and evaluation, and it was employed in one study~\cite{liu2022effective}.

As presented in Table \ref{tab:summaryTechniqueWise}, BiLSTM emerged as the most extensively utilized sequential model, appearing in 19 studies. Following this, standard LSTM models were the second most frequently employed model, featured in 6 studies.

Figure \ref{fig:yearWiseTechniqueWise} illustrates the year-wise distribution of models based on the techniques employed. As shown in the figure, sequential models reached their peak usage in 2021, with a total of 7 studies. However, by 2023, their adoption had declined, with only 3 studies utilizing these models. This trend indicates a potential shift in research focus towards integrating sequential models with other architectures and exploring strategies to enhance their performance in vulnerability detection tasks.\\

\textbf{(2) Graph-based Models}: 
This category includes models specifically designed to analyze source code in the form of graph-structured data. A total of 35 papers employed graph-based approaches to detect vulnerabilities in software systems.

\textbf{Basic Graph Neural Networks (GNNs)} are models designed to understand relationships between nodes in a network, much like identifying connections within a web of friends. In the context of source code analysis, GNN-based models have been utilized in 13 studies, where the source code is represented through various graph or tree-based structures \cite{cao2022mvd, luo2022compact, song2022hgvul, zhou2019devign, hin2022linevd, li2021acgvd, nguyen2022regvd, duan2021multicode, zeng2020efficient, de2023glice, sun2023enhanced, wu2021vulnerability, zhang2024vulnerability}. These models effectively capture structural dependencies within the code, facilitating vulnerability detection. Expanding on basic GNNs, \textbf{Gated Graph Sequence Neural Networks (GGNNs)} are a specialized variant that focuses on processing information over sequences. This approach is comparable to tracking messages exchanged among friends over time to understand evolving relationships. GGNN-based vulnerability detection models have been developed in five studies, where gated mechanisms are explored to enhance graph processing and improve detection performance \cite{wang2020combining, csahin2023semantic, wu2022inductive, wu2023learning, yang2024tensor}.

Another important variant is the \textbf{Relational Graph Convolutional Network (RGCN)}, which aims to analyze different types of relationships between graph nodes. This can be likened to understanding various social connections, such as family, work, and friendship networks within a community. In our review, RGCN-based models were incorporated in two studies to capture and leverage complex node relationships for vulnerability detection \cite{zheng2021vu1spg, dong2023sedsvd}.

Similarly, \textbf{Graph Convolutional Networks (GCNs)} are a streamlined variant of GNNs, focusing on simplifying and strengthening relationships within a graph. An example of this would be identifying closely connected clusters in a map of cities. Our analysis revealed that GCNs were employed in three studies, primarily for static vulnerability detection tasks \cite{cheng2019static, wang2023deepvd, quan2023xgv}.

In contrast, the \textbf{Self-Attention Readout Graph Isomorphism Network (SAR-GIN)} is a more advanced GNN variant that emphasizes the most critical parts of the network. This approach is analogous to identifying the most influential individuals in a social group. Among the papers we reviewed, this model was utilized in only one study, reflecting its niche but potentially impactful role in vulnerability detection \cite{xia2021source}.

Beyond these commonly used models, there are \textbf{Other Graph-Based Models} that were each employed in a single study. These include the Behavior Graph Model \cite{yuan2023enhancing}, Graph Attention Network \cite{zhang2023cpvd}, Bidirectional Graph Neural Network (BGNN) \cite{cao2021bgnn4vd}, Deep Graph Convolutional Neural Network (DGCNN) \cite{pereira2022use}, Heterogeneous Graph Neural Network (HetGNN) \cite{cheng4567888vulnerability}, Context-Aware Graph-Based Model \cite{li2023commit}, Hierarchical Embedded Graph Model \cite{hao2023vd}, Jump Graph Attention Network \cite{zhang2023static}, and the Unified Code Property Graph (UCPG) \cite{li2023vulnerability}. Although these models are less frequently applied, they showcase the diversity of graph-based techniques being explored for vulnerability detection.

As shown in Table \ref{tab:summaryTechniqueWise}, basic GNNs were the most popular graph-based approach, utilized in 13 out of 35 papers. Figure \ref{fig:yearWiseTechniqueWise} highlights that graph-based methods first emerged in 2019 and have seen a steady increase in usage over time, with a peak in 2023 when 17 papers adopted this approach for software vulnerability detection.\\
  
\textbf{(3) Transformer-based Models}: This category leverages the transformer architecture, renowned for its effectiveness in processing sequential and structured data. Below, we summarize the specific transformer-based models employed in the reviewed efforts:

\textbf{BERT} is a widely used model for capturing bidirectional context in text data and has been applied in several studies \cite{kim2022vuldebert, ziems2021security, ding2023leveraging, wu2023cdnm}. Building on BERT, \textbf{CodeBERT} is a pre-trained model specifically designed for source code analysis and was employed in one study \cite{quan2023xgv}. Similarly, \textbf{RoBERTa}, an optimized variant of BERT, has been utilized to enhance performance in vulnerability detection tasks \cite{hanif2022vulberta}. Another significant model is the \textbf{Transformer-based Language Model}, which leverages transformers to understand and predict text patterns, as demonstrated in \cite{wu2021self}.

In addition to these models, the \textbf{Heterogeneous Graph Transformer (HGT)} is designed to process heterogeneous graph-structured data, capable of handling different types of nodes and relationships, such as mapping a city's roads and landmarks. This model was developed in one study within our review \cite{yang2022source}. Furthermore, the \textbf{Transformer/VulD-transformer} is a flexible architecture for understanding data patterns, with a specialized version tailored for vulnerability detection~\cite{zhang2023vuld}. 

The \textbf{Structure-Aware Transformer (SAT)} goes a step further by incorporating the structural aspects of data, similar to understanding the layout of a building blueprint, and was featured in one study \cite{xue2023vulsat}. Additionally, the \textbf{Hierarchical Compression Encoder Model} is a specialized transformer designed for compressing and processing hierarchical information in a structured format, and its application was proposed in one work \cite{liu2022cpgbert}. 

Finally, the \textbf{Large Language Model (LLM)} represents a highly advanced class of AI models trained on vast datasets to understand and generate human-like language. LLMs were employed for vulnerability detection in one study \cite{purba2023software}.

Among the 14 transformer-based works, BERT was the most widely used, appearing in four of the studies, as shown in Table~\ref{tab:summaryTechniqueWise}. The other models were each utilized in a single paper. 

From Figure~\ref{fig:yearWiseTechniqueWise}, we observe that transformer-based approaches first emerged in the SVD domain in 2021, with three works incorporating these models. Their adoption has grown steadily, reaching a peak in 2023 with seven primary studies employing transformer-based techniques.\\

\textbf{(4) Convolutional Neural Networks (CNN)}: CNNs have also been employed in SVD tasks. CNNs, a type of deep learning model, are specifically designed to process data with a grid-like structure, such as images or sequences. They excel in identifying and learning hierarchical patterns through their architecture, which typically includes convolutional layers that apply filters to detect features, pooling layers that reduce dimensionality while retaining critical information, and fully connected layers that integrate these features for classification. 
  
 \textbf{Basic Convolutional Neural Networks (CNNs)} are fundamental deep learning models designed to scan data, such as code sequences, to identify patterns that help classify vulnerable and non-vulnerable components. This model has been widely adopted in several studies \cite{wu2022vulcnn, watson2022detecting, cao2020ftclnet, tang2022sevuldet, peng2023cevuldet, zhang2023vulnerability, li2020automated, bilgin2020vulnerability, russell2018automated, liu2022effective, zhang2023vulgai, zhang2024vulnerability, mim2023impact}. Expanding on basic CNNs, the \textbf{Dynamic Graph Convolutional Neural Network (DGCNN)} is specifically designed for graph-structured data, enabling it to dynamically learn relationships within the graph. This model was developed for vulnerability detection~\cite{xuan2023new}. 

Similarly, the \textbf{Text Convolutional Neural Network (TextCNN)} is optimized for processing text data, such as code snippets, to uncover patterns in sequential data, as demonstrated in \cite{du2023cross, cai2023software}. To further enhance the performance of CNNs, \textbf{Convolutional Pooling Layers} combine convolution and pooling operations, allowing the model to focus on the most critical features~\cite{zhang2023static}. 

In addition to these models, \textbf{AlexNet}, a classic CNN architecture originally designed for image classification, has been adapted for code vulnerability detection~\cite{xiaomeng2018cpgva}. Alongside AlexNet, \textbf{LeNet}, one of the earliest CNN models known for its simplicity and effectiveness in pattern detection, has also been applied in the same study \cite{xiaomeng2018cpgva}. Moreover, the \textbf{Task-specific CNN (TCNN)} is tailored specifically for software vulnerability detection, further demonstrating its utility~\cite{xiaomeng2018cpgva}. 

This range of CNN-based models highlights the versatility of convolutional architectures in addressing different aspects of vulnerability detection, from basic pattern recognition to sophisticated graph and text analysis.

Among all CNN-based approaches, the basic CNN model was the most widely adopted, appearing in 13 primary studies as shown in Table \ref{tab:summaryTechniqueWise}. Figure \ref{fig:yearWiseTechniqueWise} indicates that CNNs were first introduced in SVD tasks in 2018 within the scope of our reviewed efforts. In 2023, CNN-based approaches were employed by 9 studies, making CNNs the second most popular technique after graph-based approaches.\\

\textbf{(5) Other Approaches} encompass a diverse set of models utilized in vulnerability detection, each leveraging unique methodologies to address specific challenges. One example is the \textbf{Quantum Neural Network}, which harnesses quantum computing principles to enhance computational efficiency and tackle complex patterns~\cite{zhou2022new}. Another notable approach is the \textbf{Metric Transfer Learning Framework (MTLF)}, which applies transfer learning techniques to adapt metric-based features from one domain to another, thereby improving vulnerability detection performance \cite{liu2020cd}. Additionally, the \textbf{Hierarchical Attention Network} captures hierarchical relationships within data and assigns attention weights to various information levels, facilitating precise vulnerability identification \cite{gu2022hierarchical}.

Moreover, the \textbf{Cross-Modal Feature Enhancement and Fusion} method integrates features from different modalities, such as code structure and textual information, to improve detection accuracy \cite{tao2023vulnerability}. Approaches like \textbf{Serialization-Based and Graph-Based Neural Networks} transform source code into serialized or graph-based representations for more effective analysis, with serialization-based techniques~\cite{sun2023enhanced} and graph-based methods~\cite{tian2023learning}. The \textbf{Multi-Layer Perceptron (MLP)}, a simple yet powerful feedforward neural network, has also been applied for vulnerability detection through dense source code representations in three studies~\cite{zhang2023vulnerability, wu2021vulnerability, alenezi2021efficient}.

In addition, \textbf{Sequence and Structure Fusion-Based Models} enhance detection capabilities by combining sequential and structural features of code \cite{tian2023learning}, while \textbf{Curriculum Learning} introduces tasks in an incremental manner, starting from simple to complex, to optimize model training \cite{du2023automated}. To address privacy concerns, \textbf{Horizontal Federated Learning} enables multiple parties to collaboratively train models without sharing raw data~\cite{zhang2024vulnerability}. Furthermore, \textbf{Meta-Learning} focuses on training models to adapt quickly to new tasks with minimal data, improving detection efficiency \cite{sun2023software}.

For handling imbalanced datasets, \textbf{Positive and Unlabeled (PU) Learning} utilizes only positive and unlabeled examples during training~\cite{wen2023less}. Enhancing model interpretability, \textbf{Integrated Gradients Enhanced with Saliency (IGS)} combines integrated gradients with saliency maps to explain and improve model predictions \cite{peng2023ptlvd}. Additionally, \textbf{Pathfinding and Heuristic-Based Pruning} employs pathfinding algorithms alongside heuristic pruning to streamline code analysis for vulnerabilities \cite{gear2023software}, while \textbf{Path-Flow-Based Models} analyze execution flow paths within code to detect potential security issues \cite{cheng2022path}.

Lastly, the \textbf{Attention Neural Network} applies attention mechanisms to focus on the most relevant parts of the input data, enhancing vulnerability detection performance \cite{duan2019vulsniper}. Collectively, these models highlight the diversity and innovative approaches adopted in the field of software vulnerability detection.

This category encompasses a diverse range of models that do not fit neatly into the previously defined categories. As we can observe from Table \ref{tab:summaryTechniqueWise}, among these models, MLP stands out as a widely used approach. As depicted in Figure \ref{fig:yearWiseTechniqueWise}, there is a trend of increasing popularity in this category over recent years. This suggests that researchers are increasingly exploring a variety of techniques, indicating potential for future improvements in vulnerability detection methodologies.

\begin{figure}[tbh!]
\centering
\vspace{-5pt}
    \includegraphics[scale=0.3]{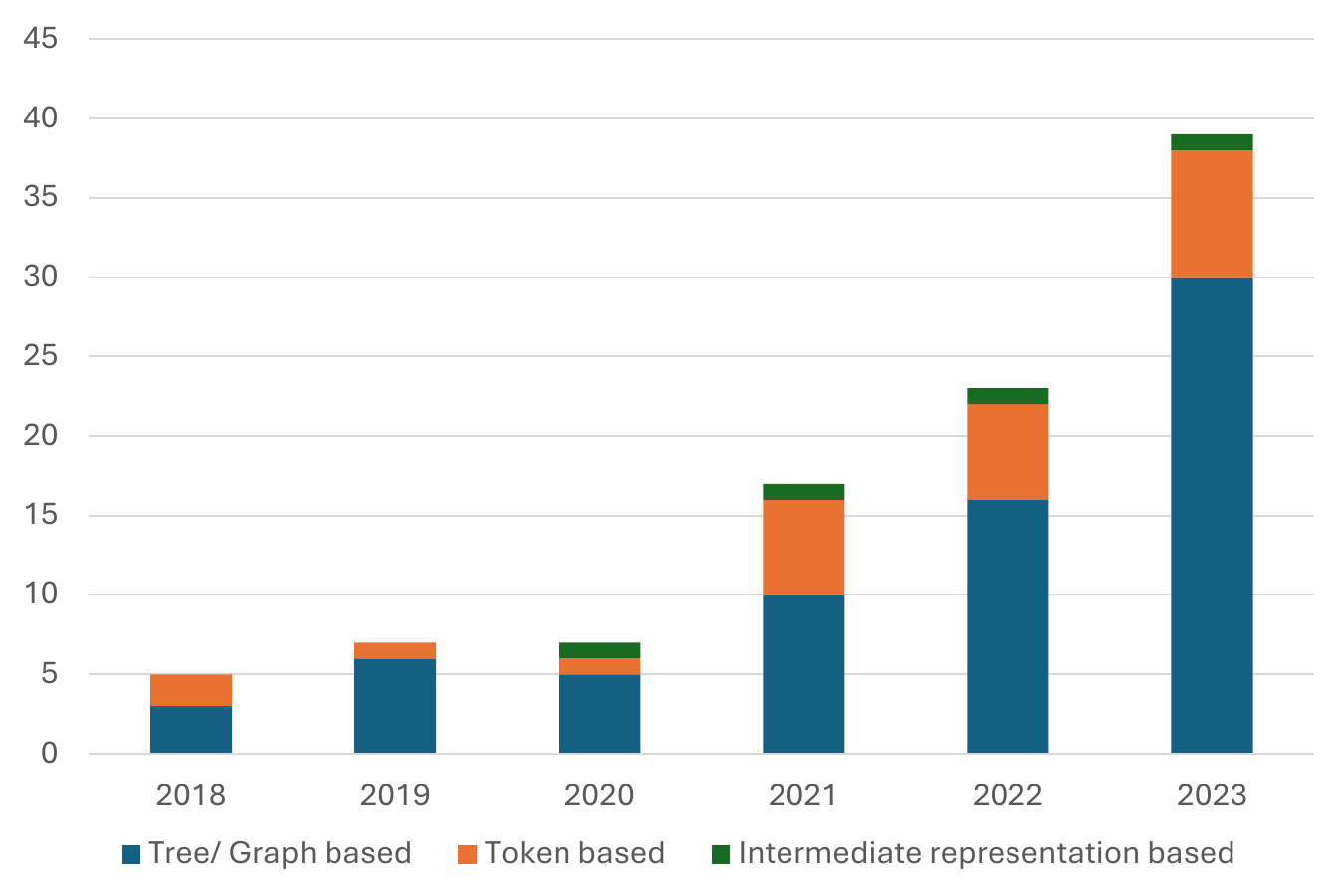}
\vspace{-5pt}
\vspace{-5pt}
 \caption{Year-wise distribution of feature representation techniques in DL-based papers}
    \label{fig:distYearFeature}
\end{figure}

\subsubsection{\textbf{Feature Representation}}
In this subsection, we categorize existing DL-based works based on the feature representation techniques employed. Source code can be represented in multiple ways, including structurally, semantically, or through intermediate representations. Accordingly, we divide feature representation techniques into three major categories. Figure \ref{fig:distYearFeature} illustrates the year-wise distribution of papers for each subcategory, highlighting the usage trends. Table \ref{tab:summaryFeatureRepresentation} provides a detailed list of models within each category and their respective paper counts.

\textbf{(1) Graph-Based Representations}:  
    In this category, source code is represented as structures such as graphs or trees to capture dependencies, control flows, and program behaviors. One of the foundational techniques in this category is the \textbf{Control Flow Graph (CFG)}, which captures the control flow within a program, illustrating how the execution moves between different code segments \cite{salimi2020improving, li2021acgvd, cheng2019static, cao2021bgnn4vd, wu2021vulnerability, pereira2022use, ding2023leveraging, wu2023learning, yang2024tensor}. Complementing CFGs, the \textbf{Program Dependence Graph (PDG)} models dependencies between program statements, providing insights into both control and data relationships within the code \cite{salimi2020improving, li2021sysevr, li2021vulnerability, zou2019mu, cao2022mvd, wu2022vulcnn, hin2022linevd, watson2022detecting, zeng2020efficient, wang2023deepvd, peng2023cevuldet, zhang2023vuld, xue2023vulsat, zhang2023vulgai, li2023commit, wu2023cdnm, sun2023software, wu2023learning, mim2023impact, peng2023ptlvd, gear2023software, tao2023vulnerability}.

Building upon these, the \textbf{Code Property Graph (CPG)} integrates multiple representations, including abstract syntax trees (ASTs), control flow graphs, and program dependence graphs, to form a comprehensive code analysis structure \cite{guo2022hyvuldect, xiaomeng2018cpgva, zhang2023vulnerability, xuan2023new, chang2023vdda, cai2023software, zhang22023vulnerability, dong2023sedsvd, duan2019vulsniper, zhang2023cpvd, liu2022cpgbert, cheng4567888vulnerability, zhang2024vulnerability, gear2023software, song2022hgvul, zhang2023static}. Additionally, specialized models like the \textbf{Control Dependence Graph (CDG)} and the \textbf{Data Dependence Graph (DDG)} focus on representing control and data dependencies, respectively \cite{yang2022source}. The \textbf{Program Control Dependence Graph (PCDG)} further integrates both types of dependencies to provide a holistic view of program behavior \cite{wang2020combining}.

To track the movement of data within applications, the \textbf{Data Flow Graph (DFG)} is employed, mapping the flow of data across different program components \cite{li2021acgvd, cao2021bgnn4vd, yang2024tensor}, while the \textbf{Execution Flow Graph (EFG)} captures the runtime execution paths of programs \cite{wang2023deepvd}. For behavior-specific analysis, the \textbf{Behavior Graph} focuses on identifying distinct behavioral patterns within the code \cite{yuan2023enhancing}.

The \textbf{Abstract Syntax Tree (AST)} is the most widely adopted representation, structuring the syntactic elements of code in a hierarchical tree format to facilitate both syntax and semantic analysis \cite{liu2019deepbalance, li2021sysevr, lin2019software, mao2020explainable, yang2022source, li2021vuldeelocator, lin2018cross, gu2022hierarchical, xia2021source, zhou2019devign, wang2020combining, csahin2023semantic, liu2020cd, dam2018automatic, duan2021multicode, feng2020efficient, wang2023deepvd, zhang22023vulnerability, wu2021vulnerability, liu2022effective, tian2023learning, hao2023vd, yang2024tensor, gear2023software, tao2023vulnerability, tian2024enhancing}. Variations like the \textbf{Control Flow Abstract Syntax Tree (CFAST)} combine control flow information with AST structures for enhanced code representation \cite{zhang2021isvsf}, while the \textbf{Binary AST} adapts AST techniques for binary-level code analysis \cite{bilgin2020vulnerability, cao2021bgnn4vd}.

Further extensions include the \textbf{Slice Property Graph (SPG)}, which integrates program slicing with graph representations to isolate relevant code segments for analysis \cite{zheng2021vu1spg}, and the \textbf{Attributed Control Flow Graph (ACFG)}, which enriches CFGs with additional attributes to capture more granular program details \cite{tian2023learning}. The \textbf{Unified Code Property Graph (UCPG)} consolidates multiple code representations into a unified analytical structure \cite{li2023vulnerability}, while models like \textbf{Path-Flow} focus specifically on tracing execution paths \cite{cheng2022path}.

Lastly, the \textbf{Natural Code Sequence (NCS)} approach treats source code as sequences of natural language tokens, enabling the application of natural language processing techniques for code analysis \cite{yang2024tensor, song2022hgvul}. Among these models, AST is the most widely adopted, appearing in 26 studies. Additionally, as depicted in Figure \ref{fig:distYearFeature}, graph and tree-based representations have maintained consistent popularity, with 30 papers employing these methods in 2023.

    Among these, the Abstract Syntax Tree (AST) is the most widely adopted model, appearing in 26 papers. Figure \ref{fig:distYearFeature} indicates that Graph/Tree-based representations have been consistently popular, with 30 papers using this method in 2023.\\

\textbf{(2) Token-Based Representations}:  
    Here, source code is treated as a sequence of tokens, focusing on semantic properties while disregarding structural elements.  
   \textbf{Code Representations} in vulnerability detection vary based on the level of abstraction and the specific analysis goals. One of the simplest forms is using the \textbf{Source Code} directly, without converting it into any structural format, treating it as a plain text sequence. This approach allows for straightforward processing and has been applied in several studies \cite{cao2020ftclnet, ziems2021security, nguyen2021information, russell2018automated, wen2023less}. Moving to more granular representations, the \textbf{Code Gadget} focuses on small, functional segments of code that are often used to analyze specific vulnerabilities. This method enables targeted analysis and has been employed in many studies~\cite{li2018vuldeepecker, zhou2022new, kim2022vuldebert, tang2022sevuldet, du2023automated, quan2023xgv, purba2023software}.

Another commonly used representation is the \textbf{Code Slice}, which extracts specific portions of the code based on defined slicing criteria, such as data or control dependencies. This method helps isolate relevant code fragments for vulnerability detection and has been utilized in multiple efforts~\cite{tian2021bbreglocator, wu2022inductive, wu2021self, de2023glice, sun2023enhanced, alenezi2021efficient, jeon2021autovas}. In contrast, the \textbf{Code Snippet} represents small, contiguous segments of code, often extracted for focused analyses on particular code regions \cite{du2023cross}.

Finally, at the most atomic level, the \textbf{Token} representation treats individual code tokens as discrete units of analysis, ignoring their placement within the overall code structure. This approach facilitates fine-grained analysis~\cite{saccente2019project, chen2022hlt, hanif2022vulberta}. Collectively, these representations provide diverse perspectives for analyzing source code, enabling a wide range of techniques for vulnerability detection.

    Token-based representations are less popular compared to Graph/Tree-based methods, as reflected in Table \ref{tab:summaryFeatureRepresentation}. However, specific techniques like Code Gadgets and Code Slices have seen broader utilization. Figure \ref{fig:distYearFeature} shows a steady rise in the use of token-based methods, though their total adoption remains limited.\\

\textbf{(3) Intermediate Representation-Based}:  
This category centers on representations used during the intermediate stages of code analysis, providing a simplified and abstract view of the source code. These techniques are designed to transform source code into formats that facilitate more in-depth and comprehensive analysis.

 One widely used form of intermediate representation for source code analysis is the \textbf{LLVM Intermediate Representation (LLVM IR)}, which is a low-level, platform-independent format designed to facilitate program analysis and transformations. This representation enables detailed examination of program behavior and has been utilized in one effort~\cite{li2021vuldeelocator}. In contrast, the \textbf{Source-Level Intermediate Representation (SIR)} operates at a higher level, preserving semantic information from the original source code, making it effective for analyses that require a closer connection to source-level constructs \cite{song2022hgvul}.

For more compact representations, the \textbf{Minimum Intermediate Representation (MIR)} is employed. MIR focuses on reducing redundancy while maintaining the essential features of the program, providing an efficient format for vulnerability detection tasks \cite{li2020automated}. Finally, at the lowest level, \textbf{Assembly} representation is used, which offers a detailed view of the code close to machine instructions. This representation is particularly useful for low-level security analyses and has been applied in one effort~\cite{tao2023vulnerability}.

Together, these intermediate representations provide diverse perspectives for analyzing software, from high-level semantic structures to low-level machine-oriented details, enabling comprehensive approaches to vulnerability detection.

    Intermediate representation-based approaches are the least adopted, with only four papers leveraging these methods. 

Table \ref{tab:summaryFeatureRepresentation} provides a concise overview of the models used for each feature representation technique, along with the corresponding paper count for each model. Notably, the AST emerges as the most widely utilized model in the Graph/Tree-based category, appearing in 26 papers. Furthermore, Figure \ref{fig:distYearFeature} presents the year-wise distribution of each feature representation technique, highlighting the sustained popularity of graph/tree-based representations throughout the survey period. In 2023, for instance, 30 papers adopted graph/tree-based structures for feature representation.

In comparison, the Token-based approach appears less prevalent, as evident by the data in Table \ref{tab:summaryFeatureRepresentation}. However, terms like "code gadget," "code slice," and "code snippet" are frequently used interchangeably to describe non-consecutive code segments, with both code gadgets and code slices being widely employed in several studies. As shown in Figure \ref{fig:distYearFeature}, while the number of papers employing text-based feature representation techniques has gradually increased over time, the overall paper count remains considerably lower than that of studies using graph/tree-based techniques.

\begin{table}[tbh!]
\renewcommand{\arraystretch}{1.10}
\footnotesize

\caption{Feature representation methods used in DL-based techniques}
%\vspace{-15pt}

\centering

\begin{tabular}{ p{3cm} p{1.3cm} p{5.2cm} p{2cm} }
\hline
\textbf{Category} & \textbf{\# Unique Papers} & \textbf{Models (Total Papers)} & \textbf{Most Popular Model} \\
\hline
Graph/Tree Based & 70 & CFG (9), PDG (23), CPG (13), CDG (1), DDG (1), PCDG (1), DFG (3), EFG (1), behavior graph (1), AST (26), CFAST (1), binary AST (2), slice property graph (1), value-flow path (1), abstract graph (1), graph (1), attributed control flow graph (1), CG (1), IG (1), NCS (2), unified code property graph (1), path-flow (1) & AST (26) \\
%\hline
Text-based & 23 & code gadget (7), code slice (7), source code (5), code snippet (1), token (3) & code gadget (7), code slice (7) \\
%\hline
Intermediate representation based & 4 & LLVM IR (1), SIR (1), minimum intermediate representation (1), assembly (1) & - \\
\hline
\end{tabular}
\label{tab:summaryFeatureRepresentation}
\end{table}

\begin{table}[tbh!]
\renewcommand{\arraystretch}{1.10}
\footnotesize

\caption{Embedding methods used in DL-based techniques}

\centering

\begin{tabular}{ p{3cm} p{1.3cm} p{5.2cm} p{2cm} }
\hline
\textbf{Category} & \textbf{\# Unique Papers} & \textbf{Models (Total Papers)} & \textbf{Most Popular Model} \\
\hline
Text-based & 73 & word2Vec (44), gloVe (3), doc2Vec (5), CBoW (3), sent2vec (6), GRU (1), FastText (3), PACE (1), n-gram (1), Keras library (1), one hot encoding (2) & word2Vec (44), own (12) \\

BERT Based & 13 & BERT (4), CodeBERT (8), RobertA (1) & CodeBERT (8) \\

Graph based & 7 & node2vec (2), knowledge graph embedding (1), graphToVec (1), struct2vec (1),label-GCN (1), GPT-GNN (1), graph embedding (1) & node2vec (2) \\

Other & 4 & large language model embedding (1), MPNet (1), value flow path embedding (1), code2vec (1) & - \\
\hline
\end{tabular}
\label{tab:summaryEmbeddingMethods}
\end{table}

\begin{figure}[tbh!]
\centering
\vspace{-5pt}
    \includegraphics[scale=0.3]{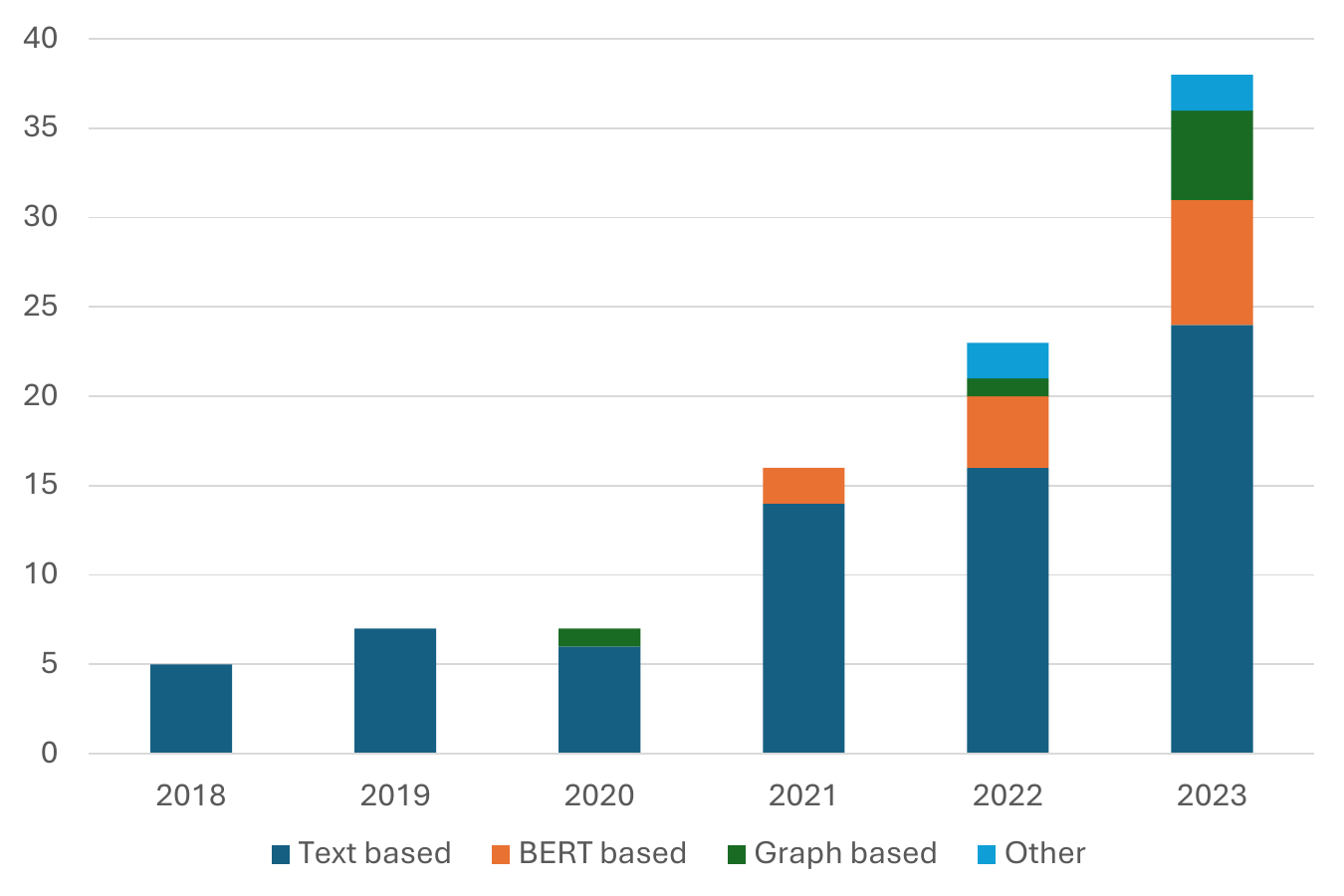}
\vspace{-5pt}
\vspace{-5pt}
 \caption{Year-wise distribution of the embedding techniques used in DL-based papers}
    \label{fig:EmbeddingYearwise}
\end{figure}

\subsubsection{\textbf{Embeddings}}
Embeddings are essential for converting raw code into continuous vector representations, enabling the code to be effectively input into deep learning models. In this section, we classify the existing deep learning-based approaches according to the embedding techniques used for code representation. The embedding methods identified in the literature are outlined below:

    \textbf{(1) Text-based Embedding}: This category includes text-based embedding methods where various versions of source code are passed through embedding models to generate vector representations.

One of the most commonly used methods is \textbf{Word2Vec}, which learns vector representations of words by predicting the context in which words appear, effectively capturing semantic relationships. This technique has been extensively utilized in numerous studies \cite{li2018vuldeepecker, liu2019deepbalance, lin2019software, zou2019mu, chen2022hlt, guo2022hyvuldect, yang2022source, li2021vuldeelocator, tian2021bbreglocator, song2022hgvul, lin2018cross, xiaomeng2018cpgva, gu2022hierarchical, xia2021source, wang2020combining, li2021acgvd, zheng2021vu1spg, csahin2023semantic, cao2020ftclnet, wu2022inductive, feng2020efficient, wu2021self, tang2022sevuldet, de2023glice, xuan2023new, tao2023vulnerability, chang2023vdda, du2023cross, cai2023software, dong2023sedsvd, cao2021bgnn4vd, jeon2021autovas, russell2018automated, zhang2023cpvd, tian2023learning, du2023automated, cheng4567888vulnerability, zhang2024vulnerability, wu2023cdnm, yang2024tensor, liu2020cd, zhang2023static, peng2023cevuldet}. Complementing Word2Vec, \textbf{GloVe} generates word embeddings by factoring a word co-occurrence matrix, capturing global statistical information, and has been applied in three studies~\cite{li2021vulnerability, zhang2023vulnerability, jeon2021autovas}.

Further extending embedding capabilities, \textbf{doc2Vec} learns fixed-length vector representations for entire documents or code snippets, preserving context across longer code fragments \cite{chang2023vdda, jeon2021autovas, li2023vulnerability, cao2022mvd, cheng2019static}. Another related method, the \textbf{Continuous Bag of Words (CBoW)}, predicts a target word based on its surrounding context, effectively representing words through their neighboring tokens \cite{lin2019software, lin2018cross, li2020automated}. For capturing semantic meaning at the sentence or code fragment level, \textbf{Sent2Vec} has been employed, which generates continuous vector representations that reflect the overall context \cite{wu2022vulcnn, watson2022detecting, jeon2021autovas, mim2023impact, xue2023vulsat, pereira2022use}.

In addition to traditional embeddings, recurrent models like the \textbf{Gated Recurrent Unit (GRU)} are used to learn sequential dependencies within code, leveraging gated mechanisms to control information flow \cite{li2021vulnerability}. \textbf{FastText}, an extension of Word2Vec, represents words as bags of character n-grams, allowing it to capture subword-level information and handle rare words more effectively \cite{chang2023vdda, zhang2023vuld, jeon2021autovas}. Moreover, character-level embedding methods such as \textbf{Position-Aware Character Embedding (PACE)} focus on the positional information of characters within code fragments, enhancing syntax-related feature learning \cite{duan2021multicode}.

Other notable techniques include \textbf{n-gram Based Embedding}, which captures local dependencies in the code by embedding sequences of consecutive words or characters \cite{alenezi2021efficient}, and the \textbf{Keras Library Embedding}, a custom embedding approach implemented using the Keras framework to generate vector representations based on specific code patterns \cite{saccente2019project}. Simpler approaches like \textbf{One-Hot Encoding} represent tokens as binary vectors, indicating the presence or absence of specific elements in the vocabulary \cite{du2023automated, zhang2023static}. 

Finally, several studies have proposed \textbf{Custom Embedding Methods} tailored to specific code analysis tasks, leveraging novel techniques to transform source code into meaningful vector representations \cite{zhou2022new, li2021sysevr, dam2018automatic, bilgin2020vulnerability, mao2020explainable, nguyen2021information, wu2021vulnerability, chang2023vdda, liu2022cpgbert, hao2023vd, tian2024enhancing}. Collectively, these embedding methods form the foundation for feature extraction in source code analysis, enabling machine learning models to perform effective vulnerability detection.

    Table \ref{tab:summaryEmbeddingMethods} provides a summarized overview of all the embedding methods mentioned. Notably, Word2Vec stands out as the most widely used technique among text-based methods, incorporated in 44 studies. Figure \ref{fig:yearWiseTechniqueWise} shows the year-wise distribution of embedding techniques used in the studies included in our survey. Text-based embedding methods have consistently been popular, with 21 papers employing them in 2023 alone.\\

    \textbf{(2) BERT-Based Embedding}: Techniques in this category leverage models pre-trained on large corpora of text data, which are then fine-tuned for specific downstream tasks, including vulnerability detection in source code. One of the most widely used models in this category is \textbf{BERT}, a transformer-based architecture that captures deep contextual relationships by analyzing bidirectional dependencies within the data. This model has been effectively applied to code analysis in several studies \cite{zhang2021isvsf, kim2022vuldebert, ziems2021security, ding2023leveraging}. Building on the success of BERT, \textbf{CodeBERT} was introduced as a variant specifically designed for programming languages. Pre-trained on a large corpus of source code, CodeBERT has been fine-tuned for various tasks, including vulnerability detection, and has demonstrated strong performance in multiple studies \cite{hin2022linevd, nguyen2022regvd, zhang22023vulnerability, sun2023enhanced, wen2023less, peng2023ptlvd, quan2023xgv, yuan2023enhancing}.

In addition to BERT and CodeBERT, \textbf{RoBERTa} represents a robustly optimized version of BERT. By employing a larger batch size, longer training periods, and more extensive datasets, RoBERTa has achieved improved performance across a range of NLP and code-related tasks, including vulnerability detection \cite{hanif2022vulberta}. Collectively, these BERT-based embedding models offer advanced capabilities for understanding both natural language and programming code, making them highly effective tools for software vulnerability analysis.

    Table \ref{tab:summaryEmbeddingMethods} offers an overview of the embedding methods discussed. BERT-based models have become increasingly popular in recent studies due to their powerful representation capabilities. The table clearly shows that, among BERT-based embedding techniques, CodeBERT is the most widely used method, appearing in eight papers. Furthermore, Figure \ref{fig:yearWiseTechniqueWise} illustrates that BERT-based models were first introduced in SVD in 2021, and since then, the number of papers utilizing BERT-based embeddings has steadily increased, although it remains significantly lower compared to the adoption of text-based methods.\\

    \textbf{(3) Graph-based Embedding}: These techniques leverage graph representations of code to capture complex relationships between different elements and structures within a program. These approaches model code as graphs, where nodes represent entities such as functions or variables, and edges denote relationships like dependencies or control flows. 

One widely used technique in this category is \textbf{Node2vec}, which learns vector representations for nodes by optimizing a neighborhood-preserving objective, allowing it to effectively capture local graph structures \cite{wang2023deepvd, zhang2023vulgai}. Complementing this, \textbf{Knowledge Graph Embedding} maps knowledge graphs into vector spaces, enabling the model to capture semantic relationships between entities and their attributes, which is particularly useful for code analysis tasks \cite{cao2022mvd}.

For capturing the overall structure of code, \textbf{GraphToVec} is employed to learn fixed-size vector representations of entire graphs, facilitating the analysis of code interactions and structural patterns \cite{zeng2020efficient}. Similarly, \textbf{Struct2vec} focuses on encoding structural information by considering the global context of nodes and edges within the graph, thereby enhancing the representation of code hierarchies \cite{zhang2023vulgai}. Additionally, \textbf{Label-GCN}, a variant of the graph convolutional network (GCN), incorporates label information into the graph learning process, improving code representation and vulnerability detection performance \cite{li2023commit}.

Another notable model is \textbf{GPT-GNN}, which combines the language modeling capabilities of GPT with the relational learning power of graph neural networks, effectively capturing both sequential code patterns and graph-based relationships for task-specific applications \cite{sun2023software}. Finally, \textbf{Graph Embedding} refers to general techniques that map graph structures, such as control flow graphs or program dependence graphs, into continuous vector spaces, enabling downstream analysis for tasks like vulnerability detection \cite{wu2023learning}.

Collectively, these graph-based embedding methods provide robust frameworks for representing code in a way that captures both syntactic and semantic relationships, facilitating more accurate and comprehensive vulnerability analysis.

Table \ref{tab:summaryEmbeddingMethods} indicates that several graph-based embedding methods were utilized, with node2vec appearing in two papers, while the remaining approaches were each used in just one paper. As shown in Figure \ref{fig:yearWiseTechniqueWise}, graph-based embeddings are less popular than text-based methods, although their use has gradually increased over time. In 2023, a total of five papers employed graph-based embedding techniques.\\

\textbf{(4) Other Embedding Techniques:} This category encompass a variety of approaches that do not fit directly into traditional categories but offer unique advantages for source code representation and vulnerability detection. 
    
One such technique is \textbf{LLM Embedding}, which leverages large pre-trained language models to generate rich, contextual embeddings for source code, capturing deep semantic relationships and code patterns \cite{purba2023software}. Complementing this, \textbf{MPNet} is a transformer-based model designed to generate embeddings by learning contextual relationships within code snippets, enhancing the understanding of code semantics and dependencies \cite{luo2022compact}.

Another notable method is \textbf{Value Flow Path Embedding}, which focuses on capturing the flow of values through a program, encoding these paths into representations that can be effectively utilized for downstream tasks such as vulnerability detection \cite{cheng2022path}. Additionally, \textbf{Code2Vec} offers a distinctive approach by learning code embeddings based on abstract syntax trees (ASTs), mapping source code into continuous vector spaces for more effective and interpretable representations \cite{gear2023software}.

Collectively, these embedding techniques broaden the scope of code representation methods, offering specialized solutions that enhance the performance of code analysis models in various software engineering tasks.

\section{Insights, Challenges, and Future Directions}
\label{sec:Future}

In the preceding sections, we analyzed and summarized key characteristics of existing vulnerability detection techniques based on the selected literature. Building on this analysis, thid section explores the core insights, challenges, limitations, and open problems in software vulnerability detection, while also outlining potential directions for future research—thereby addressing \textbf{RQ4}.

\subsection{Datset Issues}
Issues related to datasets are pivotal concerns in vulnerability detection, particularly with the advent of ML and DL-based approaches, which demand substantial data for effective training. While several datasets exist, as discussed in Section \ref{sec:dataset}, certain limitations persist among them. These limitations impact the performance of recent vulnerability detection models, as they are directly influenced by the quality and characteristics of the datasets utilized. The major dataset issues are outlined below.  

\subsubsection{Lack of Real-World, Large-Scale data}
A significant challenge with current datasets is the scarcity of real-world data suitable for training purposes. For instance, the widely used SARD dataset\footnote{\url{https://samate.nist.gov/SARD}} generates synthetic samples that may not accurately reflect real-world scenarios. For example, Chakrabarty et al. \cite{chakraborty2021deep} demonstrated, with some examples, that real-world examples are more complex than the synthetic ones.  
It is worth mentioning that although another public dataset NVD\footnote{\url{https://nvd.nist.gov/}} contains around 220K records, only a limited number of samples are usable after applying the pre-processing on data \cite{shimmi2024vulsim}.

\subsubsection{Imbalanced Data}
The current datasets contain more non-vulnerable records than vulnerable ones as we observed in the primary papers used in our work in Table \ref{tab:datasetInPapers}. The same observation is also mentioned by Ghaffarian and Shahriar~\cite{ghaffarian2017software} where they worked with different sets of papers.  When a model is trained on such an imbalanced dataset, it is biased towards non-vulnerable examples~\cite{chakraborty2021deep}. One proposed solution by Chakraborty et al. \cite{chakraborty2021deep} is to use synthetic minority oversampling technique (SMOTE) where we super sample the minority class until all classes have the same frequency. 

\subsubsection{Inaccurate Labels}

According to Croft et al. \cite{croft2023data}, existing datasets suffer from inaccurate labeling. Their analysis of 70 random samples reveals that accuracy values for real-world datasets Devign, Big-Vul, and D2A are 0.80, 0.543, and 0.282 respectively. 

\subsubsection{Lack of Uniqueness}
Croft et al. \cite{croft2023data} also mention that over 94\% of data in the D2A dataset contains type-1 code clones. The uniqueness values for Devign, Big-Vul, D2A, and Juliet were 0.899, 0.830, 0.021, and 0.163 respectively. In another study, Chakraborty et al. \cite{chakraborty2021deep} demonstrated, semi-synthetic datasets like NVD, SARD,
and Juliet result
in a large number of duplicates (more than 60\%). If the model is not trained on a dataset that contains unique samples, that can have an impact on the overall performance of the model. Both inter-set and intra-set duplicate samples can lead to improper performance measurement \cite{chakraborty2021deep}. 

\subsubsection{Lack of Multi-Language Datasets}
The current datasets are mostly derived from C/C++ language as we noticed in Table \ref{tab:datasetInPapers}. The recent efforts therefore focus mainly on C/C++ as we can see from the pie chart in Figure~\ref{fig:pieChartSourceCode}(c). More datasets are required in multiple languages to facilitate work on other programming languages.

\subsubsection{Coarse Granularity Levels} As shown in Table \ref{tab:datasetInPapers} and Figure~\ref{fig:pieChartSourceCode}, most datasets—and consequently, recent research efforts—focus primarily on the function level granularity. However, finer-grained datasets are essential to support techniques capable of precisely pinpointing vulnerabilities.

\subsubsection{Limited Vulnerability Type Coverage}
Our analysis shows that most existing techniques are either vulnerability-type agnostic or narrowly focused on a limited set of vulnerability types. This is inadequate for real-world applications, where practitioners must address a wide range of vulnerabilities, and the specific type being detected is crucial context for effective remediation.

\subsection{Unavailability of Data/Model }
One of the major issues we identified is the lack of publicly available models or datasets. As shown in Figure~\ref{fig:pieChartSourceCode}(a), over 50\% of the reviewed efforts do not provide access to either, making reproducibility a significant challenge. This observation aligns with findings by Nong et al. \cite{nong2022open}, who analyzed 55 papers for adherence to open-science principles, including availability, executability, reproducibility, and applicability. Their study found that only 25.5\% of the papers released their tools publicly, and of those, 54.5\% lacked proper documentation. Additionally, 27.3\% had incomplete or non-functional implementations, rendering them effectively unreproducible. Even among functional tools, only 87.5\% were reproducible, and just 14.3\% remained replicable when tested on different datasets. Reproducibility is critical not only for validating published results but also for fostering a continuous, self-correcting research process~\cite{albertoni2023reproducibility}.

\subsection{Language Dependency}
Another key observation is that most existing models are language-specific. As noted in our discussion of dataset limitations, these constraints often carry over into the models trained on them. C/C++ is by far the most commonly used language, while other programming languages remain significantly underrepresented. Moreover, few efforts have produced language-agnostic models. To support the development of more versatile and adaptable solutions, there is a clear need for language- and platform-independent approaches capable of addressing vulnerabilities across a wider range of software systems.

\subsection{Insufficient Comparative Analysis Across Approaches}
In this work, we introduced a taxonomy for source-code-based vulnerability detection (SVD) methods. Future research could build upon this framework by systematically evaluating and comparing the performance of techniques in various categories. While some comparative studies exist~\cite{zhang2023comparing,napier2023empirical,steenhoek2023empirical}, the field still lacks comprehensive and consistent benchmarking.

For example, Zhang et al.\cite{zhang2023comparing} found that certain models performed well on real-world datasets but underperformed on manually crafted ones, highlighting the sensitivity of model performance to dataset characteristics. Moreover, there remains a lack of clarity regarding the relative effectiveness of different neural network architectures within SVD\cite{zhu2023application}. For instance, Convolutional Neural Networks (CNNs) are generally better suited for local feature extraction, whereas Bidirectional LSTM (BiLSTM) networks are more effective at capturing long-range dependencies~\cite{zhu2023application}.

Embedding and feature representation techniques also significantly influence model performance. According to Zhu et al.~\cite{zhu2023application}, a BiLSTM model performed poorly when using GloVe or Doc2Vec embeddings compared to Word2Vec and FastText. Interestingly, CodeBERT embeddings also underperformed relative to Word2Vec and FastText, despite outperforming GloVe and Doc2Vec. Code2Vec, trained specifically on source code, captured richer semantic information and outperformed the other general-purpose embeddings.

These variations suggest a need for deeper investigation into the interplay between model architectures and embedding techniques. Future research should aim to identify optimal combinations or, at the very least, understand the trade-offs involved in different design choices. Our proposed taxonomy offers a structure for guiding such studies, enabling researchers to systematically test various configurations and identify the most effective strategies for vulnerability detection.

\subsection{Lack of Interpretability}
A growing trend in vulnerability detection is the adoption of deep learning (DL)-based approaches over traditional or classical machine learning (ML) methods. While this shift is promising, DL models are inherently black-box in nature~\cite{ribeiro2016should}, meaning their decision-making processes are often opaque. This lack of transparency poses a significant challenge for the SVD domain, where interpretability is critical—not only for understanding why a vulnerability was detected but also for effectively addressing and resolving it.

Among the studies we reviewed, only a few explicitly addressed interpretability. For example, Li et al.~\cite{li2021vulnerability} utilized GNNExplainer\cite{ying2019gnnexplainer}, an interpretable graph neural network framework that helps explain the model’s predictions. Similarly, Gu et al.~\cite{gu2022hierarchical} introduced a model with attention mechanisms at both the line and token levels, offering more insight into how the model processes source code.

Despite these efforts, most existing DL-based models remain largely uninterpretable. Greater attention should be directed toward integrating explainable AI techniques or human-in-the-loop methods to enhance model transparency. Improving interpretability will be essential for building trust in automated vulnerability detection systems and supporting effective remediation strategies.

\subsection{Granularity Limitations}

One significant limitation we observed is that most existing models are not designed for fine-grained vulnerability detection. As highlighted in the pie charts in Figure~\ref{fig:pieChartSourceCode}, the majority of current approaches operate at the function level. While a few studies have explored finer-grained detection, such as identifying vulnerabilities at the line or statement level, these efforts remain limited in number.

Fine-grained detection is necessary for providing developers with precise insights into vulnerable code segments. By pinpointing the exact statements responsible for vulnerabilities, such models can greatly assist in efficient debugging and remediation. Therefore, future research should focus on developing and promoting fine-grained approaches to better support practical vulnerability mitigation.

\subsection{Greater Adoption of Emerging Deep Learning Techniques is Needed}

An analysis of our taxonomy reveals that several emerging DL techniques remain underutilized in the SVD domain, often grouped under the "Other" category. Among these, federated learning (FL) stands out as a promising but largely unexplored approach.

Originally introduced by Google in 2017, FL enables the training of machine learning models across decentralized devices or servers while keeping data localized and private. Although FL has been successfully applied in various cybersecurity domains—including intrusion detection~\cite{rahman2020internet, li2020deepfed}, anomaly detection~\cite{alrashdi2019ad}, malicious attack detection~\cite{hei2020trusted}, and malware detection~\cite{rey2022federated}—its application in the SVD domain remains minimal. To date, we identified only a single study~\cite{zhang2024vulnerability} that applied FL for vulnerability detection. Given its ability to preserve privacy while leveraging distributed data, FL represents a highly relevant and underexplored direction for future SVD research.

Another emerging approach that has seen limited application in this field is quantum neural networks (QNNs). QNNs integrate the principles of quantum computing with neural network architectures, offering enhanced computational power and potential improvements in learning capacity. They have already demonstrated promise in a range of security-focused areas, including network anomaly detection~\cite{10020813}, supply chain attack detection~\cite{10020813}, intrusion detection~\cite{kadry2023intrusion}, objectionable content filtering~\cite{patel2019advanced}, and hardware security~\cite{beaudoin2022quantum}. However, in the context of software vulnerability detection, their application is nearly nonexistent, with only one known study~\cite{zhou2022new} exploring this avenue.

Given the demonstrated success of both FL and QNNs in related security domains, further exploration of these techniques in SVD is warranted. Leveraging these emerging technologies could significantly enhance the scalability, privacy, and accuracy of future vulnerability detection models.

\subsection{Poor Performance on Real-World Data}

One of the major challenges in software vulnerability detection (SVD) is the poor generalization of existing models to real-world scenarios. Chakraborty et al.~\cite{chakraborty2021deep} conducted experiments using several widely adopted vulnerability detection models and found a significant performance drop when these models were applied to real-world datasets. Specifically, they observed that when a pre-trained model was directly tested on real-world data, its performance dropped by approximately 70\%. Even after retraining the models with real-world data, the performance decline remained substantial—around 54\%.

For instance, the VulDeePecker model~\cite{li2018vuldeepecker} originally reported a precision of 87\%. However, when evaluated on real-world data using the pre-trained model, the precision dropped drastically to 11\%. Even after retraining, the precision only reached 18\%, which is still considered low and inadequate for practical deployment.

To address this issue, future work should focus on strategies that enhance model robustness and adaptability. Potential approaches include data augmentation, transfer learning, fine-tuning with diverse real-world data, and the integration of explainable AI techniques to improve model interpretability and generalization.

\section{Conclusion}
\label{sec:conclusion}

This study provides a comprehensive review of recent advancements in software vulnerability detection (SVD) using AI-based techniques. By systematically analyzing research published between 2018 and 2023, we developed a detailed taxonomy that captures key dimensions of source-code-based SVD approaches, including detection techniques, feature representation methods, and embedding strategies. In addition to this taxonomy, we documented the core characteristics of existing datasets and models, highlighted current limitations, and proposed future research directions to advance the field.

Our analysis shows that over 96\% of the reviewed studies employed deep learning (DL) methods, with graph-based techniques emerging as the most commonly used for both feature extraction and embedding generation. Despite notable advancements, several critical challenges persist, including limited datasets, reproducibility issues, granularity limitations, and a lack of interpretability. Granularity limitations are particularly problematic; without detailed, fine-grained information, practitioners struggle to effectively identify and remediate vulnerabilities. Addressing these challenges should be a primary focus of future research in this field. We also identified a need for greater exploration of emerging approaches, such as federated learning and quantum neural networks, which have shown promise in other security domains.

By addressing these gaps and embracing innovative methodologies, the research community can significantly improve the robustness, scalability, and practical applicability of vulnerability detection systems.
\bibliographystyle{ACM-Reference-Format}
\footnotesize
\bibliography{acmart}


\begin{thebibliography}{179}

%%% the Web address.

\ifx \showCODEN    \undefined \def \showCODEN     #1{\unskip}     \fi
\ifx \showDOI      \undefined \def \showDOI       #1{#1}\fi
\ifx \showISBNx    \undefined \def \showISBNx     #1{\unskip}     \fi
\ifx \showISBNxiii \undefined \def \showISBNxiii  #1{\unskip}     \fi
\ifx \showISSN     \undefined \def \showISSN      #1{\unskip}     \fi
\ifx \showLCCN     \undefined \def \showLCCN      #1{\unskip}     \fi
\ifx \shownote     \undefined \def \shownote      #1{#1}          \fi
\ifx \showarticletitle \undefined \def \showarticletitle #1{#1}   \fi
\ifx \showURL      \undefined \def \showURL       {\relax}        \fi
% invisible to TeX
\providecommand\bibfield[2]{#2}
\providecommand\bibinfo[2]{#2}
\providecommand\natexlab[1]{#1}
\providecommand\showeprint[2][]{arXiv:#2}

\bibitem[cve({[n.\,d.]})]%
        {cvedetails}
 \bibinfo{year}{[n.\,d.]}\natexlab{}.
\newblock \bibinfo{title}{CVE Details}.
\newblock \bibinfo{howpublished}{https://www.cvedetails.com/}.
\newblock


\bibitem[vul(2004)]%
        {vulcode-db}
 \bibinfo{year}{2004}\natexlab{}.
\newblock \bibinfo{title}{Vulnerable Code Database}.
\newblock \bibinfo{howpublished}{\url{https://www.vulncode-db.com/}}.
\newblock
\newblock
\shownote{Accessed: 2021-06-04}.


\bibitem[Afrose et~al\mbox{.}(2022)]%
        {afrose2022evaluation}
\bibfield{author}{\bibinfo{person}{Sharmin Afrose}, \bibinfo{person}{Ya Xiao}, \bibinfo{person}{Sazzadur Rahaman}, \bibinfo{person}{Barton~P Miller}, {and} \bibinfo{person}{Danfeng Yao}.} \bibinfo{year}{2022}\natexlab{}.
\newblock \showarticletitle{Evaluation of static vulnerability detection tools with Java cryptographic API benchmarks}.
\newblock \bibinfo{journal}{\emph{IEEE Transactions on Software Engineering}} \bibinfo{volume}{49}, \bibinfo{number}{2} (\bibinfo{year}{2022}), \bibinfo{pages}{485--497}.
\newblock


\bibitem[Akram and Luo(2021)]%
        {akram2021sqvdt}
\bibfield{author}{\bibinfo{person}{Junaid Akram} {and} \bibinfo{person}{Ping Luo}.} \bibinfo{year}{2021}\natexlab{}.
\newblock \showarticletitle{SQVDT: A scalable quantitative vulnerability detection technique for source code security assessment}.
\newblock \bibinfo{journal}{\emph{Software: Practice and Experience}} \bibinfo{volume}{51}, \bibinfo{number}{2} (\bibinfo{year}{2021}), \bibinfo{pages}{294--318}.
\newblock


\bibitem[Akter et~al\mbox{.}(2022)]%
        {10020813}
\bibfield{author}{\bibinfo{person}{Mst~Shapna Akter}, \bibinfo{person}{Md~Jobair~Hossain Faruk}, \bibinfo{person}{Nafisa Anjum}, \bibinfo{person}{Mohammad Masum}, \bibinfo{person}{Hossain Shahriar}, \bibinfo{person}{Nazmus Sakib}, \bibinfo{person}{Akond Rahman}, \bibinfo{person}{Fan Wu}, {and} \bibinfo{person}{Alfredo Cuzzocrea}.} \bibinfo{year}{2022}\natexlab{}.
\newblock \showarticletitle{Software Supply Chain Vulnerabilities Detection in Source Code: Performance Comparison between Traditional and Quantum Machine Learning Algorithms}. In \bibinfo{booktitle}{\emph{2022 IEEE International Conference on Big Data (Big Data)}}. \bibinfo{pages}{5639--5645}.
\newblock
\urldef\tempurl%
\url{https://doi.org/10.1109/BigData55660.2022.10020813}
\showDOI{\tempurl}


\bibitem[Al~Debeyan et~al\mbox{.}(2022)]%
        {al2022improving}
\bibfield{author}{\bibinfo{person}{Fahad Al~Debeyan}, \bibinfo{person}{Tracy Hall}, {and} \bibinfo{person}{David Bowes}.} \bibinfo{year}{2022}\natexlab{}.
\newblock \showarticletitle{Improving the performance of code vulnerability prediction using abstract syntax tree information}. In \bibinfo{booktitle}{\emph{Proceedings of the 18th International Conference on Predictive Models and Data Analytics in Software Engineering}}. \bibinfo{pages}{2--11}.
\newblock


\bibitem[Albertoni et~al\mbox{.}(2023)]%
        {albertoni2023reproducibility}
\bibfield{author}{\bibinfo{person}{Riccardo Albertoni}, \bibinfo{person}{Sara Colantonio}, \bibinfo{person}{Piotr Skrzypczy{\'n}ski}, {and} \bibinfo{person}{Jerzy Stefanowski}.} \bibinfo{year}{2023}\natexlab{}.
\newblock \showarticletitle{Reproducibility of machine learning: Terminology, recommendations and open issues}.
\newblock \bibinfo{journal}{\emph{arXiv preprint arXiv:2302.12691}} (\bibinfo{year}{2023}).
\newblock


\bibitem[Alenezi et~al\mbox{.}(2021)]%
        {alenezi2021efficient}
\bibfield{author}{\bibinfo{person}{Mamdouh Alenezi}, \bibinfo{person}{Mohammed Zagane}, {and} \bibinfo{person}{Yasir Javed}.} \bibinfo{year}{2021}\natexlab{}.
\newblock \showarticletitle{Efficient deep features learning for vulnerability detection using character n-gram embedding}.
\newblock \bibinfo{journal}{\emph{Jordanian Journal of Computers and Information Technology (JJCIT)}} \bibinfo{volume}{7}, \bibinfo{number}{01} (\bibinfo{year}{2021}).
\newblock


\bibitem[Alrashdi et~al\mbox{.}(2019)]%
        {alrashdi2019ad}
\bibfield{author}{\bibinfo{person}{Ibrahim Alrashdi}, \bibinfo{person}{Ali Alqazzaz}, \bibinfo{person}{Esam Aloufi}, \bibinfo{person}{Raed Alharthi}, \bibinfo{person}{Mohamed Zohdy}, {and} \bibinfo{person}{Hua Ming}.} \bibinfo{year}{2019}\natexlab{}.
\newblock \showarticletitle{Ad-iot: Anomaly detection of iot cyberattacks in smart city using machine learning}. In \bibinfo{booktitle}{\emph{2019 IEEE 9th Annual Computing and Communication Workshop and Conference (CCWC)}}. IEEE, \bibinfo{pages}{0305--0310}.
\newblock


\bibitem[Alves et~al\mbox{.}(2016)]%
        {alves2016software}
\bibfield{author}{\bibinfo{person}{Henrique Alves}, \bibinfo{person}{Baldoino Fonseca}, {and} \bibinfo{person}{Nuno Antunes}.} \bibinfo{year}{2016}\natexlab{}.
\newblock \showarticletitle{Software metrics and security vulnerabilities: dataset and exploratory study}. In \bibinfo{booktitle}{\emph{2016 12th European Dependable Computing Conference (EDCC)}}. IEEE, \bibinfo{pages}{37--44}.
\newblock


\bibitem[Ashawa and Morris(2019)]%
        {ashawa2019analysis}
\bibfield{author}{\bibinfo{person}{Moses~Aprofin Ashawa} {and} \bibinfo{person}{Sarah Morris}.} \bibinfo{year}{2019}\natexlab{}.
\newblock \showarticletitle{Analysis of android malware detection techniques: a systematic review}.
\newblock  (\bibinfo{year}{2019}).
\newblock


\bibitem[Beaudoin et~al\mbox{.}(2022)]%
        {beaudoin2022quantum}
\bibfield{author}{\bibinfo{person}{Collin Beaudoin}, \bibinfo{person}{Satwik Kundu}, \bibinfo{person}{Rasit~Onur Topaloglu}, {and} \bibinfo{person}{Swaroop Ghosh}.} \bibinfo{year}{2022}\natexlab{}.
\newblock \showarticletitle{Quantum machine learning for material synthesis and hardware security}. In \bibinfo{booktitle}{\emph{Proceedings of the 41st IEEE/ACM International Conference on Computer-Aided Design}}. \bibinfo{pages}{1--7}.
\newblock


\bibitem[Bhandari et~al\mbox{.}(2021)]%
        {bhandari2021cvefixes}
\bibfield{author}{\bibinfo{person}{Guru Bhandari}, \bibinfo{person}{Amara Naseer}, {and} \bibinfo{person}{Leon Moonen}.} \bibinfo{year}{2021}\natexlab{}.
\newblock \showarticletitle{CVEfixes: automated collection of vulnerabilities and their fixes from open-source software}. In \bibinfo{booktitle}{\emph{Proceedings of the 17th International Conference on Predictive Models and Data Analytics in Software Engineering}}. \bibinfo{pages}{30--39}.
\newblock


\bibitem[Bilgin et~al\mbox{.}(2020)]%
        {bilgin2020vulnerability}
\bibfield{author}{\bibinfo{person}{Zeki Bilgin}, \bibinfo{person}{Mehmet~Akif Ersoy}, \bibinfo{person}{Elif~Ustundag Soykan}, \bibinfo{person}{Emrah Tomur}, \bibinfo{person}{Pinar {\c{C}}omak}, {and} \bibinfo{person}{Leyli Kara{\c{c}}ay}.} \bibinfo{year}{2020}\natexlab{}.
\newblock \showarticletitle{Vulnerability prediction from source code using machine learning}.
\newblock \bibinfo{journal}{\emph{IEEE Access}}  \bibinfo{volume}{8} (\bibinfo{year}{2020}), \bibinfo{pages}{150672--150684}.
\newblock


\bibitem[Bowman and Huang(2020)]%
        {bowman2020vgraph}
\bibfield{author}{\bibinfo{person}{Benjamin Bowman} {and} \bibinfo{person}{H~Howie Huang}.} \bibinfo{year}{2020}\natexlab{}.
\newblock \showarticletitle{VGRAPH: A robust vulnerable code clone detection system using code property triplets}. In \bibinfo{booktitle}{\emph{2020 IEEE European Symposium on Security and Privacy (EuroS\&P)}}. IEEE, \bibinfo{pages}{53--69}.
\newblock


\bibitem[Brown et~al\mbox{.}(2020)]%
        {brown2020gpt3}
\bibfield{author}{\bibinfo{person}{Tom~B Brown}, \bibinfo{person}{Benjamin Mann}, \bibinfo{person}{Nick Ryder}, \bibinfo{person}{Melanie Subbiah}, \bibinfo{person}{Jared Kaplan}, \bibinfo{person}{Prafulla Dhariwal}, \bibinfo{person}{Arvind Neelakantan}, \bibinfo{person}{Pranav Shyam}, \bibinfo{person}{Girish Sastry}, \bibinfo{person}{Amanda Askell}, {et~al\mbox{.}}} \bibinfo{year}{2020}\natexlab{}.
\newblock \showarticletitle{Language models are few-shot learners}.
\newblock \bibinfo{journal}{\emph{Advances in Neural Information Processing Systems}}  \bibinfo{volume}{33} (\bibinfo{year}{2020}), \bibinfo{pages}{1877--1901}.
\newblock


\bibitem[Cai et~al\mbox{.}(2023)]%
        {cai2023software}
\bibfield{author}{\bibinfo{person}{Wenjing Cai}, \bibinfo{person}{Junlin Chen}, \bibinfo{person}{Jiaping Yu}, {and} \bibinfo{person}{Lipeng Gao}.} \bibinfo{year}{2023}\natexlab{}.
\newblock \showarticletitle{A software vulnerability detection method based on deep learning with complex network analysis and subgraph partition}.
\newblock \bibinfo{journal}{\emph{Information and Software Technology}}  \bibinfo{volume}{164} (\bibinfo{year}{2023}), \bibinfo{pages}{107328}.
\newblock


\bibitem[Cao et~al\mbox{.}(2020)]%
        {cao2020ftclnet}
\bibfield{author}{\bibinfo{person}{Defu Cao}, \bibinfo{person}{Jing Huang}, \bibinfo{person}{Xuanyu Zhang}, {and} \bibinfo{person}{Xianhua Liu}.} \bibinfo{year}{2020}\natexlab{}.
\newblock \showarticletitle{FTCLNet: Convolutional LSTM with Fourier transform for vulnerability detection}. In \bibinfo{booktitle}{\emph{2020 IEEE 19th International Conference on Trust, Security and Privacy in Computing and Communications (TrustCom)}}. IEEE, \bibinfo{pages}{539--546}.
\newblock


\bibitem[Cao et~al\mbox{.}(2021)]%
        {cao2021bgnn4vd}
\bibfield{author}{\bibinfo{person}{Sicong Cao}, \bibinfo{person}{Xiaobing Sun}, \bibinfo{person}{Lili Bo}, \bibinfo{person}{Ying Wei}, {and} \bibinfo{person}{Bin Li}.} \bibinfo{year}{2021}\natexlab{}.
\newblock \showarticletitle{Bgnn4vd: Constructing bidirectional graph neural-network for vulnerability detection}.
\newblock \bibinfo{journal}{\emph{Information and Software Technology}}  \bibinfo{volume}{136} (\bibinfo{year}{2021}), \bibinfo{pages}{106576}.
\newblock


\bibitem[Cao et~al\mbox{.}(2022)]%
        {cao2022mvd}
\bibfield{author}{\bibinfo{person}{Sicong Cao}, \bibinfo{person}{Xiaobing Sun}, \bibinfo{person}{Lili Bo}, \bibinfo{person}{Rongxin Wu}, \bibinfo{person}{Bin Li}, {and} \bibinfo{person}{Chuanqi Tao}.} \bibinfo{year}{2022}\natexlab{}.
\newblock \showarticletitle{MVD: memory-related vulnerability detection based on flow-sensitive graph neural networks}. In \bibinfo{booktitle}{\emph{Proceedings of the 44th International Conference on Software Engineering}}. \bibinfo{pages}{1456--1468}.
\newblock


\bibitem[Chakraborty et~al\mbox{.}(2021)]%
        {chakraborty2021deep}
\bibfield{author}{\bibinfo{person}{Saikat Chakraborty}, \bibinfo{person}{Rahul Krishna}, \bibinfo{person}{Yangruibo Ding}, {and} \bibinfo{person}{Baishakhi Ray}.} \bibinfo{year}{2021}\natexlab{}.
\newblock \showarticletitle{Deep learning based vulnerability detection: Are we there yet}.
\newblock \bibinfo{journal}{\emph{IEEE Transactions on Software Engineering}} (\bibinfo{year}{2021}).
\newblock


\bibitem[Chakraborty et~al\mbox{.}(2022)]%
        {9448435}
\bibfield{author}{\bibinfo{person}{Saikat Chakraborty}, \bibinfo{person}{Rahul Krishna}, \bibinfo{person}{Yangruibo Ding}, {and} \bibinfo{person}{Baishakhi Ray}.} \bibinfo{year}{2022}\natexlab{}.
\newblock \showarticletitle{Deep Learning Based Vulnerability Detection: Are We There Yet?}
\newblock \bibinfo{journal}{\emph{IEEE Transactions on Software Engineering}} \bibinfo{volume}{48}, \bibinfo{number}{9} (\bibinfo{year}{2022}), \bibinfo{pages}{3280--3296}.
\newblock
\urldef\tempurl%
\url{https://doi.org/10.1109/TSE.2021.3087402}
\showDOI{\tempurl}


\bibitem[Chang et~al\mbox{.}(2023)]%
        {chang2023vdda}
\bibfield{author}{\bibinfo{person}{Jiaqi Chang}, \bibinfo{person}{Zhujuan Ma}, \bibinfo{person}{Binghao Cao}, {and} \bibinfo{person}{Erzhou Zhu}.} \bibinfo{year}{2023}\natexlab{}.
\newblock \showarticletitle{VDDA: An Effective Software Vulnerability Detection Model Based on Deep Learning and Attention Mechanism}. In \bibinfo{booktitle}{\emph{2023 26th International Conference on Computer Supported Cooperative Work in Design (CSCWD)}}. IEEE, \bibinfo{pages}{474--479}.
\newblock


\bibitem[Chen et~al\mbox{.}(2023)]%
        {chen2023diversevul}
\bibfield{author}{\bibinfo{person}{Yizheng Chen}, \bibinfo{person}{Zhoujie Ding}, \bibinfo{person}{Lamya Alowain}, \bibinfo{person}{Xinyun Chen}, {and} \bibinfo{person}{David Wagner}.} \bibinfo{year}{2023}\natexlab{}.
\newblock \showarticletitle{Diversevul: A new vulnerable source code dataset for deep learning based vulnerability detection}. In \bibinfo{booktitle}{\emph{Proceedings of the 26th International Symposium on Research in Attacks, Intrusions and Defenses}}. \bibinfo{pages}{654--668}.
\newblock


\bibitem[Chen and Liu(2022)]%
        {chen2022hlt}
\bibfield{author}{\bibinfo{person}{Yupan Chen} {and} \bibinfo{person}{Zhihong Liu}.} \bibinfo{year}{2022}\natexlab{}.
\newblock \showarticletitle{HLT: A Hierarchical Vulnerability Detection Model Based on Transformer}. In \bibinfo{booktitle}{\emph{2022 4th International Conference on Data Intelligence and Security (ICDIS)}}. IEEE, \bibinfo{pages}{50--54}.
\newblock


\bibitem[Cheng et~al\mbox{.}({[n.\,d.]})]%
        {cheng4567888vulnerability}
\bibfield{author}{\bibinfo{person}{Ge Cheng}, \bibinfo{person}{Qifan Luo}, {and} \bibinfo{person}{Yun Zhang}.} \bibinfo{year}{[n.\,d.]}\natexlab{}.
\newblock \showarticletitle{Vulnerability Detection with Feature Fusion and Learnable Edge-Type Embedding Graph Neural Network}.
\newblock \bibinfo{journal}{\emph{Available at SSRN 4567888}} (\bibinfo{year}{[n.\,d.]}).
\newblock


\bibitem[Cheng et~al\mbox{.}(2019)]%
        {cheng2019static}
\bibfield{author}{\bibinfo{person}{Xiao Cheng}, \bibinfo{person}{Haoyu Wang}, \bibinfo{person}{Jiayi Hua}, \bibinfo{person}{Miao Zhang}, \bibinfo{person}{Guoai Xu}, \bibinfo{person}{Li Yi}, {and} \bibinfo{person}{Yulei Sui}.} \bibinfo{year}{2019}\natexlab{}.
\newblock \showarticletitle{Static detection of control-flow-related vulnerabilities using graph embedding}. In \bibinfo{booktitle}{\emph{2019 24th International Conference on Engineering of Complex Computer Systems (ICECCS)}}. IEEE, \bibinfo{pages}{41--50}.
\newblock


\bibitem[Cheng et~al\mbox{.}(2022)]%
        {cheng2022path}
\bibfield{author}{\bibinfo{person}{Xiao Cheng}, \bibinfo{person}{Guanqin Zhang}, \bibinfo{person}{Haoyu Wang}, {and} \bibinfo{person}{Yulei Sui}.} \bibinfo{year}{2022}\natexlab{}.
\newblock \showarticletitle{Path-sensitive code embedding via contrastive learning for software vulnerability detection}. In \bibinfo{booktitle}{\emph{Proceedings of the 31st ACM SIGSOFT International Symposium on Software Testing and Analysis}}. \bibinfo{pages}{519--531}.
\newblock


\bibitem[Croft et~al\mbox{.}(2023)]%
        {croft2023data}
\bibfield{author}{\bibinfo{person}{Roland Croft}, \bibinfo{person}{M~Ali Babar}, {and} \bibinfo{person}{M~Mehdi Kholoosi}.} \bibinfo{year}{2023}\natexlab{}.
\newblock \showarticletitle{Data quality for software vulnerability datasets}. In \bibinfo{booktitle}{\emph{2023 IEEE/ACM 45th International Conference on Software Engineering (ICSE)}}. IEEE, \bibinfo{pages}{121--133}.
\newblock


\bibitem[Cui et~al\mbox{.}(2020)]%
        {cui2020vuldetector}
\bibfield{author}{\bibinfo{person}{Lei Cui}, \bibinfo{person}{Zhiyu Hao}, \bibinfo{person}{Yang Jiao}, \bibinfo{person}{Haiqiang Fei}, {and} \bibinfo{person}{Xiaochun Yun}.} \bibinfo{year}{2020}\natexlab{}.
\newblock \showarticletitle{Vuldetector: Detecting vulnerabilities using weighted feature graph comparison}.
\newblock \bibinfo{journal}{\emph{IEEE Transactions on Information Forensics and Security}}  \bibinfo{volume}{16} (\bibinfo{year}{2020}), \bibinfo{pages}{2004--2017}.
\newblock


\bibitem[Dam et~al\mbox{.}(2018)]%
        {dam2018automatic}
\bibfield{author}{\bibinfo{person}{Hoa~Khanh Dam}, \bibinfo{person}{Truyen Tran}, \bibinfo{person}{Trang Pham}, \bibinfo{person}{Shien~Wee Ng}, \bibinfo{person}{John Grundy}, {and} \bibinfo{person}{Aditya Ghose}.} \bibinfo{year}{2018}\natexlab{}.
\newblock \showarticletitle{Automatic feature learning for predicting vulnerable software components}.
\newblock \bibinfo{journal}{\emph{IEEE Transactions on Software Engineering}} \bibinfo{volume}{47}, \bibinfo{number}{1} (\bibinfo{year}{2018}), \bibinfo{pages}{67--85}.
\newblock


\bibitem[De~Kraker et~al\mbox{.}(2023)]%
        {de2023glice}
\bibfield{author}{\bibinfo{person}{Wesley De~Kraker}, \bibinfo{person}{Harald Vranken}, {and} \bibinfo{person}{Arjen Hommmersom}.} \bibinfo{year}{2023}\natexlab{}.
\newblock \showarticletitle{GLICE: Combining Graph Neural Networks and Program Slicing to Improve Software Vulnerability Detection}. In \bibinfo{booktitle}{\emph{2023 IEEE European Symposium on Security and Privacy Workshops (EuroS\&PW)}}. IEEE, \bibinfo{pages}{34--41}.
\newblock


\bibitem[Devlin et~al\mbox{.}(2018)]%
        {devlin2018bert}
\bibfield{author}{\bibinfo{person}{Jacob Devlin}, \bibinfo{person}{Ming-Wei Chang}, \bibinfo{person}{Kenton Lee}, {and} \bibinfo{person}{Kristina Toutanova}.} \bibinfo{year}{2018}\natexlab{}.
\newblock \showarticletitle{Bert: Pre-training of deep bidirectional transformers for language understanding}.
\newblock \bibinfo{journal}{\emph{arXiv preprint arXiv:1810.04805}} (\bibinfo{year}{2018}).
\newblock


\bibitem[Ding et~al\mbox{.}(2023)]%
        {ding2023leveraging}
\bibfield{author}{\bibinfo{person}{Yue Ding}, \bibinfo{person}{Qian Wu}, \bibinfo{person}{Yinzhu Li}, \bibinfo{person}{Dongdong Wang}, {and} \bibinfo{person}{Jiaxin Huang}.} \bibinfo{year}{2023}\natexlab{}.
\newblock \showarticletitle{Leveraging Deep Learning Models for Cross-function Null Pointer Risks Detection}. In \bibinfo{booktitle}{\emph{2023 IEEE International Conference On Artificial Intelligence Testing (AITest)}}. IEEE, \bibinfo{pages}{107--113}.
\newblock


\bibitem[Dong et~al\mbox{.}(2023)]%
        {dong2023sedsvd}
\bibfield{author}{\bibinfo{person}{Yukun Dong}, \bibinfo{person}{Yeer Tang}, \bibinfo{person}{Xiaotong Cheng}, \bibinfo{person}{Yufei Yang}, {and} \bibinfo{person}{Shuqi Wang}.} \bibinfo{year}{2023}\natexlab{}.
\newblock \showarticletitle{SedSVD: Statement-level software vulnerability detection based on Relational Graph Convolutional Network with subgraph embedding}.
\newblock \bibinfo{journal}{\emph{Information and Software Technology}}  \bibinfo{volume}{158} (\bibinfo{year}{2023}), \bibinfo{pages}{107168}.
\newblock


\bibitem[Du et~al\mbox{.}(2023a)]%
        {du2023cross}
\bibfield{author}{\bibinfo{person}{Gewangzi Du}, \bibinfo{person}{Liwei Chen}, \bibinfo{person}{Tongshuai Wu}, \bibinfo{person}{Xiong Zheng}, \bibinfo{person}{Ningning Cui}, {and} \bibinfo{person}{Gang Shi}.} \bibinfo{year}{2023}\natexlab{a}.
\newblock \showarticletitle{Cross Domain on Snippets: BiLSTM-TextCNN based Vulnerability Detection with Domain Adaptation}. In \bibinfo{booktitle}{\emph{2023 26th International Conference on Computer Supported Cooperative Work in Design (CSCWD)}}. IEEE, \bibinfo{pages}{1896--1901}.
\newblock


\bibitem[Du et~al\mbox{.}(2023b)]%
        {du2023automated}
\bibfield{author}{\bibinfo{person}{Qianjin Du}, \bibinfo{person}{Wei Kun}, \bibinfo{person}{Xiaohui Kuang}, \bibinfo{person}{Xiang Li}, {and} \bibinfo{person}{Gang Zhao}.} \bibinfo{year}{2023}\natexlab{b}.
\newblock \showarticletitle{Automated Software Vulnerability Detection via Curriculum Learning}. In \bibinfo{booktitle}{\emph{2023 IEEE International Conference on Multimedia and Expo (ICME)}}. IEEE, \bibinfo{pages}{2855--2860}.
\newblock


\bibitem[Duan et~al\mbox{.}(2021)]%
        {duan2021multicode}
\bibfield{author}{\bibinfo{person}{Xu Duan}, \bibinfo{person}{Jingzheng Wu}, \bibinfo{person}{Mengnan Du}, \bibinfo{person}{Tianyue Luo}, \bibinfo{person}{Mutian Yang}, {and} \bibinfo{person}{Yanjun Wu}.} \bibinfo{year}{2021}\natexlab{}.
\newblock \showarticletitle{MultiCode: A Unified Code Analysis Framework based on Multi-type and Multi-granularity Semantic Learning}. In \bibinfo{booktitle}{\emph{2021 IEEE International Symposium on Software Reliability Engineering Workshops (ISSREW)}}. IEEE, \bibinfo{pages}{359--364}.
\newblock


\bibitem[Duan et~al\mbox{.}(2019)]%
        {duan2019vulsniper}
\bibfield{author}{\bibinfo{person}{Xu Duan}, \bibinfo{person}{Jingzheng Wu}, \bibinfo{person}{Shouling Ji}, \bibinfo{person}{Zhiqing Rui}, \bibinfo{person}{Tianyue Luo}, \bibinfo{person}{Mutian Yang}, {and} \bibinfo{person}{Yanjun Wu}.} \bibinfo{year}{2019}\natexlab{}.
\newblock \showarticletitle{VulSniper: Focus Your Attention to Shoot Fine-Grained Vulnerabilities.}. In \bibinfo{booktitle}{\emph{IJCAI}}. \bibinfo{pages}{4665--4671}.
\newblock


\bibitem[Eberendu et~al\mbox{.}(2022)]%
        {eberendu2022systematic}
\bibfield{author}{\bibinfo{person}{Adanma~Cecilia Eberendu}, \bibinfo{person}{Valentine~Ikechukwu Udegbe}, \bibinfo{person}{Edmond~Onwubiko Ezennorom}, \bibinfo{person}{Anita~Chinonso Ibegbulam}, \bibinfo{person}{Titus~Ifeanyi Chinebu}, {et~al\mbox{.}}} \bibinfo{year}{2022}\natexlab{}.
\newblock \showarticletitle{A systematic literature review of software vulnerability detection}.
\newblock \bibinfo{journal}{\emph{European Journal of Computer Science and Information Technology}} \bibinfo{volume}{10}, \bibinfo{number}{1} (\bibinfo{year}{2022}), \bibinfo{pages}{23--37}.
\newblock


\bibitem[Fan et~al\mbox{.}(2020)]%
        {fan2020ac}
\bibfield{author}{\bibinfo{person}{Jiahao Fan}, \bibinfo{person}{Yi Li}, \bibinfo{person}{Shaohua Wang}, {and} \bibinfo{person}{Tien~N Nguyen}.} \bibinfo{year}{2020}\natexlab{}.
\newblock \showarticletitle{AC/C++ code vulnerability dataset with code changes and CVE summaries}. In \bibinfo{booktitle}{\emph{Proceedings of the 17th International Conference on Mining Software Repositories}}. \bibinfo{pages}{508--512}.
\newblock


\bibitem[Feng et~al\mbox{.}(2020)]%
        {feng2020efficient}
\bibfield{author}{\bibinfo{person}{Hantao Feng}, \bibinfo{person}{Xiaotong Fu}, \bibinfo{person}{Hongyu Sun}, \bibinfo{person}{He Wang}, {and} \bibinfo{person}{Yuqing Zhang}.} \bibinfo{year}{2020}\natexlab{}.
\newblock \showarticletitle{Efficient vulnerability detection based on abstract syntax tree and deep learning}. In \bibinfo{booktitle}{\emph{IEEE INFOCOM 2020-IEEE Conference on Computer Communications Workshops (INFOCOM WKSHPS)}}. IEEE, \bibinfo{pages}{722--727}.
\newblock


\bibitem[Feurer et~al\mbox{.}(2015)]%
        {feurer2015efficient}
\bibfield{author}{\bibinfo{person}{Matthias Feurer}, \bibinfo{person}{Aaron Klein}, \bibinfo{person}{Katharina Eggensperger}, \bibinfo{person}{Jost Springenberg}, \bibinfo{person}{Manuel Blum}, {and} \bibinfo{person}{Frank Hutter}.} \bibinfo{year}{2015}\natexlab{}.
\newblock \showarticletitle{Efficient and robust automated machine learning}.
\newblock \bibinfo{journal}{\emph{Advances in neural information processing systems}}  \bibinfo{volume}{28} (\bibinfo{year}{2015}).
\newblock


\bibitem[Gear et~al\mbox{.}(2023)]%
        {gear2023software}
\bibfield{author}{\bibinfo{person}{Joseph Gear}, \bibinfo{person}{Yue Xu}, \bibinfo{person}{Ernest Foo}, \bibinfo{person}{Praveen Gauravaram}, \bibinfo{person}{Zahra Jadidi}, {and} \bibinfo{person}{Leonie Simpson}.} \bibinfo{year}{2023}\natexlab{}.
\newblock \showarticletitle{Software Vulnerability Detection Using Informed Code Graph Pruning}.
\newblock \bibinfo{journal}{\emph{IEEE Access}}  \bibinfo{volume}{11} (\bibinfo{year}{2023}), \bibinfo{pages}{135626--135644}.
\newblock


\bibitem[Ghaffarian and Shahriari(2017a)]%
        {p127}
\bibfield{author}{\bibinfo{person}{Seyed~Mohammad Ghaffarian} {and} \bibinfo{person}{Hamid~Reza Shahriari}.} \bibinfo{year}{2017}\natexlab{a}.
\newblock \showarticletitle{Software vulnerability analysis and discovery using machine-learning and data-mining techniques: A survey}.
\newblock \bibinfo{journal}{\emph{ACM Computing Surveys (CSUR)}} \bibinfo{volume}{50}, \bibinfo{number}{4} (\bibinfo{year}{2017}), \bibinfo{pages}{1--36}.
\newblock


\bibitem[Ghaffarian and Shahriari(2017b)]%
        {ghaffarian2017software}
\bibfield{author}{\bibinfo{person}{Seyed~Mohammad Ghaffarian} {and} \bibinfo{person}{Hamid~Reza Shahriari}.} \bibinfo{year}{2017}\natexlab{b}.
\newblock \showarticletitle{Software vulnerability analysis and discovery using machine-learning and data-mining techniques: A survey}.
\newblock \bibinfo{journal}{\emph{ACM Computing Surveys (CSUR)}} \bibinfo{volume}{50}, \bibinfo{number}{4} (\bibinfo{year}{2017}), \bibinfo{pages}{1--36}.
\newblock


\bibitem[Gu et~al\mbox{.}(2022)]%
        {gu2022hierarchical}
\bibfield{author}{\bibinfo{person}{Mianxue Gu}, \bibinfo{person}{Hantao Feng}, \bibinfo{person}{Hongyu Sun}, \bibinfo{person}{Peng Liu}, \bibinfo{person}{Qiuling Yue}, \bibinfo{person}{Jinglu Hu}, \bibinfo{person}{Chunjie Cao}, {and} \bibinfo{person}{Yuqing Zhang}.} \bibinfo{year}{2022}\natexlab{}.
\newblock \showarticletitle{Hierarchical Attention Network for Interpretable and Fine-Grained Vulnerability Detection}. In \bibinfo{booktitle}{\emph{IEEE INFOCOM 2022-IEEE Conference on Computer Communications Workshops (INFOCOM WKSHPS)}}. IEEE, \bibinfo{pages}{1--6}.
\newblock


\bibitem[Guo et~al\mbox{.}(2022)]%
        {guo2022hyvuldect}
\bibfield{author}{\bibinfo{person}{Wenbo Guo}, \bibinfo{person}{Yong Fang}, \bibinfo{person}{Cheng Huang}, \bibinfo{person}{Haoran Ou}, \bibinfo{person}{Chun Lin}, {and} \bibinfo{person}{Yongyan Guo}.} \bibinfo{year}{2022}\natexlab{}.
\newblock \showarticletitle{HyVulDect: A hybrid semantic vulnerability mining system based on Graph Neural Network}.
\newblock \bibinfo{journal}{\emph{Computers \& Security}} (\bibinfo{year}{2022}), \bibinfo{pages}{102823}.
\newblock


\bibitem[Han et~al\mbox{.}(2019)]%
        {han2019optimized}
\bibfield{author}{\bibinfo{person}{Lansheng Han}, \bibinfo{person}{Man Zhou}, \bibinfo{person}{Yekui Qian}, \bibinfo{person}{Cai Fu}, {and} \bibinfo{person}{Deqing Zou}.} \bibinfo{year}{2019}\natexlab{}.
\newblock \showarticletitle{An optimized static propositional function model to detect software vulnerability}.
\newblock \bibinfo{journal}{\emph{IEEE Access}}  \bibinfo{volume}{7} (\bibinfo{year}{2019}), \bibinfo{pages}{143499--143510}.
\newblock


\bibitem[Hanif and Maffeis(2022)]%
        {hanif2022vulberta}
\bibfield{author}{\bibinfo{person}{Hazim Hanif} {and} \bibinfo{person}{Sergio Maffeis}.} \bibinfo{year}{2022}\natexlab{}.
\newblock \showarticletitle{Vulberta: Simplified source code pre-training for vulnerability detection}. In \bibinfo{booktitle}{\emph{2022 International joint conference on neural networks (IJCNN)}}. IEEE, \bibinfo{pages}{1--8}.
\newblock


\bibitem[Hanif et~al\mbox{.}(2021)]%
        {p126}
\bibfield{author}{\bibinfo{person}{Hazim Hanif}, \bibinfo{person}{Mohd Hairul Nizam~Md Nasir}, \bibinfo{person}{Mohd~Faizal Ab~Razak}, \bibinfo{person}{Ahmad Firdaus}, {and} \bibinfo{person}{Nor~Badrul Anuar}.} \bibinfo{year}{2021}\natexlab{}.
\newblock \showarticletitle{The rise of software vulnerability: Taxonomy of software vulnerabilities detection and machine learning approaches}.
\newblock \bibinfo{journal}{\emph{Journal of Network and Computer Applications}}  \bibinfo{volume}{179} (\bibinfo{year}{2021}), \bibinfo{pages}{103009}.
\newblock


\bibitem[Hao et~al\mbox{.}(2023)]%
        {hao2023vd}
\bibfield{author}{\bibinfo{person}{Jingwei Hao}, \bibinfo{person}{Senlin Luo}, \bibinfo{person}{Limin Pan}, {and} \bibinfo{person}{Chuantao Chen}.} \bibinfo{year}{2023}\natexlab{}.
\newblock \showarticletitle{VD-HEN: Capturing Semantic Dependencies for Source Code Vulnerability Detection With a Hierarchical Embedding Network}.
\newblock \bibinfo{journal}{\emph{Computer}} \bibinfo{volume}{56}, \bibinfo{number}{10} (\bibinfo{year}{2023}), \bibinfo{pages}{49--61}.
\newblock


\bibitem[Hei et~al\mbox{.}(2020)]%
        {hei2020trusted}
\bibfield{author}{\bibinfo{person}{Xinhong Hei}, \bibinfo{person}{Xinyue Yin}, \bibinfo{person}{Yichuan Wang}, \bibinfo{person}{Ju Ren}, {and} \bibinfo{person}{Lei Zhu}.} \bibinfo{year}{2020}\natexlab{}.
\newblock \showarticletitle{A trusted feature aggregator federated learning for distributed malicious attack detection}.
\newblock \bibinfo{journal}{\emph{Computers \& Security}}  \bibinfo{volume}{99} (\bibinfo{year}{2020}), \bibinfo{pages}{102033}.
\newblock


\bibitem[Hin et~al\mbox{.}(2022)]%
        {hin2022linevd}
\bibfield{author}{\bibinfo{person}{David Hin}, \bibinfo{person}{Andrey Kan}, \bibinfo{person}{Huaming Chen}, {and} \bibinfo{person}{M~Ali Babar}.} \bibinfo{year}{2022}\natexlab{}.
\newblock \showarticletitle{LineVD: Statement-level vulnerability detection using graph neural networks}. In \bibinfo{booktitle}{\emph{Proceedings of the 19th International Conference on Mining Software Repositories}}. \bibinfo{pages}{596--607}.
\newblock


\bibitem[Hong et~al\mbox{.}(2022)]%
        {hong2022xvdb}
\bibfield{author}{\bibinfo{person}{Hyunji Hong}, \bibinfo{person}{Seunghoon Woo}, \bibinfo{person}{Eunjin Choi}, \bibinfo{person}{Jihyun Choi}, {and} \bibinfo{person}{Heejo Lee}.} \bibinfo{year}{2022}\natexlab{}.
\newblock \showarticletitle{xVDB: A high-coverage approach for constructing a vulnerability database}.
\newblock \bibinfo{journal}{\emph{IEEE Access}}  \bibinfo{volume}{10} (\bibinfo{year}{2022}), \bibinfo{pages}{85050--85063}.
\newblock


\bibitem[IBM({[n.\,d.]})]%
        {ibm_self_supervised}
\bibfield{author}{\bibinfo{person}{IBM}.} \bibinfo{year}{[n.\,d.]}\natexlab{}.
\newblock \bibinfo{title}{What is Self-Supervised Learning?}
\newblock
\newblock
\urldef\tempurl%
\url{https://www.ibm.com/think/topics/self-supervised-learning}
\showURL{%
\tempurl}
\newblock
\shownote{Accessed: 2025-01-24}.


\bibitem[Jeon and Kim(2021)]%
        {jeon2021autovas}
\bibfield{author}{\bibinfo{person}{Sanghoon Jeon} {and} \bibinfo{person}{Huy~Kang Kim}.} \bibinfo{year}{2021}\natexlab{}.
\newblock \showarticletitle{AutoVAS: An automated vulnerability analysis system with a deep learning approach}.
\newblock \bibinfo{journal}{\emph{Computers \& Security}}  \bibinfo{volume}{106} (\bibinfo{year}{2021}), \bibinfo{pages}{102308}.
\newblock


\bibitem[Jie et~al\mbox{.}(2016)]%
        {7866201}
\bibfield{author}{\bibinfo{person}{Gong Jie}, \bibinfo{person}{Kuang Xiao-Hui}, {and} \bibinfo{person}{Liu Qiang}.} \bibinfo{year}{2016}\natexlab{}.
\newblock \showarticletitle{Survey on Software Vulnerability Analysis Method Based on Machine Learning}. In \bibinfo{booktitle}{\emph{2016 IEEE First International Conference on Data Science in Cyberspace (DSC)}}. \bibinfo{pages}{642--647}.
\newblock
\urldef\tempurl%
\url{https://doi.org/10.1109/DSC.2016.33}
\showDOI{\tempurl}


\bibitem[Kadry et~al\mbox{.}(2023)]%
        {kadry2023intrusion}
\bibfield{author}{\bibinfo{person}{Heba Kadry}, \bibinfo{person}{Ahmed Farouk}, \bibinfo{person}{Elnomery~A Zanaty}, {and} \bibinfo{person}{Omar Reyad}.} \bibinfo{year}{2023}\natexlab{}.
\newblock \showarticletitle{Intrusion detection model using optimized quantum neural network and elliptical curve cryptography for data security}.
\newblock \bibinfo{journal}{\emph{Alexandria Engineering Journal}}  \bibinfo{volume}{71} (\bibinfo{year}{2023}), \bibinfo{pages}{491--500}.
\newblock


\bibitem[Kang et~al\mbox{.}(2022)]%
        {kang2022tracer}
\bibfield{author}{\bibinfo{person}{Wooseok Kang}, \bibinfo{person}{Byoungho Son}, {and} \bibinfo{person}{Kihong Heo}.} \bibinfo{year}{2022}\natexlab{}.
\newblock \showarticletitle{TRACER: signature-based static analysis for detecting recurring vulnerabilities}. In \bibinfo{booktitle}{\emph{Proceedings of the 2022 ACM SIGSAC Conference on Computer and Communications Security}}. \bibinfo{pages}{1695--1708}.
\newblock


\bibitem[Kim et~al\mbox{.}(2022)]%
        {kim2022vuldebert}
\bibfield{author}{\bibinfo{person}{Soolin Kim}, \bibinfo{person}{Jusop Choi}, \bibinfo{person}{Muhammad~Ejaz Ahmed}, \bibinfo{person}{Surya Nepal}, {and} \bibinfo{person}{Hyoungshick Kim}.} \bibinfo{year}{2022}\natexlab{}.
\newblock \showarticletitle{VulDeBERT: A Vulnerability Detection System Using BERT}. In \bibinfo{booktitle}{\emph{2022 IEEE International Symposium on Software Reliability Engineering Workshops (ISSREW)}}. IEEE, \bibinfo{pages}{69--74}.
\newblock


\bibitem[Kluban et~al\mbox{.}(2022)]%
        {kluban2022measuring}
\bibfield{author}{\bibinfo{person}{Maryna Kluban}, \bibinfo{person}{Mohammad Mannan}, {and} \bibinfo{person}{Amr Youssef}.} \bibinfo{year}{2022}\natexlab{}.
\newblock \showarticletitle{On measuring vulnerable javascript functions in the wild}. In \bibinfo{booktitle}{\emph{Proceedings of the 2022 ACM on Asia Conference on Computer and Communications Security}}. \bibinfo{pages}{917--930}.
\newblock


\bibitem[Li et~al\mbox{.}(2020b)]%
        {li2020deepfed}
\bibfield{author}{\bibinfo{person}{Beibei Li}, \bibinfo{person}{Yuhao Wu}, \bibinfo{person}{Jiarui Song}, \bibinfo{person}{Rongxing Lu}, \bibinfo{person}{Tao Li}, {and} \bibinfo{person}{Liang Zhao}.} \bibinfo{year}{2020}\natexlab{b}.
\newblock \showarticletitle{DeepFed: Federated deep learning for intrusion detection in industrial cyber--physical systems}.
\newblock \bibinfo{journal}{\emph{IEEE Transactions on Industrial Informatics}} \bibinfo{volume}{17}, \bibinfo{number}{8} (\bibinfo{year}{2020}), \bibinfo{pages}{5615--5624}.
\newblock


\bibitem[Li et~al\mbox{.}(2021a)]%
        {li2021acgvd}
\bibfield{author}{\bibinfo{person}{Min Li}, \bibinfo{person}{Chunfang Li}, \bibinfo{person}{Shuailou Li}, \bibinfo{person}{Yanna Wu}, \bibinfo{person}{Boyang Zhang}, {and} \bibinfo{person}{Yu Wen}.} \bibinfo{year}{2021}\natexlab{a}.
\newblock \showarticletitle{Acgvd: Vulnerability detection based on comprehensive graph via graph neural network with attention}. In \bibinfo{booktitle}{\emph{Information and Communications Security: 23rd International Conference, ICICS 2021, Chongqing, China, November 19-21, 2021, Proceedings, Part I 23}}. Springer, \bibinfo{pages}{243--259}.
\newblock


\bibitem[Li et~al\mbox{.}(2023a)]%
        {li2023vulnerability}
\bibfield{author}{\bibinfo{person}{Wei Li}, \bibinfo{person}{Xiang Li}, \bibinfo{person}{Wanzheng Feng}, \bibinfo{person}{Guanglu Jin}, \bibinfo{person}{Zhihan Liu}, {and} \bibinfo{person}{Jing Jia}.} \bibinfo{year}{2023}\natexlab{a}.
\newblock \showarticletitle{Vulnerability Detection Based on Unified Code Property Graph}. In \bibinfo{booktitle}{\emph{International Conference on Web Information Systems and Applications}}. Springer, \bibinfo{pages}{359--370}.
\newblock


\bibitem[Li et~al\mbox{.}(2020a)]%
        {li2020automated}
\bibfield{author}{\bibinfo{person}{Xin Li}, \bibinfo{person}{Lu Wang}, \bibinfo{person}{Yang Xin}, \bibinfo{person}{Yixian Yang}, {and} \bibinfo{person}{Yuling Chen}.} \bibinfo{year}{2020}\natexlab{a}.
\newblock \showarticletitle{Automated vulnerability detection in source code using minimum intermediate representation learning}.
\newblock \bibinfo{journal}{\emph{Applied Sciences}} \bibinfo{volume}{10}, \bibinfo{number}{5} (\bibinfo{year}{2020}), \bibinfo{pages}{1692}.
\newblock


\bibitem[Li et~al\mbox{.}(2021b)]%
        {li2021vulnerability}
\bibfield{author}{\bibinfo{person}{Yi Li}, \bibinfo{person}{Shaohua Wang}, {and} \bibinfo{person}{Tien~N Nguyen}.} \bibinfo{year}{2021}\natexlab{b}.
\newblock \showarticletitle{Vulnerability detection with fine-grained interpretations}. In \bibinfo{booktitle}{\emph{Proceedings of the 29th ACM Joint Meeting on European Software Engineering Conference and Symposium on the Foundations of Software Engineering}}. \bibinfo{pages}{292--303}.
\newblock


\bibitem[Li et~al\mbox{.}(2023b)]%
        {li2023commit}
\bibfield{author}{\bibinfo{person}{Yi Li}, \bibinfo{person}{Aashish Yadavally}, \bibinfo{person}{Jiaxing Zhang}, \bibinfo{person}{Shaohua Wang}, {and} \bibinfo{person}{Tien~N Nguyen}.} \bibinfo{year}{2023}\natexlab{b}.
\newblock \showarticletitle{Commit-Level, Neural Vulnerability Detection and Assessment}. In \bibinfo{booktitle}{\emph{Proceedings of the 31st ACM Joint European Software Engineering Conference and Symposium on the Foundations of Software Engineering}}. \bibinfo{pages}{1024--1036}.
\newblock


\bibitem[Li et~al\mbox{.}(2021c)]%
        {li2021vuldeelocator}
\bibfield{author}{\bibinfo{person}{Zhen Li}, \bibinfo{person}{Deqing Zou}, \bibinfo{person}{Shouhuai Xu}, \bibinfo{person}{Zhaoxuan Chen}, \bibinfo{person}{Yawei Zhu}, {and} \bibinfo{person}{Hai Jin}.} \bibinfo{year}{2021}\natexlab{c}.
\newblock \showarticletitle{Vuldeelocator: a deep learning-based fine-grained vulnerability detector}.
\newblock \bibinfo{journal}{\emph{IEEE Transactions on Dependable and Secure Computing}} \bibinfo{volume}{19}, \bibinfo{number}{4} (\bibinfo{year}{2021}), \bibinfo{pages}{2821--2837}.
\newblock


\bibitem[Li et~al\mbox{.}(2021d)]%
        {li2021sysevr}
\bibfield{author}{\bibinfo{person}{Zhen Li}, \bibinfo{person}{Deqing Zou}, \bibinfo{person}{Shouhuai Xu}, \bibinfo{person}{Hai Jin}, \bibinfo{person}{Yawei Zhu}, {and} \bibinfo{person}{Zhaoxuan Chen}.} \bibinfo{year}{2021}\natexlab{d}.
\newblock \showarticletitle{Sysevr: A framework for using deep learning to detect software vulnerabilities}.
\newblock \bibinfo{journal}{\emph{IEEE Transactions on Dependable and Secure Computing}} \bibinfo{volume}{19}, \bibinfo{number}{4} (\bibinfo{year}{2021}), \bibinfo{pages}{2244--2258}.
\newblock


\bibitem[Li et~al\mbox{.}(2018)]%
        {li2018vuldeepecker}
\bibfield{author}{\bibinfo{person}{Zhen Li}, \bibinfo{person}{Deqing Zou}, \bibinfo{person}{Shouhuai Xu}, \bibinfo{person}{Xinyu Ou}, \bibinfo{person}{Hai Jin}, \bibinfo{person}{Sujuan Wang}, \bibinfo{person}{Zhijun Deng}, {and} \bibinfo{person}{Yuyi Zhong}.} \bibinfo{year}{2018}\natexlab{}.
\newblock \showarticletitle{Vuldeepecker: A deep learning-based system for vulnerability detection}.
\newblock \bibinfo{journal}{\emph{arXiv preprint arXiv:1801.01681}} (\bibinfo{year}{2018}).
\newblock


\bibitem[Liang et~al\mbox{.}(2018)]%
        {liang2018fuzzing}
\bibfield{author}{\bibinfo{person}{Hongliang Liang}, \bibinfo{person}{Xiaoxiao Pei}, \bibinfo{person}{Xiaodong Jia}, \bibinfo{person}{Wuwei Shen}, {and} \bibinfo{person}{Jian Zhang}.} \bibinfo{year}{2018}\natexlab{}.
\newblock \showarticletitle{Fuzzing: State of the art}.
\newblock \bibinfo{journal}{\emph{IEEE Transactions on Reliability}} \bibinfo{volume}{67}, \bibinfo{number}{3} (\bibinfo{year}{2018}), \bibinfo{pages}{1199--1218}.
\newblock


\bibitem[Lin et~al\mbox{.}(2019)]%
        {lin2019software}
\bibfield{author}{\bibinfo{person}{Guanjun Lin}, \bibinfo{person}{Jun Zhang}, \bibinfo{person}{Wei Luo}, \bibinfo{person}{Lei Pan}, \bibinfo{person}{Olivier De~Vel}, \bibinfo{person}{Paul Montague}, {and} \bibinfo{person}{Yang Xiang}.} \bibinfo{year}{2019}\natexlab{}.
\newblock \showarticletitle{Software vulnerability discovery via learning multi-domain knowledge bases}.
\newblock \bibinfo{journal}{\emph{IEEE Transactions on Dependable and Secure Computing}} \bibinfo{volume}{18}, \bibinfo{number}{5} (\bibinfo{year}{2019}), \bibinfo{pages}{2469--2485}.
\newblock


\bibitem[Lin et~al\mbox{.}(2018)]%
        {lin2018cross}
\bibfield{author}{\bibinfo{person}{Guanjun Lin}, \bibinfo{person}{Jun Zhang}, \bibinfo{person}{Wei Luo}, \bibinfo{person}{Lei Pan}, \bibinfo{person}{Yang Xiang}, \bibinfo{person}{Olivier De~Vel}, {and} \bibinfo{person}{Paul Montague}.} \bibinfo{year}{2018}\natexlab{}.
\newblock \showarticletitle{Cross-project transfer representation learning for vulnerable function discovery}.
\newblock \bibinfo{journal}{\emph{IEEE Transactions on Industrial Informatics}} \bibinfo{volume}{14}, \bibinfo{number}{7} (\bibinfo{year}{2018}), \bibinfo{pages}{3289--3297}.
\newblock


\bibitem[Liu et~al\mbox{.}(2012)]%
        {6405650}
\bibfield{author}{\bibinfo{person}{Bingchang Liu}, \bibinfo{person}{Liang Shi}, \bibinfo{person}{Zhuhua Cai}, {and} \bibinfo{person}{Min Li}.} \bibinfo{year}{2012}\natexlab{}.
\newblock \showarticletitle{Software Vulnerability Discovery Techniques: A Survey}. In \bibinfo{booktitle}{\emph{2012 Fourth International Conference on Multimedia Information Networking and Security}}. \bibinfo{pages}{152--156}.
\newblock
\urldef\tempurl%
\url{https://doi.org/10.1109/MINES.2012.202}
\showDOI{\tempurl}


\bibitem[Liu et~al\mbox{.}(2022)]%
        {liu2022cpgbert}
\bibfield{author}{\bibinfo{person}{Jingqiang Liu}, \bibinfo{person}{Xiaoxi Zhu}, \bibinfo{person}{Chaoge Liu}, \bibinfo{person}{Xiang Cui}, {and} \bibinfo{person}{Qixu Liu}.} \bibinfo{year}{2022}\natexlab{}.
\newblock \showarticletitle{CPGBERT: An Effective Model for Defect Detection by Learning Program Semantics via Code Property Graph}. In \bibinfo{booktitle}{\emph{2022 IEEE International Conference on Trust, Security and Privacy in Computing and Communications (TrustCom)}}. IEEE, \bibinfo{pages}{274--282}.
\newblock


\bibitem[Liu et~al\mbox{.}(2019b)]%
        {liu2019survey}
\bibfield{author}{\bibinfo{person}{Miao Liu}, \bibinfo{person}{Boyu Zhang}, \bibinfo{person}{Wenbin Chen}, {and} \bibinfo{person}{Xunlai Zhang}.} \bibinfo{year}{2019}\natexlab{b}.
\newblock \showarticletitle{A survey of exploitation and detection methods of XSS vulnerabilities}.
\newblock \bibinfo{journal}{\emph{IEEE access}}  \bibinfo{volume}{7} (\bibinfo{year}{2019}), \bibinfo{pages}{182004--182016}.
\newblock


\bibitem[Liu et~al\mbox{.}(2019a)]%
        {liu2019deepbalance}
\bibfield{author}{\bibinfo{person}{Shigang Liu}, \bibinfo{person}{Guanjun Lin}, \bibinfo{person}{Qing-Long Han}, \bibinfo{person}{Sheng Wen}, \bibinfo{person}{Jun Zhang}, {and} \bibinfo{person}{Yang Xiang}.} \bibinfo{year}{2019}\natexlab{a}.
\newblock \showarticletitle{DeepBalance: Deep-learning and fuzzy oversampling for vulnerability detection}.
\newblock \bibinfo{journal}{\emph{IEEE Transactions on Fuzzy Systems}} \bibinfo{volume}{28}, \bibinfo{number}{7} (\bibinfo{year}{2019}), \bibinfo{pages}{1329--1343}.
\newblock


\bibitem[Liu et~al\mbox{.}(2020)]%
        {liu2020cd}
\bibfield{author}{\bibinfo{person}{Shigang Liu}, \bibinfo{person}{Guanjun Lin}, \bibinfo{person}{Lizhen Qu}, \bibinfo{person}{Jun Zhang}, \bibinfo{person}{Olivier De~Vel}, \bibinfo{person}{Paul Montague}, {and} \bibinfo{person}{Yang Xiang}.} \bibinfo{year}{2020}\natexlab{}.
\newblock \showarticletitle{CD-VulD: Cross-domain vulnerability discovery based on deep domain adaptation}.
\newblock \bibinfo{journal}{\emph{IEEE Transactions on Dependable and Secure Computing}} \bibinfo{volume}{19}, \bibinfo{number}{1} (\bibinfo{year}{2020}), \bibinfo{pages}{438--451}.
\newblock


\bibitem[Liu and Wang(2022)]%
        {liu2022effective}
\bibfield{author}{\bibinfo{person}{Yuankun Liu} {and} \bibinfo{person}{Yu Wang}.} \bibinfo{year}{2022}\natexlab{}.
\newblock \showarticletitle{An Effective Software Vulnerability Detection Method Based On Devised Deep-Learning Model To Fix The Vague Separation}. In \bibinfo{booktitle}{\emph{Proceedings of the 2022 3rd International Symposium on Big Data and Artificial Intelligence}}. \bibinfo{pages}{90--95}.
\newblock


\bibitem[Lomio et~al\mbox{.}(2022)]%
        {lomio2022just}
\bibfield{author}{\bibinfo{person}{Francesco Lomio}, \bibinfo{person}{Emanuele Iannone}, \bibinfo{person}{Andrea De~Lucia}, \bibinfo{person}{Fabio Palomba}, {and} \bibinfo{person}{Valentina Lenarduzzi}.} \bibinfo{year}{2022}\natexlab{}.
\newblock \showarticletitle{Just-in-time software vulnerability detection: Are we there yet?}
\newblock \bibinfo{journal}{\emph{Journal of Systems and Software}} (\bibinfo{year}{2022}), \bibinfo{pages}{111283}.
\newblock


\bibitem[Lu et~al\mbox{.}(2021)]%
        {codexGluePaper}
\bibfield{author}{\bibinfo{person}{Shuai Lu}, \bibinfo{person}{Daya Guo}, \bibinfo{person}{Shuo Ren}, \bibinfo{person}{Junjie Huang}, \bibinfo{person}{Alexey Svyatkovskiy}, \bibinfo{person}{Ambrosio Blanco}, \bibinfo{person}{Colin~B. Clement}, \bibinfo{person}{Dawn Drain}, \bibinfo{person}{Daxin Jiang}, \bibinfo{person}{Duyu Tang}, \bibinfo{person}{Ge Li}, \bibinfo{person}{Lidong Zhou}, \bibinfo{person}{Linjun Shou}, \bibinfo{person}{Long Zhou}, \bibinfo{person}{Michele Tufano}, \bibinfo{person}{Ming Gong}, \bibinfo{person}{Ming Zhou}, \bibinfo{person}{Nan Duan}, \bibinfo{person}{Neel Sundaresan}, \bibinfo{person}{Shao~Kun Deng}, \bibinfo{person}{Shengyu Fu}, {and} \bibinfo{person}{Shujie Liu}.} \bibinfo{year}{2021}\natexlab{}.
\newblock \showarticletitle{CodeXGLUE: {A} Machine Learning Benchmark Dataset for Code Understanding and Generation}.
\newblock \bibinfo{journal}{\emph{CoRR}}  \bibinfo{volume}{abs/2102.04664} (\bibinfo{year}{2021}).
\newblock


\bibitem[Luo et~al\mbox{.}(2022)]%
        {luo2022compact}
\bibfield{author}{\bibinfo{person}{Yu Luo}, \bibinfo{person}{Weifeng Xu}, {and} \bibinfo{person}{Dianxiang Xu}.} \bibinfo{year}{2022}\natexlab{}.
\newblock \showarticletitle{Compact Abstract Graphs for Detecting Code Vulnerability with GNN Models}. In \bibinfo{booktitle}{\emph{Proceedings of the 38th Annual Computer Security Applications Conference}}. \bibinfo{pages}{497--507}.
\newblock


\bibitem[Malhotra(2015)]%
        {MALHOTRA2015504}
\bibfield{author}{\bibinfo{person}{Ruchika Malhotra}.} \bibinfo{year}{2015}\natexlab{}.
\newblock \showarticletitle{A systematic review of machine learning techniques for software fault prediction}.
\newblock \bibinfo{journal}{\emph{Applied Soft Computing}}  \bibinfo{volume}{27} (\bibinfo{year}{2015}), \bibinfo{pages}{504--518}.
\newblock
\showISSN{1568-4946}
\urldef\tempurl%
\url{https://doi.org/10.1016/j.asoc.2014.11.023}
\showDOI{\tempurl}


\bibitem[Mao et~al\mbox{.}(2020)]%
        {mao2020explainable}
\bibfield{author}{\bibinfo{person}{Yi Mao}, \bibinfo{person}{Yun Li}, \bibinfo{person}{Jiatai Sun}, {and} \bibinfo{person}{Yixin Chen}.} \bibinfo{year}{2020}\natexlab{}.
\newblock \showarticletitle{Explainable software vulnerability detection based on attention-based bidirectional recurrent neural networks}. In \bibinfo{booktitle}{\emph{2020 IEEE International Conference on Big Data (Big Data)}}. IEEE, \bibinfo{pages}{4651--4656}.
\newblock


\bibitem[McMahan et~al\mbox{.}(2017)]%
        {mcmahan2017communication}
\bibfield{author}{\bibinfo{person}{Brendan McMahan}, \bibinfo{person}{Eider Moore}, \bibinfo{person}{Daniel Ramage}, \bibinfo{person}{Seth Hampson}, {and} \bibinfo{person}{Blaise~Aguera y Arcas}.} \bibinfo{year}{2017}\natexlab{}.
\newblock \showarticletitle{Communication-efficient learning of deep networks from decentralized data}. In \bibinfo{booktitle}{\emph{Artificial intelligence and statistics}}. PMLR, \bibinfo{pages}{1273--1282}.
\newblock


\bibitem[Medeiros et~al\mbox{.}(2020)]%
        {medeiros2020vulnerable}
\bibfield{author}{\bibinfo{person}{Nadia Medeiros}, \bibinfo{person}{Naghmeh Ivaki}, \bibinfo{person}{Pedro Costa}, {and} \bibinfo{person}{Marco Vieira}.} \bibinfo{year}{2020}\natexlab{}.
\newblock \showarticletitle{Vulnerable code detection using software metrics and machine learning}.
\newblock \bibinfo{journal}{\emph{IEEE Access}}  \bibinfo{volume}{8} (\bibinfo{year}{2020}), \bibinfo{pages}{219174--219198}.
\newblock


\bibitem[Mendeley(2004)]%
        {Mendeley}
\bibfield{author}{\bibinfo{person}{Mendeley}.} \bibinfo{year}{2004}\natexlab{}.
\newblock \bibinfo{title}{NIST Software Assurance Reference Dataset}.
\newblock \bibinfo{howpublished}{\url{https://www.mendeley.com/}}.
\newblock
\newblock
\shownote{Accessed: 2024-01-28}.


\bibitem[Mim et~al\mbox{.}(2023)]%
        {mim2023impact}
\bibfield{author}{\bibinfo{person}{Rabaya~Sultana Mim}, \bibinfo{person}{Afrina Khatun}, \bibinfo{person}{Toukir Ahammed}, {and} \bibinfo{person}{Kazi Sakib}.} \bibinfo{year}{2023}\natexlab{}.
\newblock \showarticletitle{Impact of Centrality on Automated Vulnerability Detection Using Convolutional Neural Network}. In \bibinfo{booktitle}{\emph{2023 International Conference on Information and Communication Technology for Sustainable Development (ICICT4SD)}}. IEEE, \bibinfo{pages}{331--335}.
\newblock


\bibitem[{MITRE}(1999)]%
        {cve}
\bibfield{author}{\bibinfo{person}{{MITRE}}.} \bibinfo{year}{1999}\natexlab{}.
\newblock \bibinfo{title}{Common Vulnerabilities and Exposures}.
\newblock \bibinfo{howpublished}{\url{https://cve.mitre.org/cve/}}.
\newblock
\newblock
\shownote{Accessed: 2023-10-20}.


\bibitem[Moher et~al\mbox{.}(2009)]%
        {primsma}
\bibfield{author}{\bibinfo{person}{David Moher}, \bibinfo{person}{Alessandro Liberati}, \bibinfo{person}{Jennifer Tetzlaff}, \bibinfo{person}{Douglas~G Altman}, {and} \bibinfo{person}{the PRISMA~Group*}.} \bibinfo{year}{2009}\natexlab{}.
\newblock \showarticletitle{Preferred reporting items for systematic reviews and meta-analyses: the PRISMA statement}.
\newblock \bibinfo{journal}{\emph{Annals of internal medicine}} \bibinfo{volume}{151}, \bibinfo{number}{4} (\bibinfo{year}{2009}), \bibinfo{pages}{264--269}.
\newblock


\bibitem[Mosolyg{\'o} et~al\mbox{.}(2022)]%
        {mosolygo2022line}
\bibfield{author}{\bibinfo{person}{Bal{\'a}zs Mosolyg{\'o}}, \bibinfo{person}{Norbert V{\'a}ndor}, \bibinfo{person}{P{\'e}ter Heged{\H{u}}s}, {and} \bibinfo{person}{Rudolf Ferenc}.} \bibinfo{year}{2022}\natexlab{}.
\newblock \showarticletitle{A Line-Level Explainable Vulnerability Detection Approach for Java}. In \bibinfo{booktitle}{\emph{International Conference on Computational Science and Its Applications}}. Springer, \bibinfo{pages}{106--122}.
\newblock


\bibitem[Napier et~al\mbox{.}(2023)]%
        {napier2023empirical}
\bibfield{author}{\bibinfo{person}{Kollin Napier}, \bibinfo{person}{Tanmay Bhowmik}, {and} \bibinfo{person}{Shaowei Wang}.} \bibinfo{year}{2023}\natexlab{}.
\newblock \showarticletitle{An empirical study of text-based machine learning models for vulnerability detection}.
\newblock \bibinfo{journal}{\emph{Empirical Software Engineering}} \bibinfo{volume}{28}, \bibinfo{number}{2} (\bibinfo{year}{2023}), \bibinfo{pages}{38}.
\newblock


\bibitem[Nazim et~al\mbox{.}(2022)]%
        {nazim2022systematic}
\bibfield{author}{\bibinfo{person}{Mohammad Taneem~Bin Nazim}, \bibinfo{person}{Md~Jobair~Hossain Faruk}, \bibinfo{person}{Hossain Shahriar}, \bibinfo{person}{Md~Abdullah Khan}, \bibinfo{person}{Mohammad Masum}, \bibinfo{person}{Nazmus Sakib}, {and} \bibinfo{person}{Fan Wu}.} \bibinfo{year}{2022}\natexlab{}.
\newblock \showarticletitle{Systematic analysis of deep learning model for vulnerable code detection}. In \bibinfo{booktitle}{\emph{2022 IEEE 46th Annual Computers, Software, and Applications Conference (COMPSAC)}}. IEEE, \bibinfo{pages}{1768--1773}.
\newblock


\bibitem[Nguyen et~al\mbox{.}(2021)]%
        {nguyen2021information}
\bibfield{author}{\bibinfo{person}{Van Nguyen}, \bibinfo{person}{Trung Le}, \bibinfo{person}{Olivier De~Vel}, \bibinfo{person}{Paul Montague}, \bibinfo{person}{John Grundy}, {and} \bibinfo{person}{Dinh Phung}.} \bibinfo{year}{2021}\natexlab{}.
\newblock \showarticletitle{Information-theoretic source code vulnerability highlighting}. In \bibinfo{booktitle}{\emph{2021 International Joint Conference on Neural Networks (IJCNN)}}. IEEE, \bibinfo{pages}{1--8}.
\newblock


\bibitem[Nguyen et~al\mbox{.}(2022)]%
        {nguyen2022regvd}
\bibfield{author}{\bibinfo{person}{Van-Anh Nguyen}, \bibinfo{person}{Dai~Quoc Nguyen}, \bibinfo{person}{Van Nguyen}, \bibinfo{person}{Trung Le}, \bibinfo{person}{Quan~Hung Tran}, {and} \bibinfo{person}{Dinh Phung}.} \bibinfo{year}{2022}\natexlab{}.
\newblock \showarticletitle{ReGVD: Revisiting graph neural networks for vulnerability detection}. In \bibinfo{booktitle}{\emph{Proceedings of the ACM/IEEE 44th International Conference on Software Engineering: Companion Proceedings}}. \bibinfo{pages}{178--182}.
\newblock


\bibitem[Nikitopoulos et~al\mbox{.}(2021)]%
        {nikitopoulos2021crossvul}
\bibfield{author}{\bibinfo{person}{Georgios Nikitopoulos}, \bibinfo{person}{Konstantina Dritsa}, \bibinfo{person}{Panos Louridas}, {and} \bibinfo{person}{Dimitris Mitropoulos}.} \bibinfo{year}{2021}\natexlab{}.
\newblock \showarticletitle{CrossVul: a cross-language vulnerability dataset with commit data}. In \bibinfo{booktitle}{\emph{Proceedings of the 29th ACM Joint Meeting on European Software Engineering Conference and Symposium on the Foundations of Software Engineering}}. \bibinfo{pages}{1565--1569}.
\newblock


\bibitem[Nong et~al\mbox{.}(2022)]%
        {nong2022open}
\bibfield{author}{\bibinfo{person}{Yu Nong}, \bibinfo{person}{Rainy Sharma}, \bibinfo{person}{Abdelwahab Hamou-Lhadj}, \bibinfo{person}{Xiapu Luo}, {and} \bibinfo{person}{Haipeng Cai}.} \bibinfo{year}{2022}\natexlab{}.
\newblock \showarticletitle{Open science in software engineering: A study on deep learning-based vulnerability detection}.
\newblock \bibinfo{journal}{\emph{IEEE Transactions on Software Engineering}} \bibinfo{volume}{49}, \bibinfo{number}{4} (\bibinfo{year}{2022}), \bibinfo{pages}{1983--2005}.
\newblock


\bibitem[{NVD}(2024a)]%
        {CVElog1}
\bibfield{author}{\bibinfo{person}{{NVD}}.} \bibinfo{year}{2024}\natexlab{a}.
\newblock \bibinfo{title}{National Vulnerability Database}.
\newblock \bibinfo{howpublished}{\url{https://nvd.nist.gov/vuln/detail/CVE-2021-44228}}.
\newblock
\newblock
\shownote{Accessed: 2024-01-01}.


\bibitem[{NVD}(2024b)]%
        {CVElog2}
\bibfield{author}{\bibinfo{person}{{NVD}}.} \bibinfo{year}{2024}\natexlab{b}.
\newblock \bibinfo{title}{National Vulnerability Database}.
\newblock \bibinfo{howpublished}{\url{https://nvd.nist.gov/vuln/detail/CVE-2021-45046}}.
\newblock
\newblock
\shownote{Accessed: 2024-01-01}.


\bibitem[Patel et~al\mbox{.}(2019)]%
        {patel2019advanced}
\bibfield{author}{\bibinfo{person}{Om~Prakash Patel}, \bibinfo{person}{Neha Bharill}, \bibinfo{person}{Aruna Tiwari}, \bibinfo{person}{Vikram Patel}, \bibinfo{person}{Ojas Gupta}, \bibinfo{person}{Jian Cao}, \bibinfo{person}{Jun Li}, {and} \bibinfo{person}{Mukesh Prasad}.} \bibinfo{year}{2019}\natexlab{}.
\newblock \showarticletitle{Advanced quantum based neural network classifier and its application for objectionable web content filtering}.
\newblock \bibinfo{journal}{\emph{IEEE Access}}  \bibinfo{volume}{7} (\bibinfo{year}{2019}), \bibinfo{pages}{98069--98082}.
\newblock


\bibitem[Peng et~al\mbox{.}(2023b)]%
        {peng2023cevuldet}
\bibfield{author}{\bibinfo{person}{Bitao Peng}, \bibinfo{person}{Zhen Liu}, \bibinfo{person}{Jinrong Zhang}, {and} \bibinfo{person}{Pengcheng Su}.} \bibinfo{year}{2023}\natexlab{b}.
\newblock \showarticletitle{CEVulDet: A Code Edge Representation Learnable Vulnerability Detector}. In \bibinfo{booktitle}{\emph{2023 International Joint Conference on Neural Networks (IJCNN)}}. IEEE, \bibinfo{pages}{1--8}.
\newblock


\bibitem[Peng et~al\mbox{.}(2023a)]%
        {peng2023ptlvd}
\bibfield{author}{\bibinfo{person}{Tao Peng}, \bibinfo{person}{Shixu Chen}, \bibinfo{person}{Fei Zhu}, \bibinfo{person}{Junwei Tang}, \bibinfo{person}{Junping Liu}, {and} \bibinfo{person}{Xinrong Hu}.} \bibinfo{year}{2023}\natexlab{a}.
\newblock \showarticletitle{PTLVD: Program Slicing and Transformer-based Line-level Vulnerability Detection System}. In \bibinfo{booktitle}{\emph{2023 IEEE 23rd International Working Conference on Source Code Analysis and Manipulation (SCAM)}}. IEEE, \bibinfo{pages}{162--173}.
\newblock


\bibitem[Pereira et~al\mbox{.}(2022)]%
        {pereira2022use}
\bibfield{author}{\bibinfo{person}{Jos{\'e}~D'Abruzzo Pereira}, \bibinfo{person}{Nuno Louren{\c{c}}o}, {and} \bibinfo{person}{Marco Vieira}.} \bibinfo{year}{2022}\natexlab{}.
\newblock \showarticletitle{On the Use of Deep Graph CNN to Detect Vulnerable C Functions}. In \bibinfo{booktitle}{\emph{Proceedings of the 11th Latin-American Symposium on Dependable Computing}}. \bibinfo{pages}{45--50}.
\newblock


\bibitem[Pistoia et~al\mbox{.}(2007)]%
        {pistoia2007survey}
\bibfield{author}{\bibinfo{person}{Marco Pistoia}, \bibinfo{person}{Satish Chandra}, \bibinfo{person}{Stephen~J Fink}, {and} \bibinfo{person}{Eran Yahav}.} \bibinfo{year}{2007}\natexlab{}.
\newblock \showarticletitle{A survey of static analysis methods for identifying security vulnerabilities in software systems}.
\newblock \bibinfo{journal}{\emph{IBM systems journal}} \bibinfo{volume}{46}, \bibinfo{number}{2} (\bibinfo{year}{2007}), \bibinfo{pages}{265--288}.
\newblock


\bibitem[Ponta et~al\mbox{.}(2019)]%
        {ponta2019manually}
\bibfield{author}{\bibinfo{person}{Serena~Elisa Ponta}, \bibinfo{person}{Henrik Plate}, \bibinfo{person}{Antonino Sabetta}, \bibinfo{person}{Michele Bezzi}, {and} \bibinfo{person}{C{\'e}dric Dangremont}.} \bibinfo{year}{2019}\natexlab{}.
\newblock \showarticletitle{A manually-curated dataset of fixes to vulnerabilities of open-source software}. In \bibinfo{booktitle}{\emph{2019 IEEE/ACM 16th International Conference on Mining Software Repositories (MSR)}}. IEEE, \bibinfo{pages}{383--387}.
\newblock


\bibitem[Purba et~al\mbox{.}(2023)]%
        {purba2023software}
\bibfield{author}{\bibinfo{person}{Moumita~Das Purba}, \bibinfo{person}{Arpita Ghosh}, \bibinfo{person}{Benjamin~J Radford}, {and} \bibinfo{person}{Bill Chu}.} \bibinfo{year}{2023}\natexlab{}.
\newblock \showarticletitle{Software vulnerability detection using large language models}. In \bibinfo{booktitle}{\emph{2023 IEEE 34th International Symposium on Software Reliability Engineering Workshops (ISSREW)}}. IEEE, \bibinfo{pages}{112--119}.
\newblock


\bibitem[Quan et~al\mbox{.}(2023)]%
        {quan2023xgv}
\bibfield{author}{\bibinfo{person}{Vu~Le~Anh Quan}, \bibinfo{person}{Chau~Thuan Phat}, \bibinfo{person}{Kiet Van~Nguyen}, \bibinfo{person}{Phan~The Duy}, {and} \bibinfo{person}{Van-Hau Pham}.} \bibinfo{year}{2023}\natexlab{}.
\newblock \showarticletitle{XGV-BERT: Leveraging Contextualized Language Model and Graph Neural Network for Efficient Software Vulnerability Detection}.
\newblock \bibinfo{journal}{\emph{arXiv preprint arXiv:2309.14677}} (\bibinfo{year}{2023}).
\newblock


\bibitem[Rahaman et~al\mbox{.}(2019)]%
        {rahaman2019cryptoguard}
\bibfield{author}{\bibinfo{person}{Sazzadur Rahaman}, \bibinfo{person}{Ya Xiao}, \bibinfo{person}{Sharmin Afrose}, \bibinfo{person}{Fahad Shaon}, \bibinfo{person}{Ke Tian}, \bibinfo{person}{Miles Frantz}, \bibinfo{person}{Murat Kantarcioglu}, {and} \bibinfo{person}{Danfeng Yao}.} \bibinfo{year}{2019}\natexlab{}.
\newblock \showarticletitle{Cryptoguard: High precision detection of cryptographic vulnerabilities in massive-sized java projects}. In \bibinfo{booktitle}{\emph{Proceedings of the 2019 ACM SIGSAC Conference on Computer and Communications Security}}. \bibinfo{pages}{2455--2472}.
\newblock


\bibitem[Rahman et~al\mbox{.}(2020)]%
        {rahman2020internet}
\bibfield{author}{\bibinfo{person}{Sawsan~Abdul Rahman}, \bibinfo{person}{Hanine Tout}, \bibinfo{person}{Chamseddine Talhi}, {and} \bibinfo{person}{Azzam Mourad}.} \bibinfo{year}{2020}\natexlab{}.
\newblock \showarticletitle{Internet of things intrusion detection: Centralized, on-device, or federated learning?}
\newblock \bibinfo{journal}{\emph{IEEE Network}} \bibinfo{volume}{34}, \bibinfo{number}{6} (\bibinfo{year}{2020}), \bibinfo{pages}{310--317}.
\newblock


\bibitem[Rei and Abreu(2017)]%
        {rei2017database}
\bibfield{author}{\bibinfo{person}{Sofia Rei} {and} \bibinfo{person}{Rui Abreu}.} \bibinfo{year}{2017}\natexlab{}.
\newblock \showarticletitle{A database of existing vulnerabilities to enable controlled testing studies}.
\newblock \bibinfo{journal}{\emph{International Journal of Secure Software Engineering (IJSSE)}} \bibinfo{volume}{8}, \bibinfo{number}{3} (\bibinfo{year}{2017}), \bibinfo{pages}{1--23}.
\newblock


\bibitem[Reis and Abreu(2017)]%
        {reis2017secbench}
\bibfield{author}{\bibinfo{person}{Sofia Reis} {and} \bibinfo{person}{Rui Abreu}.} \bibinfo{year}{2017}\natexlab{}.
\newblock \showarticletitle{SECBENCH: A Database of Real Security Vulnerabilities.}. In \bibinfo{booktitle}{\emph{SecSE@ ESORICS}}. \bibinfo{pages}{69--85}.
\newblock


\bibitem[Reis and Abreu(2021)]%
        {reis2021ground}
\bibfield{author}{\bibinfo{person}{Sofia Reis} {and} \bibinfo{person}{Rui Abreu}.} \bibinfo{year}{2021}\natexlab{}.
\newblock \showarticletitle{A ground-truth dataset of real security patches}.
\newblock \bibinfo{journal}{\emph{arXiv preprint arXiv:2110.09635}} (\bibinfo{year}{2021}).
\newblock


\bibitem[Rey et~al\mbox{.}(2022)]%
        {rey2022federated}
\bibfield{author}{\bibinfo{person}{Valerian Rey}, \bibinfo{person}{Pedro Miguel~S{\'a}nchez S{\'a}nchez}, \bibinfo{person}{Alberto~Huertas Celdr{\'a}n}, {and} \bibinfo{person}{G{\'e}r{\^o}me Bovet}.} \bibinfo{year}{2022}\natexlab{}.
\newblock \showarticletitle{Federated learning for malware detection in IoT devices}.
\newblock \bibinfo{journal}{\emph{Computer Networks}}  \bibinfo{volume}{204} (\bibinfo{year}{2022}), \bibinfo{pages}{108693}.
\newblock


\bibitem[Ribeiro et~al\mbox{.}(2016)]%
        {ribeiro2016should}
\bibfield{author}{\bibinfo{person}{Marco~Tulio Ribeiro}, \bibinfo{person}{Sameer Singh}, {and} \bibinfo{person}{Carlos Guestrin}.} \bibinfo{year}{2016}\natexlab{}.
\newblock \showarticletitle{" Why should i trust you?" Explaining the predictions of any classifier}. In \bibinfo{booktitle}{\emph{Proceedings of the 22nd ACM SIGKDD international conference on knowledge discovery and data mining}}. \bibinfo{pages}{1135--1144}.
\newblock


\bibitem[Russell et~al\mbox{.}(2018)]%
        {russell2018automated}
\bibfield{author}{\bibinfo{person}{Rebecca Russell}, \bibinfo{person}{Louis Kim}, \bibinfo{person}{Lei Hamilton}, \bibinfo{person}{Tomo Lazovich}, \bibinfo{person}{Jacob Harer}, \bibinfo{person}{Onur Ozdemir}, \bibinfo{person}{Paul Ellingwood}, {and} \bibinfo{person}{Marc McConley}.} \bibinfo{year}{2018}\natexlab{}.
\newblock \showarticletitle{Automated vulnerability detection in source code using deep representation learning}. In \bibinfo{booktitle}{\emph{2018 17th IEEE international conference on machine learning and applications (ICMLA)}}. IEEE, \bibinfo{pages}{757--762}.
\newblock


\bibitem[Saccente et~al\mbox{.}(2019)]%
        {saccente2019project}
\bibfield{author}{\bibinfo{person}{Nicholas Saccente}, \bibinfo{person}{Josh Dehlinger}, \bibinfo{person}{Lin Deng}, \bibinfo{person}{Suranjan Chakraborty}, {and} \bibinfo{person}{Yin Xiong}.} \bibinfo{year}{2019}\natexlab{}.
\newblock \showarticletitle{Project achilles: A prototype tool for static method-level vulnerability detection of Java source code using a recurrent neural network}. In \bibinfo{booktitle}{\emph{2019 34th IEEE/ACM International Conference on Automated Software Engineering Workshop (ASEW)}}. IEEE, \bibinfo{pages}{114--121}.
\newblock


\bibitem[{\c{S}}ahin(2023)]%
        {csahin2023semantic}
\bibfield{author}{\bibinfo{person}{Canan~Batur {\c{S}}ahin}.} \bibinfo{year}{2023}\natexlab{}.
\newblock \showarticletitle{Semantic-based vulnerability detection by functional connectivity of gated graph sequence neural networks}.
\newblock \bibinfo{journal}{\emph{Soft Computing}} \bibinfo{volume}{27}, \bibinfo{number}{9} (\bibinfo{year}{2023}), \bibinfo{pages}{5703--5719}.
\newblock


\bibitem[Salimi et~al\mbox{.}(2020)]%
        {salimi2020improving}
\bibfield{author}{\bibinfo{person}{Solmaz Salimi}, \bibinfo{person}{Maryam Ebrahimzadeh}, {and} \bibinfo{person}{Mehdi Kharrazi}.} \bibinfo{year}{2020}\natexlab{}.
\newblock \showarticletitle{Improving real-world vulnerability characterization with vulnerable slices}. In \bibinfo{booktitle}{\emph{Proceedings of the 16th ACM International Conference on Predictive Models and Data Analytics in Software Engineering}}. \bibinfo{pages}{11--20}.
\newblock


\bibitem[Scarselli et~al\mbox{.}(2008)]%
        {scarselli2008graph}
\bibfield{author}{\bibinfo{person}{Franco Scarselli}, \bibinfo{person}{Marco Gori}, \bibinfo{person}{Ah~Chung Tsoi}, \bibinfo{person}{Markus Hagenbuchner}, {and} \bibinfo{person}{Gabriele Monfardini}.} \bibinfo{year}{2008}\natexlab{}.
\newblock \showarticletitle{The graph neural network model}.
\newblock \bibinfo{journal}{\emph{IEEE transactions on neural networks}} \bibinfo{volume}{20}, \bibinfo{number}{1} (\bibinfo{year}{2008}), \bibinfo{pages}{61--80}.
\newblock


\bibitem[Semasaba et~al\mbox{.}(2020)]%
        {semasaba2020literature}
\bibfield{author}{\bibinfo{person}{Abubakar Omari~Abdallah Semasaba}, \bibinfo{person}{Wei Zheng}, \bibinfo{person}{Xiaoxue Wu}, {and} \bibinfo{person}{Samuel~Akwasi Agyemang}.} \bibinfo{year}{2020}\natexlab{}.
\newblock \showarticletitle{Literature survey of deep learning-based vulnerability analysis on source code}.
\newblock \bibinfo{journal}{\emph{IET Software}} \bibinfo{volume}{14}, \bibinfo{number}{6} (\bibinfo{year}{2020}), \bibinfo{pages}{654--664}.
\newblock


\bibitem[Senanayake et~al\mbox{.}(2023)]%
        {senanayake2023android}
\bibfield{author}{\bibinfo{person}{Janaka Senanayake}, \bibinfo{person}{Harsha Kalutarage}, \bibinfo{person}{Mhd~Omar Al-Kadri}, \bibinfo{person}{Andrei Petrovski}, {and} \bibinfo{person}{Luca Piras}.} \bibinfo{year}{2023}\natexlab{}.
\newblock \showarticletitle{Android source code vulnerability detection: a systematic literature review}.
\newblock \bibinfo{journal}{\emph{Comput. Surveys}} \bibinfo{volume}{55}, \bibinfo{number}{9} (\bibinfo{year}{2023}), \bibinfo{pages}{1--37}.
\newblock


\bibitem[Shahriar and Zulkernine(2012)]%
        {10.1145/2187671.2187673}
\bibfield{author}{\bibinfo{person}{Hossain Shahriar} {and} \bibinfo{person}{Mohammad Zulkernine}.} \bibinfo{year}{2012}\natexlab{}.
\newblock \showarticletitle{Mitigating Program Security Vulnerabilities: Approaches and Challenges}.
\newblock  \bibinfo{volume}{44}, \bibinfo{number}{3} (\bibinfo{year}{2012}).
\newblock
\showISSN{0360-0300}
\urldef\tempurl%
\url{https://doi.org/10.1145/2187671.2187673}
\showDOI{\tempurl}


\bibitem[Shimmi et~al\mbox{.}(2024)]%
        {shimmi2024vulsim}
\bibfield{author}{\bibinfo{person}{Samiha Shimmi}, \bibinfo{person}{Ashiqur Rahman}, \bibinfo{person}{Mohan Gadde}, \bibinfo{person}{Hamed Okhravi}, {and} \bibinfo{person}{Mona Rahimi}.} \bibinfo{year}{2024}\natexlab{}.
\newblock \showarticletitle{$\{$VulSim$\}$: Leveraging Similarity of $\{$Multi-Dimensional$\}$ Neighbor Embeddings for Vulnerability Detection}. In \bibinfo{booktitle}{\emph{33rd USENIX Security Symposium (USENIX Security 24)}}. \bibinfo{pages}{1777--1794}.
\newblock


\bibitem[Shiri~Harzevili et~al\mbox{.}(2024)]%
        {10.1145/3699711}
\bibfield{author}{\bibinfo{person}{Nima Shiri~Harzevili}, \bibinfo{person}{Alvine Boaye~Belle}, \bibinfo{person}{Junjie Wang}, \bibinfo{person}{Song Wang}, \bibinfo{person}{Zhen Ming~(Jack) Jiang}, {and} \bibinfo{person}{Nachiappan Nagappan}.} \bibinfo{year}{2024}\natexlab{}.
\newblock \showarticletitle{A Systematic Literature Review on Automated Software Vulnerability Detection Using Machine Learning}.
\newblock \bibinfo{journal}{\emph{ACM Comput. Surv.}} \bibinfo{volume}{57}, \bibinfo{number}{3}, Article \bibinfo{articleno}{55} (\bibinfo{date}{Nov.} \bibinfo{year}{2024}), \bibinfo{numpages}{36}~pages.
\newblock
\showISSN{0360-0300}
\urldef\tempurl%
\url{https://doi.org/10.1145/3699711}
\showDOI{\tempurl}


\bibitem[Song et~al\mbox{.}(2019)]%
        {song2019sok}
\bibfield{author}{\bibinfo{person}{Dokyung Song}, \bibinfo{person}{Julian Lettner}, \bibinfo{person}{Prabhu Rajasekaran}, \bibinfo{person}{Yeoul Na}, \bibinfo{person}{Stijn Volckaert}, \bibinfo{person}{Per Larsen}, {and} \bibinfo{person}{Michael Franz}.} \bibinfo{year}{2019}\natexlab{}.
\newblock \showarticletitle{SoK: Sanitizing for security}. In \bibinfo{booktitle}{\emph{2019 IEEE Symposium on Security and Privacy (SP)}}. IEEE, \bibinfo{pages}{1275--1295}.
\newblock


\bibitem[Song et~al\mbox{.}(2022)]%
        {song2022hgvul}
\bibfield{author}{\bibinfo{person}{Zihua Song}, \bibinfo{person}{Junfeng Wang}, \bibinfo{person}{Shengli Liu}, \bibinfo{person}{Zhiyang Fang}, \bibinfo{person}{Kaiyuan Yang}, {et~al\mbox{.}}} \bibinfo{year}{2022}\natexlab{}.
\newblock \showarticletitle{HGVul: A code vulnerability detection method based on heterogeneous source-level intermediate representation}.
\newblock \bibinfo{journal}{\emph{Security and Communication Networks}}  \bibinfo{volume}{2022} (\bibinfo{year}{2022}).
\newblock


\bibitem[Steenhoek et~al\mbox{.}(2023)]%
        {steenhoek2023empirical}
\bibfield{author}{\bibinfo{person}{Benjamin Steenhoek}, \bibinfo{person}{Md~Mahbubur Rahman}, \bibinfo{person}{Richard Jiles}, {and} \bibinfo{person}{Wei Le}.} \bibinfo{year}{2023}\natexlab{}.
\newblock \showarticletitle{An empirical study of deep learning models for vulnerability detection}. In \bibinfo{booktitle}{\emph{2023 IEEE/ACM 45th International Conference on Software Engineering (ICSE)}}. IEEE, \bibinfo{pages}{2237--2248}.
\newblock


\bibitem[Sun et~al\mbox{.}(2023a)]%
        {sun2023software}
\bibfield{author}{\bibinfo{person}{Hao Sun}, \bibinfo{person}{Zhe Bu}, \bibinfo{person}{Yang Xiao}, \bibinfo{person}{Chengsheng Zhou}, \bibinfo{person}{Zhiyu Hao}, {and} \bibinfo{person}{Hongsong Zhu}.} \bibinfo{year}{2023}\natexlab{a}.
\newblock \showarticletitle{Software Vulnerability Detection Using an Enhanced Generalization Strategy}. In \bibinfo{booktitle}{\emph{International Symposium on Dependable Software Engineering: Theories, Tools, and Applications}}. Springer, \bibinfo{pages}{226--242}.
\newblock


\bibitem[Sun et~al\mbox{.}(2023b)]%
        {sun2023enhanced}
\bibfield{author}{\bibinfo{person}{Hao Sun}, \bibinfo{person}{Yongji Liu}, \bibinfo{person}{Zhenquan Ding}, \bibinfo{person}{Yang Xiao}, \bibinfo{person}{Zhiyu Hao}, {and} \bibinfo{person}{Hongsong Zhu}.} \bibinfo{year}{2023}\natexlab{b}.
\newblock \showarticletitle{An Enhanced Vulnerability Detection in Software Using a Heterogeneous Encoding Ensemble}. In \bibinfo{booktitle}{\emph{2023 IEEE Symposium on Computers and Communications (ISCC)}}. IEEE, \bibinfo{pages}{1214--1220}.
\newblock


\bibitem[Tang et~al\mbox{.}(2022)]%
        {tang2022sevuldet}
\bibfield{author}{\bibinfo{person}{Zhiquan Tang}, \bibinfo{person}{Qiao Hu}, \bibinfo{person}{Yupeng Hu}, \bibinfo{person}{Wenxin Kuang}, {and} \bibinfo{person}{Jiongyi Chen}.} \bibinfo{year}{2022}\natexlab{}.
\newblock \showarticletitle{SEVulDet: A Semantics-Enhanced Learnable Vulnerability Detector}. In \bibinfo{booktitle}{\emph{2022 52nd Annual IEEE/IFIP International Conference on Dependable Systems and Networks (DSN)}}. IEEE, \bibinfo{pages}{150--162}.
\newblock


\bibitem[Tao et~al\mbox{.}(2023)]%
        {tao2023vulnerability}
\bibfield{author}{\bibinfo{person}{Wenxin Tao}, \bibinfo{person}{Xiaohong Su}, \bibinfo{person}{Jiayuan Wan}, \bibinfo{person}{Hongwei Wei}, {and} \bibinfo{person}{Weining Zheng}.} \bibinfo{year}{2023}\natexlab{}.
\newblock \showarticletitle{Vulnerability Detection Through Cross-modal Feature Enhancement and Fusion}.
\newblock \bibinfo{journal}{\emph{Computers \& Security}} (\bibinfo{year}{2023}), \bibinfo{pages}{103341}.
\newblock


\bibitem[Tian et~al\mbox{.}(2021)]%
        {tian2021bbreglocator}
\bibfield{author}{\bibinfo{person}{Junfeng Tian}, \bibinfo{person}{Junkun Zhang}, {and} \bibinfo{person}{Fanming Liu}.} \bibinfo{year}{2021}\natexlab{}.
\newblock \showarticletitle{Bbreglocator: A vulnerability detection system based on bounding box regression}. In \bibinfo{booktitle}{\emph{2021 51st Annual IEEE/IFIP International Conference on Dependable Systems and Networks Workshops (DSN-W)}}. IEEE, \bibinfo{pages}{93--100}.
\newblock


\bibitem[Tian et~al\mbox{.}(2023)]%
        {tian2023learning}
\bibfield{author}{\bibinfo{person}{Zhenzhou Tian}, \bibinfo{person}{Binhui Tian}, \bibinfo{person}{Jiajun Lv}, {and} \bibinfo{person}{Lingwei Chen}.} \bibinfo{year}{2023}\natexlab{}.
\newblock \showarticletitle{Learning and fusing multi-view code representations for function vulnerability detection}.
\newblock \bibinfo{journal}{\emph{Electronics}} \bibinfo{volume}{12}, \bibinfo{number}{11} (\bibinfo{year}{2023}), \bibinfo{pages}{2495}.
\newblock


\bibitem[Tian et~al\mbox{.}(2024)]%
        {tian2024enhancing}
\bibfield{author}{\bibinfo{person}{Zhenzhou Tian}, \bibinfo{person}{Binhui Tian}, \bibinfo{person}{Jiajun Lv}, \bibinfo{person}{Yanping Chen}, {and} \bibinfo{person}{Lingwei Chen}.} \bibinfo{year}{2024}\natexlab{}.
\newblock \showarticletitle{Enhancing vulnerability detection via AST decomposition and neural sub-tree encoding}.
\newblock \bibinfo{journal}{\emph{Expert Systems with Applications}}  \bibinfo{volume}{238} (\bibinfo{year}{2024}), \bibinfo{pages}{121865}.
\newblock


\bibitem[Vaswani et~al\mbox{.}(2017)]%
        {vaswani2017attention}
\bibfield{author}{\bibinfo{person}{Ashish Vaswani}, \bibinfo{person}{Noam Shazeer}, \bibinfo{person}{Niki Parmar}, \bibinfo{person}{Jakob Uszkoreit}, \bibinfo{person}{Llion Jones}, \bibinfo{person}{Aidan~N Gomez}, \bibinfo{person}{{\L}ukasz Kaiser}, {and} \bibinfo{person}{Illia Polosukhin}.} \bibinfo{year}{2017}\natexlab{}.
\newblock \showarticletitle{Attention is all you need}.
\newblock \bibinfo{journal}{\emph{Advances in neural information processing systems}}  \bibinfo{volume}{30} (\bibinfo{year}{2017}).
\newblock


\bibitem[Wang et~al\mbox{.}(2020)]%
        {wang2020combining}
\bibfield{author}{\bibinfo{person}{Huanting Wang}, \bibinfo{person}{Guixin Ye}, \bibinfo{person}{Zhanyong Tang}, \bibinfo{person}{Shin~Hwei Tan}, \bibinfo{person}{Songfang Huang}, \bibinfo{person}{Dingyi Fang}, \bibinfo{person}{Yansong Feng}, \bibinfo{person}{Lizhong Bian}, {and} \bibinfo{person}{Zheng Wang}.} \bibinfo{year}{2020}\natexlab{}.
\newblock \showarticletitle{Combining graph-based learning with automated data collection for code vulnerability detection}.
\newblock \bibinfo{journal}{\emph{IEEE Transactions on Information Forensics and Security}}  \bibinfo{volume}{16} (\bibinfo{year}{2020}), \bibinfo{pages}{1943--1958}.
\newblock


\bibitem[Wang et~al\mbox{.}(2023)]%
        {wang2023deepvd}
\bibfield{author}{\bibinfo{person}{Wenbo Wang}, \bibinfo{person}{Tien~N Nguyen}, \bibinfo{person}{Shaohua Wang}, \bibinfo{person}{Yi Li}, \bibinfo{person}{Jiyuan Zhang}, {and} \bibinfo{person}{Aashish Yadavally}.} \bibinfo{year}{2023}\natexlab{}.
\newblock \showarticletitle{DeepVD: Toward Class-Separation Features for Neural Network Vulnerability Detection}. In \bibinfo{booktitle}{\emph{2023 IEEE/ACM 45th International Conference on Software Engineering (ICSE)}}. IEEE, \bibinfo{pages}{2249--2261}.
\newblock


\bibitem[Watson et~al\mbox{.}(2022)]%
        {watson2022detecting}
\bibfield{author}{\bibinfo{person}{Anne Watson}, \bibinfo{person}{Ekincan Ufuktepe}, {and} \bibinfo{person}{Kannappan Palaniappan}.} \bibinfo{year}{2022}\natexlab{}.
\newblock \showarticletitle{Detecting Software Code Vulnerabilities Using 2D Convolutional Neural Networks with Program Slicing Feature Maps}. In \bibinfo{booktitle}{\emph{2022 IEEE Applied Imagery Pattern Recognition Workshop (AIPR)}}. IEEE, \bibinfo{pages}{1--9}.
\newblock


\bibitem[Wen et~al\mbox{.}(2023)]%
        {wen2023less}
\bibfield{author}{\bibinfo{person}{Xin-Cheng Wen}, \bibinfo{person}{Xinchen Wang}, \bibinfo{person}{Cuiyun Gao}, \bibinfo{person}{Shaohua Wang}, \bibinfo{person}{Yang Liu}, {and} \bibinfo{person}{Zhaoquan Gu}.} \bibinfo{year}{2023}\natexlab{}.
\newblock \showarticletitle{When Less is Enough: Positive and Unlabeled Learning Model for Vulnerability Detection}. In \bibinfo{booktitle}{\emph{2023 38th IEEE/ACM International Conference on Automated Software Engineering (ASE)}}. IEEE, \bibinfo{pages}{345--357}.
\newblock


\bibitem[Wu et~al\mbox{.}(2023b)]%
        {wu2023learning}
\bibfield{author}{\bibinfo{person}{Bozhi Wu}, \bibinfo{person}{Shangqing Liu}, \bibinfo{person}{Yang Xiao}, \bibinfo{person}{Zhiming Li}, \bibinfo{person}{Jun Sun}, {and} \bibinfo{person}{Shang-Wei Lin}.} \bibinfo{year}{2023}\natexlab{b}.
\newblock \showarticletitle{Learning Program Semantics for Vulnerability Detection via Vulnerability-Specific Inter-procedural Slicing}. In \bibinfo{booktitle}{\emph{Proceedings of the 31st ACM Joint European Software Engineering Conference and Symposium on the Foundations of Software Engineering}}. \bibinfo{pages}{1371--1383}.
\newblock


\bibitem[Wu et~al\mbox{.}(2020)]%
        {wu2020graph}
\bibfield{author}{\bibinfo{person}{Peng Wu}, \bibinfo{person}{Liangze Yin}, \bibinfo{person}{Xiang Du}, \bibinfo{person}{Liyuan Jia}, {and} \bibinfo{person}{Wei Dong}.} \bibinfo{year}{2020}\natexlab{}.
\newblock \showarticletitle{Graph-based vulnerability detection via extracting features from sliced code}. In \bibinfo{booktitle}{\emph{2020 IEEE 20th International Conference on Software Quality, Reliability and Security Companion (QRS-C)}}. IEEE, \bibinfo{pages}{38--45}.
\newblock


\bibitem[Wu et~al\mbox{.}(2022a)]%
        {wu2022inductive}
\bibfield{author}{\bibinfo{person}{Tongshuai Wu}, \bibinfo{person}{Liwei Chen}, \bibinfo{person}{Gewangzi Du}, \bibinfo{person}{Chenguang Zhu}, \bibinfo{person}{Ningning Cui}, {and} \bibinfo{person}{Gang Shi}.} \bibinfo{year}{2022}\natexlab{a}.
\newblock \showarticletitle{Inductive Vulnerability Detection via Gated Graph Neural Network}. In \bibinfo{booktitle}{\emph{2022 IEEE 25th International Conference on Computer Supported Cooperative Work in Design (CSCWD)}}. IEEE, \bibinfo{pages}{519--524}.
\newblock


\bibitem[Wu et~al\mbox{.}(2023a)]%
        {wu2023cdnm}
\bibfield{author}{\bibinfo{person}{Tongshuai Wu}, \bibinfo{person}{Liwei Chen}, \bibinfo{person}{Gewangzi Du}, \bibinfo{person}{Chenguang Zhu}, \bibinfo{person}{Ningning Cui}, {and} \bibinfo{person}{Gang Shi}.} \bibinfo{year}{2023}\natexlab{a}.
\newblock \showarticletitle{CDNM: Clustering-Based Data Normalization Method For Automated Vulnerability Detection}.
\newblock \bibinfo{journal}{\emph{Comput. J.}} (\bibinfo{year}{2023}), \bibinfo{pages}{bxad080}.
\newblock


\bibitem[Wu et~al\mbox{.}(2021a)]%
        {wu2021self}
\bibfield{author}{\bibinfo{person}{Tongshuai Wu}, \bibinfo{person}{Liwei Chen}, \bibinfo{person}{Gewangzi Du}, \bibinfo{person}{Chenguang Zhu}, {and} \bibinfo{person}{Gang Shi}.} \bibinfo{year}{2021}\natexlab{a}.
\newblock \showarticletitle{Self-attention based automated vulnerability detection with effective data representation}. In \bibinfo{booktitle}{\emph{2021 IEEE Intl Conf on Parallel \& Distributed Processing with Applications, Big Data \& Cloud Computing, Sustainable Computing \& Communications, Social Computing \& Networking (ISPA/BDCloud/SocialCom/SustainCom)}}. IEEE, \bibinfo{pages}{892--899}.
\newblock


\bibitem[Wu et~al\mbox{.}(2021b)]%
        {wu2021vulnerability}
\bibfield{author}{\bibinfo{person}{Yuelong Wu}, \bibinfo{person}{Jintian Lu}, \bibinfo{person}{Yunyi Zhang}, {and} \bibinfo{person}{Shuyuan Jin}.} \bibinfo{year}{2021}\natexlab{b}.
\newblock \showarticletitle{Vulnerability detection in c/c++ source code with graph representation learning}. In \bibinfo{booktitle}{\emph{2021 IEEE 11th Annual Computing and Communication Workshop and Conference (CCWC)}}. IEEE, \bibinfo{pages}{1519--1524}.
\newblock


\bibitem[Wu et~al\mbox{.}(2022b)]%
        {wu2022vulcnn}
\bibfield{author}{\bibinfo{person}{Yueming Wu}, \bibinfo{person}{Deqing Zou}, \bibinfo{person}{Shihan Dou}, \bibinfo{person}{Wei Yang}, \bibinfo{person}{Duo Xu}, {and} \bibinfo{person}{Hai Jin}.} \bibinfo{year}{2022}\natexlab{b}.
\newblock \showarticletitle{VulCNN: An image-inspired scalable vulnerability detection system}. In \bibinfo{booktitle}{\emph{Proceedings of the 44th International Conference on Software Engineering}}. \bibinfo{pages}{2365--2376}.
\newblock


\bibitem[Xia et~al\mbox{.}(2021)]%
        {xia2021source}
\bibfield{author}{\bibinfo{person}{Xiaoling Xia}, \bibinfo{person}{Yu Wang}, {and} \bibinfo{person}{Ye Yang}.} \bibinfo{year}{2021}\natexlab{}.
\newblock \showarticletitle{Source Code Vulnerability Detection Based On SAR-GIN}. In \bibinfo{booktitle}{\emph{2021 2nd International Conference on Electronics, Communications and Information Technology (CECIT)}}. IEEE, \bibinfo{pages}{1144--1149}.
\newblock


\bibitem[Xiaomeng et~al\mbox{.}(2018)]%
        {xiaomeng2018cpgva}
\bibfield{author}{\bibinfo{person}{Wang Xiaomeng}, \bibinfo{person}{Zhang Tao}, \bibinfo{person}{Wu Runpu}, \bibinfo{person}{Xin Wei}, {and} \bibinfo{person}{Hou Changyu}.} \bibinfo{year}{2018}\natexlab{}.
\newblock \showarticletitle{CPGVA: Code property graph based vulnerability analysis by deep learning}. In \bibinfo{booktitle}{\emph{2018 10th International Conference on Advanced Infocomm Technology (ICAIT)}}. IEEE, \bibinfo{pages}{184--188}.
\newblock


\bibitem[Xuan(2023)]%
        {xuan2023new}
\bibfield{author}{\bibinfo{person}{Cho~Do Xuan}.} \bibinfo{year}{2023}\natexlab{}.
\newblock \showarticletitle{A new approach to software vulnerability detection based on CPG analysis}.
\newblock \bibinfo{journal}{\emph{Cogent Engineering}} \bibinfo{volume}{10}, \bibinfo{number}{1} (\bibinfo{year}{2023}), \bibinfo{pages}{2221962}.
\newblock


\bibitem[Xue et~al\mbox{.}(2023)]%
        {xue2023vulsat}
\bibfield{author}{\bibinfo{person}{Jintao Xue}, \bibinfo{person}{Zihan Yu}, \bibinfo{person}{Yubo Song}, \bibinfo{person}{Zhongyuan Qin}, \bibinfo{person}{Xin Sun}, {and} \bibinfo{person}{Wen Wang}.} \bibinfo{year}{2023}\natexlab{}.
\newblock \showarticletitle{VulSAT: Source Code Vulnerability Detection Scheme Based on SAT Structure}. In \bibinfo{booktitle}{\emph{2023 8th International Conference on Signal and Image Processing (ICSIP)}}. IEEE, \bibinfo{pages}{639--644}.
\newblock


\bibitem[Yang et~al\mbox{.}(2022)]%
        {yang2022source}
\bibfield{author}{\bibinfo{person}{Hongyu Yang}, \bibinfo{person}{Haiyun Yang}, \bibinfo{person}{Liang Zhang}, {and} \bibinfo{person}{Xiang Cheng}.} \bibinfo{year}{2022}\natexlab{}.
\newblock \showarticletitle{Source Code Vulnerability Detection Using Vulnerability Dependency Representation Graph}. In \bibinfo{booktitle}{\emph{2022 IEEE International Conference on Trust, Security and Privacy in Computing and Communications (TrustCom)}}. IEEE, \bibinfo{pages}{457--464}.
\newblock


\bibitem[Yang et~al\mbox{.}(2024)]%
        {yang2024tensor}
\bibfield{author}{\bibinfo{person}{Jia Yang}, \bibinfo{person}{Ou Ruan}, {and} \bibinfo{person}{JiXin Zhang}.} \bibinfo{year}{2024}\natexlab{}.
\newblock \showarticletitle{Tensor-based gated graph neural network for automatic vulnerability detection in source code}.
\newblock \bibinfo{journal}{\emph{Software Testing, Verification and Reliability}} (\bibinfo{year}{2024}), \bibinfo{pages}{e1867}.
\newblock


\bibitem[Ying et~al\mbox{.}(2019)]%
        {ying2019gnnexplainer}
\bibfield{author}{\bibinfo{person}{Zhitao Ying}, \bibinfo{person}{Dylan Bourgeois}, \bibinfo{person}{Jiaxuan You}, \bibinfo{person}{Marinka Zitnik}, {and} \bibinfo{person}{Jure Leskovec}.} \bibinfo{year}{2019}\natexlab{}.
\newblock \showarticletitle{Gnnexplainer: Generating explanations for graph neural networks}.
\newblock \bibinfo{journal}{\emph{Advances in neural information processing systems}}  \bibinfo{volume}{32} (\bibinfo{year}{2019}).
\newblock


\bibitem[Yuan et~al\mbox{.}(2023)]%
        {yuan2023enhancing}
\bibfield{author}{\bibinfo{person}{Bin Yuan}, \bibinfo{person}{Yifan Lu}, \bibinfo{person}{Yilin Fang}, \bibinfo{person}{Yueming Wu}, \bibinfo{person}{Deqing Zou}, \bibinfo{person}{Zhen Li}, \bibinfo{person}{Zhi Li}, {and} \bibinfo{person}{Hai Jin}.} \bibinfo{year}{2023}\natexlab{}.
\newblock \showarticletitle{Enhancing Deep Learning-based Vulnerability Detection by Building Behavior Graph Model}. In \bibinfo{booktitle}{\emph{2023 IEEE/ACM 45th International Conference on Software Engineering (ICSE)}}. IEEE, \bibinfo{pages}{2262--2274}.
\newblock


\bibitem[Zaazaa and El~Bakkali(2020)]%
        {zaazaa2020dynamic}
\bibfield{author}{\bibinfo{person}{Oualid Zaazaa} {and} \bibinfo{person}{Hanan El~Bakkali}.} \bibinfo{year}{2020}\natexlab{}.
\newblock \showarticletitle{Dynamic vulnerability detection approaches and tools: State of the Art}. In \bibinfo{booktitle}{\emph{2020 Fourth International Conference On Intelligent Computing in Data Sciences (ICDS)}}. IEEE, \bibinfo{pages}{1--6}.
\newblock


\bibitem[Zagane et~al\mbox{.}(2020)]%
        {zagane2020new}
\bibfield{author}{\bibinfo{person}{Mohammed Zagane}, \bibinfo{person}{Mustapha~Kamel Abdi}, {and} \bibinfo{person}{Mamdouh Alenezi}.} \bibinfo{year}{2020}\natexlab{}.
\newblock \showarticletitle{A new approach to locate software vulnerabilities using code metrics}.
\newblock \bibinfo{journal}{\emph{International Journal of Software Innovation (IJSI)}} \bibinfo{volume}{8}, \bibinfo{number}{3} (\bibinfo{year}{2020}), \bibinfo{pages}{82--95}.
\newblock


\bibitem[Zeng et~al\mbox{.}(2020b)]%
        {zeng2020efficient}
\bibfield{author}{\bibinfo{person}{Jingxiang Zeng}, \bibinfo{person}{Xiaofan Nie}, \bibinfo{person}{Liwei Chen}, \bibinfo{person}{Jinfeng Li}, \bibinfo{person}{Gewangzi Du}, {and} \bibinfo{person}{Gang Shi}.} \bibinfo{year}{2020}\natexlab{b}.
\newblock \showarticletitle{An efficient vulnerability extrapolation using similarity of graph kernel of pdgs}. In \bibinfo{booktitle}{\emph{2020 IEEE 19th International Conference on Trust, Security and Privacy in Computing and Communications (TrustCom)}}. IEEE, \bibinfo{pages}{1664--1671}.
\newblock


\bibitem[Zeng et~al\mbox{.}(2020a)]%
        {zeng2020software}
\bibfield{author}{\bibinfo{person}{Peng Zeng}, \bibinfo{person}{Guanjun Lin}, \bibinfo{person}{Lei Pan}, \bibinfo{person}{Yonghang Tai}, {and} \bibinfo{person}{Jun Zhang}.} \bibinfo{year}{2020}\natexlab{a}.
\newblock \showarticletitle{Software vulnerability analysis and discovery using deep learning techniques: A survey}.
\newblock \bibinfo{journal}{\emph{IEEE Access}}  \bibinfo{volume}{8} (\bibinfo{year}{2020}), \bibinfo{pages}{197158--197172}.
\newblock


\bibitem[Zhang et~al\mbox{.}(2023c)]%
        {zhang2023cpvd}
\bibfield{author}{\bibinfo{person}{Chunyong Zhang}, \bibinfo{person}{Bin Liu}, \bibinfo{person}{Yang Xin}, {and} \bibinfo{person}{Liangwei Yao}.} \bibinfo{year}{2023}\natexlab{c}.
\newblock \showarticletitle{CPVD: Cross Project Vulnerability Detection Based On Graph Attention Network And Domain Adaptation}.
\newblock \bibinfo{journal}{\emph{IEEE Transactions on Software Engineering}} (\bibinfo{year}{2023}).
\newblock


\bibitem[Zhang and Xin(2023a)]%
        {zhang2023static}
\bibfield{author}{\bibinfo{person}{Chunyong Zhang} {and} \bibinfo{person}{Yang Xin}.} \bibinfo{year}{2023}\natexlab{a}.
\newblock \showarticletitle{Static vulnerability detection based on class separation}.
\newblock \bibinfo{journal}{\emph{Journal of Systems and Software}}  \bibinfo{volume}{206} (\bibinfo{year}{2023}), \bibinfo{pages}{111832}.
\newblock


\bibitem[Zhang and Xin(2023b)]%
        {zhang2023vulgai}
\bibfield{author}{\bibinfo{person}{Chunyong Zhang} {and} \bibinfo{person}{Yang Xin}.} \bibinfo{year}{2023}\natexlab{b}.
\newblock \showarticletitle{VulGAI: vulnerability detection based on graphs and images}.
\newblock \bibinfo{journal}{\emph{Computers \& Security}}  \bibinfo{volume}{135} (\bibinfo{year}{2023}), \bibinfo{pages}{103501}.
\newblock


\bibitem[Zhang et~al\mbox{.}(2024)]%
        {zhang2024vulnerability}
\bibfield{author}{\bibinfo{person}{Chunyong Zhang}, \bibinfo{person}{Tianxiang Yu}, \bibinfo{person}{Bin Liu}, {and} \bibinfo{person}{Yang Xin}.} \bibinfo{year}{2024}\natexlab{}.
\newblock \showarticletitle{Vulnerability detection based on federated learning}.
\newblock \bibinfo{journal}{\emph{Information and Software Technology}}  \bibinfo{volume}{167} (\bibinfo{year}{2024}), \bibinfo{pages}{107371}.
\newblock


\bibitem[Zhang et~al\mbox{.}(2021)]%
        {zhang2021isvsf}
\bibfield{author}{\bibinfo{person}{Haibin Zhang}, \bibinfo{person}{Yifei Bi}, \bibinfo{person}{Hongzhi Guo}, \bibinfo{person}{Wen Sun}, {and} \bibinfo{person}{Jianpeng Li}.} \bibinfo{year}{2021}\natexlab{}.
\newblock \showarticletitle{ISVSF: Intelligent vulnerability detection against Java via sentence-level pattern exploring}.
\newblock \bibinfo{journal}{\emph{IEEE Systems Journal}} \bibinfo{volume}{16}, \bibinfo{number}{1} (\bibinfo{year}{2021}), \bibinfo{pages}{1032--1043}.
\newblock


\bibitem[Zhang et~al\mbox{.}(2023a)]%
        {zhang2023vulnerability}
\bibfield{author}{\bibinfo{person}{Junwei Zhang}, \bibinfo{person}{Zhongxin Liu}, \bibinfo{person}{Xing Hu}, \bibinfo{person}{Xin Xia}, {and} \bibinfo{person}{Shanping Li}.} \bibinfo{year}{2023}\natexlab{a}.
\newblock \showarticletitle{Vulnerability Detection by Learning from Syntax-Based Execution Paths of Code}.
\newblock \bibinfo{journal}{\emph{IEEE Transactions on Software Engineering}} (\bibinfo{year}{2023}).
\newblock


\bibitem[Zhang et~al\mbox{.}(2023b)]%
        {zhang22023vulnerability}
\bibfield{author}{\bibinfo{person}{Junwei Zhang}, \bibinfo{person}{Zhongxin Liu}, \bibinfo{person}{Xing Hu}, \bibinfo{person}{Xin Xia}, {and} \bibinfo{person}{Shanping Li}.} \bibinfo{year}{2023}\natexlab{b}.
\newblock \showarticletitle{Vulnerability Detection by Learning from Syntax-Based Execution Paths of Code}.
\newblock \bibinfo{journal}{\emph{IEEE Transactions on Software Engineering}} (\bibinfo{year}{2023}).
\newblock


\bibitem[Zhang et~al\mbox{.}(2023d)]%
        {zhang2023vuld}
\bibfield{author}{\bibinfo{person}{Xuejun Zhang}, \bibinfo{person}{Fenghe Zhang}, \bibinfo{person}{Bo Zhao}, \bibinfo{person}{Bo Zhou}, {and} \bibinfo{person}{Boyang Xiao}.} \bibinfo{year}{2023}\natexlab{d}.
\newblock \showarticletitle{VulD-Transformer: Source Code Vulnerability Detection via Transformer}. In \bibinfo{booktitle}{\emph{Proceedings of the 14th Asia-Pacific Symposium on Internetware}}. \bibinfo{pages}{185--193}.
\newblock


\bibitem[Zhang et~al\mbox{.}(2022)]%
        {zhang2022example}
\bibfield{author}{\bibinfo{person}{Ying Zhang}, \bibinfo{person}{Ya Xiao}, \bibinfo{person}{Md~Mahir~Asef Kabir}, \bibinfo{person}{Danfeng Yao}, {and} \bibinfo{person}{Na Meng}.} \bibinfo{year}{2022}\natexlab{}.
\newblock \showarticletitle{Example-based vulnerability detection and repair in java code}. In \bibinfo{booktitle}{\emph{Proceedings of the 30th IEEE/ACM International Conference on Program Comprehension}}. \bibinfo{pages}{190--201}.
\newblock


\bibitem[Zhang et~al\mbox{.}(2023e)]%
        {zhang2023comparing}
\bibfield{author}{\bibinfo{person}{Yuting Zhang}, \bibinfo{person}{Jiahao Zhu}, \bibinfo{person}{Yixin Yang}, \bibinfo{person}{Ming Wen}, {and} \bibinfo{person}{Hai Jin}.} \bibinfo{year}{2023}\natexlab{e}.
\newblock \showarticletitle{Comparing the Performance of Different Code Representations for Learning-based Vulnerability Detection}. In \bibinfo{booktitle}{\emph{Proceedings of the 14th Asia-Pacific Symposium on Internetware}}. \bibinfo{pages}{174--184}.
\newblock


\bibitem[Zheng et~al\mbox{.}(2021a)]%
        {zheng2021vu1spg}
\bibfield{author}{\bibinfo{person}{Weining Zheng}, \bibinfo{person}{Yuan Jiang}, {and} \bibinfo{person}{Xiaohong Su}.} \bibinfo{year}{2021}\natexlab{a}.
\newblock \showarticletitle{Vu1SPG: Vulnerability detection based on slice property graph representation learning}. In \bibinfo{booktitle}{\emph{2021 IEEE 32nd International Symposium on Software Reliability Engineering (ISSRE)}}. IEEE, \bibinfo{pages}{457--467}.
\newblock


\bibitem[Zheng et~al\mbox{.}(2021c)]%
        {zheng2021representation}
\bibfield{author}{\bibinfo{person}{Wei Zheng}, \bibinfo{person}{Abubakar Omari~Abdallah Semasaba}, \bibinfo{person}{Xiaoxue Wu}, \bibinfo{person}{Samuel~Akwasi Agyemang}, \bibinfo{person}{Tao Liu}, {and} \bibinfo{person}{Yuan Ge}.} \bibinfo{year}{2021}\natexlab{c}.
\newblock \showarticletitle{Representation vs. Model: What Matters Most for Source Code Vulnerability Detection}. In \bibinfo{booktitle}{\emph{2021 IEEE International Conference on Software Analysis, Evolution and Reengineering (SANER)}}. IEEE, \bibinfo{pages}{647--653}.
\newblock


\bibitem[Zheng et~al\mbox{.}(2021b)]%
        {zheng2021d2a}
\bibfield{author}{\bibinfo{person}{Yunhui Zheng}, \bibinfo{person}{Saurabh Pujar}, \bibinfo{person}{Burn Lewis}, \bibinfo{person}{Luca Buratti}, \bibinfo{person}{Edward Epstein}, \bibinfo{person}{Bo Yang}, \bibinfo{person}{Jim Laredo}, \bibinfo{person}{Alessandro Morari}, {and} \bibinfo{person}{Zhong Su}.} \bibinfo{year}{2021}\natexlab{b}.
\newblock \showarticletitle{D2a: A dataset built for ai-based vulnerability detection methods using differential analysis}. In \bibinfo{booktitle}{\emph{2021 IEEE/ACM 43rd International Conference on Software Engineering: Software Engineering in Practice (ICSE-SEIP)}}. IEEE, \bibinfo{pages}{111--120}.
\newblock


\bibitem[Zhou et~al\mbox{.}(2022)]%
        {zhou2022new}
\bibfield{author}{\bibinfo{person}{Xin Zhou}, \bibinfo{person}{Jianmin Pang}, \bibinfo{person}{Feng Yue}, \bibinfo{person}{Fudong Liu}, \bibinfo{person}{Jiayu Guo}, \bibinfo{person}{Wenfu Liu}, \bibinfo{person}{Zhihui Song}, \bibinfo{person}{Guoqiang Shu}, \bibinfo{person}{Bing Xia}, {and} \bibinfo{person}{Zheng Shan}.} \bibinfo{year}{2022}\natexlab{}.
\newblock \showarticletitle{A new method of software vulnerability detection based on a quantum neural network}.
\newblock \bibinfo{journal}{\emph{Scientific Reports}} \bibinfo{volume}{12}, \bibinfo{number}{1} (\bibinfo{year}{2022}), \bibinfo{pages}{8053}.
\newblock


\bibitem[Zhou et~al\mbox{.}(2019a)]%
        {devign}
\bibfield{author}{\bibinfo{person}{Yaqin Zhou}, \bibinfo{person}{Shangqing Liu}, \bibinfo{person}{Jingkai Siow}, \bibinfo{person}{Xiaoning Du}, {and} \bibinfo{person}{Yang Liu}.} \bibinfo{year}{2019}\natexlab{a}.
\newblock \showarticletitle{Devign: Effective Vulnerability Identification by Learning Comprehensive Program Semantics via Graph Neural Networks}. In \bibinfo{booktitle}{\emph{Advances in Neural Information Processing Systems}}, \bibfield{editor}{\bibinfo{person}{H.~Wallach}, \bibinfo{person}{H.~Larochelle}, \bibinfo{person}{A.~Beygelzimer}, \bibinfo{person}{F.~d\textquotesingle Alch\'{e}-Buc}, \bibinfo{person}{E.~Fox}, {and} \bibinfo{person}{R.~Garnett}} (Eds.), Vol.~\bibinfo{volume}{32}. \bibinfo{publisher}{Curran Associates, Inc.}
\newblock
\urldef\tempurl%
\url{https://proceedings.neurips.cc/paper/2019/file/49265d2447bc3bbfe9e76306ce40a31f-Paper.pdf}
\showURL{%
\tempurl}


\bibitem[Zhou et~al\mbox{.}(2019b)]%
        {zhou2019devign}
\bibfield{author}{\bibinfo{person}{Yaqin Zhou}, \bibinfo{person}{Shangqing Liu}, \bibinfo{person}{Jingkai Siow}, \bibinfo{person}{Xiaoning Du}, {and} \bibinfo{person}{Yang Liu}.} \bibinfo{year}{2019}\natexlab{b}.
\newblock \showarticletitle{Devign: Effective vulnerability identification by learning comprehensive program semantics via graph neural networks}.
\newblock \bibinfo{journal}{\emph{Advances in neural information processing systems}}  \bibinfo{volume}{32} (\bibinfo{year}{2019}).
\newblock


\bibitem[Zhu et~al\mbox{.}(2022)]%
        {zhu2022application}
\bibfield{author}{\bibinfo{person}{Yuhui Zhu}, \bibinfo{person}{Guanjun Lin}, \bibinfo{person}{Lipeng Song}, {and} \bibinfo{person}{Jun Zhang}.} \bibinfo{year}{2022}\natexlab{}.
\newblock \showarticletitle{The application of neural network for software vulnerability detection: a review}.
\newblock \bibinfo{journal}{\emph{Neural Computing and Applications}} (\bibinfo{year}{2022}), \bibinfo{pages}{1--23}.
\newblock


\bibitem[Zhu et~al\mbox{.}(2023)]%
        {zhu2023application}
\bibfield{author}{\bibinfo{person}{Yuhui Zhu}, \bibinfo{person}{Guanjun Lin}, \bibinfo{person}{Lipeng Song}, {and} \bibinfo{person}{Jun Zhang}.} \bibinfo{year}{2023}\natexlab{}.
\newblock \showarticletitle{The application of neural network for software vulnerability detection: a review}.
\newblock \bibinfo{journal}{\emph{Neural Computing and Applications}} \bibinfo{volume}{35}, \bibinfo{number}{2} (\bibinfo{year}{2023}), \bibinfo{pages}{1279--1301}.
\newblock


\bibitem[Ziems and Wu(2021)]%
        {ziems2021security}
\bibfield{author}{\bibinfo{person}{Noah Ziems} {and} \bibinfo{person}{Shaoen Wu}.} \bibinfo{year}{2021}\natexlab{}.
\newblock \showarticletitle{Security vulnerability detection using deep learning natural language processing}. In \bibinfo{booktitle}{\emph{IEEE INFOCOM 2021-IEEE Conference on Computer Communications Workshops (INFOCOM WKSHPS)}}. IEEE, \bibinfo{pages}{1--6}.
\newblock


\bibitem[Zou et~al\mbox{.}(2019)]%
        {zou2019mu}
\bibfield{author}{\bibinfo{person}{Deqing Zou}, \bibinfo{person}{Sujuan Wang}, \bibinfo{person}{Shouhuai Xu}, \bibinfo{person}{Zhen Li}, {and} \bibinfo{person}{Hai Jin}.} \bibinfo{year}{2019}\natexlab{}.
\newblock \showarticletitle{$\mu$ VulDeePecker: A Deep Learning-Based System for Multiclass Vulnerability Detection}.
\newblock \bibinfo{journal}{\emph{IEEE Transactions on Dependable and Secure Computing}} \bibinfo{volume}{18}, \bibinfo{number}{5} (\bibinfo{year}{2019}), \bibinfo{pages}{2224--2236}.
\newblock


\end{thebibliography}

\end{document}